\documentclass[journal=jctcce,manuscript=article]{achemso}

\usepackage[T1]{fontenc}

\usepackage{geometry}
\usepackage{setspace}

\usepackage{graphicx}
\usepackage{float}
\newfloat{scheme}{htbp}{los}
\floatname{scheme}{Scheme}
\floatname{chart}{Chart}
\newfloat{graph}{htbp}{loh}

\usepackage{chemformula} 
\usepackage[version = 4]{mhchem} 

\usepackage{rotating}
\usepackage{enumerate}
\usepackage{multirow}
\usepackage{booktabs}

\author{Xiaoli Wang}
\affiliation{Qingdao Institute for Theoretical and Computational Sciences, Center for Optics Research and Engineering, Shandong University, Qingdao, Shandong, P. R. China}

\author{Xingwen Wang}
\affiliation{Qingdao Institute for Theoretical and Computational Sciences, Center for Optics Research and Engineering, Shandong University, Qingdao, Shandong, P. R. China}

\author{Zikuan Wang}\email{wzkchem5@sdu.edu.cn}
\affiliation{Qingdao Institute for Theoretical and Computational Sciences, Center for Optics Research and Engineering, Shandong University, Qingdao, Shandong, P. R. China}

\author{Wenjian Liu}\email{liuwj@sdu.edu.cn}
\affiliation{Qingdao Institute for Theoretical and Computational Sciences, Center for Optics Research and Engineering, Shandong University, Qingdao, Shandong, P. R. China}

\title{Analytic first-order non-adiabatic coupling matrix elements of spin-adapted open-shell time-dependent density functional theory}

\begin{document}

\maketitle

\begin{abstract}
While spin-adapted time-dependent density functional theory (TDDFT) approaches significantly improve the excitation energies and gradients of open-shell molecules, the effect of spin-adaptation on non-adiabatic coupling matrix elements (NACMEs) remains unknown for spin-conserving excitations. In this article, we report the derivation, implementation and benchmark studies of the ground state-excited state and excited state-excited state NACMEs of our spin-adapted TDDFT method, X-TDDFT; to our best knowledge, this represents the first implementation of the analytic NACMEs of a spin-adapted TDDFT method. Similar to the X-TDDFT analytic gradients, X-TDDFT NACMEs can be easily implemented on top of an existing U-TDDFT NACME implementation taking into account the restricted open-shell Kohn-Sham (ROKS) reference and the implicit involvement of doubly excited determinants, with acceptable computational overhead. Benchmark calculations reveal that X-TDDFT reduces the error of U-TDDFT NACMEs by 1/3$\sim$2/3 (referenced against high-level multireference NACMEs), which leads to large corrections of internal conversion rates (up to two orders of magnitude). In particular, for copper(II) porphyrin, X-TDDFT leads to qualitative revisions of the relative importance of the excited state relaxation pathways, as well as the substituent effects of the internal conversion (IC) rates, suggesting that the error of U-TDDFT NACMEs is not only large but also unsystematic. It is therefore expected that X-TDDFT NACMEs will prove useful in the photophysics/photochemistry studies of open-shell systems such as radicals and transition metal complexes.
\end{abstract}

\maketitle

\section{Introduction}\label{Introduction}

The excited state processes of organic radicals and transition metal complexes, including but not limited to photo- and electroluminescence, photocatalysis and photodegradation, have received considerable interest in recent years.\cite{2022-radical-acceptors,wenger2025radicalphotocat,lin2025TMphotocat} Although frequently possessing complex electronic structures, the sheer sizes of these molecules often necessitate the use of single-reference methods such as time-dependent density functional theory (TDDFT),\cite{tddft-Runge-Gross, tddft-Casida1995} instead of more elaborate approaches such as multireference perturbation theory (MRPT).\cite{MRPT,SDSRev} To accurately describe the electronic structures of the excited states, however, doubly excited determinants are frequently needed, which not only increases the computational cost, but also poses significant difficulties to method development since adiabatic TDDFT only includes single excitations.

Fortunately, there is one kind of double excitations that has a clear physical origin, and their coefficients can be fixed in advance. Starting from a high-spin, restricted open-shell Kohn-Sham (ROKS) reference state $\Psi_0 = |i\bar{i}j\bar{j}\cdots tu\cdots \rangle$ with $2S$ spin-parallel open-shell electrons (here and in the rest of the manuscript, we use $i,j,\ldots$, $t,u,\ldots$ and $a,b,\ldots$ to denote doubly occupied, singly occupied and virtual orbitals, respectively; molecular orbitals (MOs) without and with overbar respectively represent $\alpha$ and $\beta$ orbitals), one can construct the so-called CO-type excitations $\Psi_{\bar{i}}^{\bar{t}}$, the OV-type excitations $\Psi_{t}^{a}$, as well as the CV-type excitations $\Psi_{i}^{a}$ (CV($\alpha\alpha$)) and $\Psi_{\bar{i}}^{\bar{a}}$ (CV($\beta\beta$)) (where C, O and V denote closed-shell, open-shell and vacant-shell, respectively). While the former two classes of excitations are automatically spin-adapted, it is well-known that the CV excitations are not only not spin-adapted, but do not span a spin-complete space.\cite{XTDDFT1,XTDDFT2,XTDDFT3,XTDDFTgrad} Specifically, the in-phase combination of $\Psi_{i}^{a}$ and $\Psi_{\bar{i}}^{\bar{a}}$ (denoted the CV(0) excitation),
\begin{equation}
S_{ai}^\dag(0,0) \Psi_0 = \frac{1}{\sqrt{2}} (\Psi_i^{a} + \Psi_{\bar{i}}^{\bar{a}}), \quad S_{pq}^\dag(0,0) = \frac{1}{\sqrt{2}} (a_p^\dag a_q + a_{\bar{p}}^\dag a_{\bar{q}}), \label{CV0}
\end{equation}
is properly spin-adapted, but the corresponding out-of-phase combination (the CV(1) excitation),
\begin{equation}
T_{ai}^\dag(1,0) \Psi_0 = \frac{1}{\sqrt{2}} (\Psi_i^{a} - \Psi_{\bar{i}}^{\bar{a}}), \quad T_{pq}^\dag(1,0) = \frac{1}{\sqrt{2}} (a_p^\dag a_q - a_{\bar{p}}^\dag a_{\bar{q}}), \label{CV1}
\end{equation}
is not. Only by mixing with the double excitations $\Psi_{\bar{i}t}^{\bar{t}a}$ can Eq.~\eqref{CV1} be separated into a spin-$S$ state ($\tilde{T}_{ai}^\dag(1,0) \Psi_0$, hereafter called the spin-adapted CV(1) excitation) and a spin-$(S+1)$ state ($\bar{T}_{ai}^\dag(1,0) \Psi_0$):
\begin{eqnarray}
\tilde{T}_{ai}^\dag(1,0) \Psi_0 & = & \sqrt{\frac{S}{2(S+1)}} \left(\Psi_i^{a} -\Psi_{\bar{i}}^{\bar{a}}  - \frac{1}{S}\sum_t^{2S} \Psi_{\bar{i}t}^{\bar{t}a}\right), \label{Psi1} \\{}
\bar{T}_{ai}^\dag(1,0) \Psi_0 & = & \frac{1}{\sqrt{2(S+1)}} \left(\Psi_i^{a} - \Psi_{\bar{i}}^{\bar{a}} + \sum_t^{2S} \Psi_{\bar{i}t}^{\bar{t}a}\right). \label{Psi2}
\end{eqnarray}
Importantly, the coefficients of the double excitations are known in advance, so that the double excitations do not need to be treated as independent degrees of freedom during the calculations. In the X-TDDFT method, we therefore simply substitute the CV($\alpha\alpha$) and CV($\beta\beta$) excitations by the CV(0) and spin-adapted CV(1) excitations.\cite{XTDDFT3} This does not change the dimension of the Casida equation to be solved, but merely introduces corrections to the Casida equation, which renders X-TDDFT as cheap as the common U-TDDFT method. Note that although X-TDDFT has recently been extended to spin flip-down excitations,\cite{XSF-TDA} in this manuscript we only discuss the spin-conserving variant of X-TDDFT.

Recently, we have implemented the analytic gradients of the X-TDDFT method,\cite{XTDDFTgrad} which has for the first time enabled the TDDFT geometry optimization of large open-shell molecules in a fully spin-adapted framework. This has sparked the photophysics and photochemistry studies of numerous organic radicals and transition metal complexes, which previously often had to rely on the inaccurate U-TDDFT method. One example is our work on the doublet-quartet thermally activated delayed fluorescence (TADF) of the unsubstituted copper(II) porphyrin, CuP.\cite{wang2023CuP} However, a consistent framework of spin-conserving, spin-adapted TDDFT non-adiabatic coupling matrix element (NACME) theory is still missing, despite that rigorous U-TDDFT NACME theories have been developed by us more than a decade ago.\cite{nacme1-2014,nacme2-2014,nacme3-2021} NACMEs, defined as the matrix elements of the nuclear derivative operator between two electronic states of the same system (where $\xi$ is any of the $3N_{\mathrm{atoms}}$ nuclear coordinates of the system, and $\Psi_I~(I>0)$ is the $I$-th excited state)
\begin{equation}
g_{IJ}^{\xi} = \langle \Psi_I |\frac{d}{d\xi}| \Psi_J \rangle, \label{NACME}
\end{equation}
are related to the extent of breakdown of the Born-Oppenheimer approximation, and are indispensable for simulating internal conversion (IC) processes, no matter via non-adiabatic molecular dynamics simulations,\cite{Mai2020photochem,cui2023namd} harmonic Fermi's Golden Rule (FGR) approaches,\cite{shuai2020,ESD2018} or other methods. While one can calculate IC rates from U-TDDFT NACMEs together with X-TDDFT energies, geometries and Hessians (as we have done in the study of CuP\cite{wang2023CuP}), so as to hopefully get more accurate results than a pure U-TDDFT IC rate calculation, the error of the U-TDDFT NACMEs can potentially be large. To our best knowledge, the closest literature precedent to an analytic, spin-adapted TDDFT NACME theory is the SF-XCIS NACMEs by Thiel and coworkers for semiempirical Hamiltonians.\cite{SFXCIS_NACME} While it can in principle be extended to TDDFT calculations by mixing in DFT exchange and correlation, it is specific to spin flip-down excitations and are not suitable for the excited states of radicals or high-spin transition metal complexes. An efficient numerical NACME implementation of the MRSF-TDDFT method, an approximately spin-adapted spin flip-down TDDFT method, has recently been reported by Filatov and Choi et al.\cite{MRSFTDDFT_NACME} Still another related work is the analytic NACMEs of the TDDFT-1D method,\cite{TDDFT1D_NACME} which adds a single double excitation to the Casida equation. However, the double excitation is of the form $\Psi_{i\bar{i}}^{a\bar{a}}$ and therefore cannot be used to spin-adapt the CV(1) excitations.

In this work, we derive the analytic NACMEs of the X-TDDFT method (which is, to our best knowledge, the first analytic NACME implementation of a rigorously spin-adapted TDDFT method) using our equation-of-motion (EOM) approach of NACMEs,\cite{nacme1-2014,nacme2-2014,nacme3-2021} by replacing the usual random phase approximation (RPA) excitation and de-excitation operators with their spin-adapted counterparts. This yields corrections on top of U-TDDFT NACMEs that resemble the extra terms of X-TDDFT gradient relative to U-TDDFT gradient,\cite{XTDDFTgrad} plus a correction unique to X-TDDFT NACMEs owing to the difference of the spin-adapted CV(1) excitation (Eq.~\eqref{Psi1}) with the U-TDDFT CV(1) excitation (Eq.~\eqref{CV1}). Apart from these corrections, X-TDDFT NACME calculations use X-TDDFT excitation energies and excitation vectors, whose difference with their U-TDDFT counterparts can also yield corrections to the NACMEs. We show that, in some cases, X-TDDFT NACME vectors can be almost one order of magnitude shorter than the U-TDDFT ones, leading to a reduction of IC rates by almost two orders of magnitude. In particular, while our previous study\cite{wang2023CuP} (which used U-TDDFT NACMEs) predicted that one of the dominant relaxation pathways of the luminescent, ligand-centered excited state of the unsubstituted copper(II) porphyrin is via direct IC to the ground state, X-TDDFT predicts that the direct route is negligible, and an indirect route via ligand field states prevails. Therefore, our work shows that spin-adaptation of TDDFT is important not only for energies and gradients, but also (and perhaps even more) for NACMEs and therefore IC rates.

The paper is organized as follows. We first briefly review the theories of X-TDDFT and its analytic gradient (which bears close resemblance with X-TDDFT NACME theory) in Sec.~\ref{sec:XTDDFT}. Then, in Sec.~\ref{sec:XTDDFTNACME} we derive the working equations of the X-TDDFT NACMEs, followed by benchmark studies of small and medium-sized open-shell molecules (Sec.~\ref{sec:benchmark}). The paper is concluded by Sec.~\ref{sec:conclusion}.

\section{Recapitulation of X-TDDFT and its analytic gradient} \label{sec:XTDDFT}

The derivation of X-TDDFT consists of first deriving the spin-adapted (SA)-RPA method (also known as the X-RPA method), taking the difference of the resulting Casida $\mathbf{A}$ matrix with the usual U-RPA $\mathbf{A}$ matrix, and adding it onto the U-TDDFT $\mathbf{A}$ matrix. The X-TDDFT Casida equation therefore has the simple form\cite{XTDDFT3}
\begin{equation}
\left(\begin{array}{cc} \mathbf{A}+\Delta\mathbf{A} & \mathbf{B} \\ \mathbf{B} & \mathbf{A}+\Delta\mathbf{A} \end{array}\right) \left(\begin{array}{c} \mathbf{X}_I \\ \mathbf{Y}_I \end{array}\right) =
\omega_I^{\mathrm{X}} \left(\begin{array}{cc} \mathbf{I} & \mathbf{0} \\ \mathbf{0} & -\mathbf{I} \end{array}\right) \left(\begin{array}{c} \mathbf{X}_I \\ \mathbf{Y}_I \end{array}\right), \label{Casida}
\end{equation}
\begin{equation}
(\begin{array}{cc} \mathbf{X}_I^T & \mathbf{Y}_I^T \end{array}) \left(\begin{array}{cc} \mathbf{I} & \mathbf{0} \\ \mathbf{0} & -\mathbf{I} \end{array}\right) \left(\begin{array}{c} \mathbf{X}_J \\ \mathbf{Y}_J \end{array}\right) = \delta_{IJ}, \label{XXYYI}
\end{equation}
\begin{equation}
\Delta\mathbf{A} = \mathbf{A}^{\mathrm{SA-RPA}}-\mathbf{A}^{\mathrm{U-RPA}},
\end{equation}
where the U-TDDFT $\mathbf{A}$ and $\mathbf{B}$ matrices are given, as usual, by
\begin{eqnarray}
A_{ia\sigma,jb\tau} & = & \delta_{\sigma\tau}(\delta_{ij}F_{ab\sigma} - \delta_{ab}F_{ji\sigma}) + K_{ia\sigma{}jb\tau}, \label{UTD-Amat}\\
B_{ia\sigma,jb\tau} & = & K_{ia\sigma{}bj\tau}, \\
K_{pq\sigma,rs\tau} & = & (pq\sigma|sr\tau) + f_{pq\sigma{}sr\tau}^{\mathrm{xc}}[\rho] - c_{\mathrm{x}}\delta_{\sigma\tau}(pr\sigma|sq\sigma).\label{Kmat}
\end{eqnarray}
Here $\sigma,\tau,\ldots\in\{\alpha,\beta\}$ are spin indices (e.g. $a\sigma$ denotes a spin-$\sigma$ virtual orbital, and $ab\sigma$ is a shorthand for $a\sigma b\sigma$), $\mathbf{F}$ is the Fock matrix, $\mathbf{f}^{\mathrm{xc}}[\rho]$ is the exchange-correlation (XC) kernel, and $c_{\mathrm{x}}$ is the proportion of exact exchange. Although Eq.~\eqref{Kmat} assumes a global hybrid functional, the formulas of pure functionals and range-separated hybrids can be obtained by setting $c_{\mathrm{x}}=0$ and replacing a portion of $(pr\sigma|sq\sigma)$ by short-range/long-range exact exchange, respectively. The U-RPA method is given by setting $\mathbf{f}^{\mathrm{xc}}[\rho]=0$ and $c_{\mathrm{x}}=1$; the SA-RPA method will be detailed below.

Both the U-RPA and SA-RPA methods can be derived from EOM theory. In EOM theory, one defines an excitation operator $O_I^\dag$ that transforms the ground state ($|\Psi_0 \rangle$) into the $I$-th excited state:\cite{XTDDFT1}
\begin{equation}
| \Psi_I \rangle = O_I^\dag | \Psi_0 \rangle. \label{OIdag}
\end{equation}
One also needs the killer condition in the derivations:
\begin{equation}
O_I | \Psi_0 \rangle = 0, \label{killer}
\end{equation}
although the killer condition does not rigorously apply to the excitation operators that will be defined below. In U-RPA, $| \Psi_0 \rangle$ is the unrestricted Hartree-Fock (UHF) reference, and $O_I^\dag$ is parameterized by the excitation vectors $\mathbf{X}_I$ and $\mathbf{Y}_I$ as:
\begin{equation}
O_I^\dag = \sum_{ia\sigma} \left( (\mathbf{X}_I)_{ia\sigma} \hat{a}_{a\sigma}^\dag \hat{a}_{i\sigma} - (\mathbf{Y}_I)_{ia\sigma} \hat{a}_{i\sigma}^\dag \hat{a}_{a\sigma} \right). \label{OI}
\end{equation}
SA-RPA, on the other hand, starts from a restricted open-shell Hartree-Fock (ROHF) reference, and replaces all CV excitations of $O_I^\dag$ by their spin-adapted analogs $S_{ai}^\dag(0,0)$ and $\tilde{T}_{ai}^\dag(1,0)$, while keeping the CO and OV excitations unchanged:
\begin{eqnarray}
O_I^{\mathrm{SA},\dag} & = & \sum_{it} \left( (\mathbf{X}_I)_{\bar{i}\bar{t}} \hat{a}_{\bar{t}}^\dag \hat{a}_{\bar{i}} - (\mathbf{Y}_I)_{\bar{i}\bar{t}} \hat{a}_{\bar{i}}^\dag \hat{a}_{\bar{t}} \right) + \sum_{ta} \left( (\mathbf{X}_I)_{ta} \hat{a}_{a}^\dag \hat{a}_{t} - (\mathbf{Y}_I)_{ta} \hat{a}_{t}^\dag \hat{a}_{a} \right) \nonumber\\
& + & \sum_{ia} \left( (\mathbf{X}_I^{\mathrm{CV}(0)})_{ia} S_{ai}^\dag(0,0) - (\mathbf{Y}_I^{\mathrm{CV}(0)})_{ia} S_{ai}(0,0) \right. \nonumber\\
& + & \left. (\mathbf{X}_I^{\mathrm{CV}(1)})_{ia} \tilde{T}_{ai}^\dag(1,0) - (\mathbf{Y}_I^{\mathrm{CV}(1)})_{ia} \tilde{T}_{ai}(1,0) \right), \label{OI_SA}
\end{eqnarray}
where
\begin{equation}
\tilde{T}_{pq}^\dag(1,0) = \sqrt{\frac{S}{2(S+1)}} \left(a_p^\dag a_q - a_{\bar{p}}^\dag a_{\bar{q}} - \frac{1}{S}\sum_t^{2S}a_p^\dag a_{t} a_{\bar{t}}^\dag a_{\bar{q}} \right),
\end{equation}
and the ``spin-tensor basis'' excitation vectors are given by the usual ``spin-orbital basis'' excitation vectors as
\begin{eqnarray}
(\mathbf{X}_I^{\mathrm{CV}(0)})_{ia} & = & \frac{1}{\sqrt{2}} ((\mathbf{X}_I)_{ia} + (\mathbf{X}_I)_{\bar{i}\bar{a}}), \\
(\mathbf{X}_I^{\mathrm{CV}(1)})_{ia} & = & \frac{1}{\sqrt{2}} ((\mathbf{X}_I)_{ia} - (\mathbf{X}_I)_{\bar{i}\bar{a}}),
\end{eqnarray}
and similar for $\mathbf{Y}_I^{\mathrm{CV}(0)}$ and $\mathbf{Y}_I^{\mathrm{CV}(1)}$.

The difference of Eq.~\eqref{OI} and Eq.~\eqref{OI_SA} can be understood in two ways. On one hand, Eq.~\eqref{OI} can be transformed to the spin-tensor basis:
\begin{eqnarray}
O_I^{\dag} & = & \underbrace{\sum_{it} \left( (\mathbf{X}_I)_{\bar{i}\bar{t}} \hat{a}_{\bar{t}}^\dag \hat{a}_{\bar{i}} - (\mathbf{Y}_I)_{\bar{i}\bar{t}} \hat{a}_{\bar{i}}^\dag \hat{a}_{\bar{t}} \right)}_{O_I^{\mathrm{CO},\dag}} + \underbrace{\sum_{ta} \left( (\mathbf{X}_I)_{ta} \hat{a}_{a}^\dag \hat{a}_{t} - (\mathbf{Y}_I)_{ta} \hat{a}_{t}^\dag \hat{a}_{a} \right)}_{O_I^{\mathrm{OV},\dag}} \nonumber\\
& + & \underbrace{\sum_{ia} \left( (\mathbf{X}_I^{\mathrm{CV}(0)})_{ia} S_{ai}^\dag(0,0) - (\mathbf{Y}_I^{\mathrm{CV}(0)})_{ia} S_{ai}(0,0) \right)}_{O_I^{\mathrm{CV(0)},\dag}} \nonumber\\
& + & \underbrace{\sum_{ia} \left( (\mathbf{X}_I^{\mathrm{CV}(1)})_{ia} T_{ai}^\dag(1,0) - (\mathbf{Y}_I^{\mathrm{CV}(1)})_{ia} T_{ai}(1,0) \right)}_{O_I^{\mathrm{CV(1)},\dag}} , \label{OI_spintensor}
\end{eqnarray}
which differs from Eq.~\eqref{OI_SA} only in that the CV(1) excitation operator $T_{ai}^\dag(1,0)$ is not spin-adapted. On the other hand, we can also convert Eq.~\eqref{OI_SA} to the spin-orbital basis, and by comparing with Eq.~\eqref{OI} we obtain
\begin{eqnarray}
O_I^{\mathrm{SA},\dag} & = & O_I^{\dag} + \Delta O_I^{\dag}, \\
\Delta O_I^{\dag} & = & \frac{1}{2}\sum_{ia} \left( \left(\left(\sqrt{\frac{S}{S+1}}-1\right) (a_a^\dag a_i - a_{\bar{a}}^\dag a_{\bar{i}}) - \frac{1}{\sqrt{S(S+1)}}\sum_t^{2S}a_a^\dag a_{t} a_{\bar{t}}^\dag a_{\bar{i}}\right)((\mathbf{X}_I)_{ia} - (\mathbf{X}_I)_{\bar{i}\bar{a}})\right. \nonumber\\
& - & \left. \left(\left(\sqrt{\frac{S}{S+1}}-1\right) (a_i^\dag a_a - a_{\bar{i}}^\dag a_{\bar{a}}) - \frac{1}{\sqrt{S(S+1)}}\sum_t^{2S}a_{\bar{i}}^\dag a_{\bar{t}} a_{t}^\dag a_a\right)((\mathbf{Y}_I)_{ia} - (\mathbf{Y}_I)_{\bar{i}\bar{a}}) \right). \label{DeltaOI_SA}
\end{eqnarray}
As we shall see, $\Delta O_I^{\dag}$ gives rise to most of the extra terms of X-TDDFT NACMEs compared to U-TDDFT NACMEs, apart from those terms that arise due to the use of an ROKS reference.

The working equations of of U-RPA and SA-RPA then follow from the EOM (with $H$ being the full Hamiltonian):
\begin{equation}
\langle \Psi_0 | \frac{1}{2} ( [[ \delta O_I, H], O_I^\dag] + [\delta O_I, [ H, O_I^\dag ] ]) | \Psi_0 \rangle = \omega_I \langle \Psi_0 | [\delta O_I, O_I^\dag] | \Psi_0 \rangle, \label{Casida_EOM}
\end{equation}
for all possible variations of $O_I$, i.e.~$\delta O_I$. We will skip the subsequent derivations\cite{XTDDFT1} but merely point out that the resulting theory (called S-RPA) contains corrections to the CV(0)-CV(1), CV(1)-CV(1), CO-CV(1) and OV-CV(1) blocks of the Casida $\mathbf{A}$ and $\mathbf{B}$ matrices. However, only the corrections to the CV(0)-CV(1) and CV(1)-CV(1) blocks of $\mathbf{A}$ contribute significantly to the excitation energies. Therefore, in SA-RPA (and X-TDDFT), it was decided to only apply corrections to these two blocks of the U-RPA (U-TDDFT) $\mathbf{A}$ matrix, yielding
\begin{equation}\label{XRPAcorr1}
\Delta A_{ia\alpha,jb\alpha} = \left(1-\sqrt{\frac{S+1}{S}}+\frac{1}{2S}\right)\delta_{ij}F_{ab}^{S} + \left(-1+\sqrt{\frac{S+1}{S}}+\frac{1}{2S}\right)\delta_{ab}F_{ij}^{S},
\end{equation}
\begin{equation} \label{XRPAcorr2}
\Delta A_{ia\alpha,jb\beta}= -\frac{1}{2S}(\delta_{ij}F_{ab}^{S} + \delta_{ab}F_{ij}^{S}),
\end{equation}
\begin{equation}\label{XRPAcorr3}
\Delta A_{ia\beta,jb\beta} = \left(-1+\sqrt{\frac{S+1}{S}}+\frac{1}{2S}\right)\delta_{ij}F_{ab}^{S} + \left(1-\sqrt{\frac{S+1}{S}}+\frac{1}{2S}\right)\delta_{ab}F_{ij}^{S},
\end{equation}
with the ROHF polarization Fock matrix
\begin{equation}
F_{pq}^{S} = \frac{1}{2}(F_{pq\beta}^{\mathrm{ROHF}}-F_{pq\alpha}^{\mathrm{ROHF}}) = \frac{1}{2}\sum_t^{2S} (pt|tq). \label{FS}
\end{equation}
Importantly, the ROHF Fock matrix (instead of the ROKS Fock matrix) is used in Eq.~\eqref{FS} regardless of the actual density functional used.

Like the case of U-TDDFT,\cite{FurcheTDDFTgrad} the straightforward derivation of the X-TDDFT analytic gradient is hampered by the implicit dependence of the excitation energy $\omega^{\mathrm{X}}$ on the nuclear coordinates, via the TDDFT excitation vectors $\mathbf{X}_I,\mathbf{Y}_I$ and the MO coefficients $\mathbf{C}$.\cite{XTDDFTgrad}
The standard solution is to define a Lagrangian (where $\mathbf{S}$ is the atomic orbital (AO) overlap matrix, and $\mu,\nu,\ldots$ denote AOs; $\Omega_I,\mathbf{Z}_I,\mathbf{W}_I,\mathbf{\zeta}_I,\mathbf{\Lambda}_I$ are Lagrange multipliers)
\begin{eqnarray}
& & \mathcal{L}_I^{\mathrm{X-TDDFT}} = \mathcal{L}_I^{\mathrm{U-TDDFT/ROKS}} + \mathbf{X}_I^T\Delta\mathbf{A}\mathbf{X}_I + \mathbf{Y}_I^T\Delta\mathbf{A}\mathbf{Y}_I, \\
& & \mathcal{L}_I^{\mathrm{U-TDDFT/ROKS}}[\xi,\mathbf{X}_I,\mathbf{Y}_I,\mathbf{C},\Omega_I,\mathbf{Z}_I,\mathbf{W}_I,\mathbf{\zeta}_I,\mathbf{\Lambda}_I] \nonumber\\
& = & \mathcal{L}_I^{\mathrm{U-TDDFT}}[\xi,\mathbf{X}_I,\mathbf{Y}_I,\mathbf{C},\Omega_I,\mathbf{Z}_I,\mathbf{W}_I] + \sum_{ia}(\mathbf{\zeta}_I)_{ia}((\mathbf{Z}_I)_{ia}-(\mathbf{Z}_I)_{\bar{i}\bar{a}}) \nonumber\\
& & + \sum_{\mu{}p}(\mathbf{\Lambda}_I)_{\mu{}p}(C_{\mu{}p\alpha}-C_{\mu{}p\beta}), \\ \label{L_ROKS}
& & \mathcal{L}_I^{\mathrm{U-TDDFT}}[\xi,\mathbf{X}_I,\mathbf{Y}_I,\mathbf{C},\Omega_I,\mathbf{Z}_I,\mathbf{W}_I] \nonumber\\
& = & (\begin{array}{cc}\mathbf{X}_I^T & \mathbf{Y}_I^T \end{array}) \left(\begin{array}{cc}\mathbf{A} &\mathbf{B} \\ \mathbf{B} &\mathbf{A} \end{array}\right) \left(\begin{array}{c}\mathbf{X}_I \\ \mathbf{Y}_I \end{array}\right) - \Omega\left((\begin{array}{cc} \mathbf{X}_I^T & \mathbf{Y}_I^T \end{array}) \left(\begin{array}{cc}\mathbf{I} & \mathbf{0} \\ \mathbf{0} & -\mathbf{I} \end{array}\right) \left(\begin{array}{c} \mathbf{X}_I \\ \mathbf{Y}_I \end{array}\right) - 1\right) \nonumber\\
& & + \sum_{ia\sigma}(\mathbf{Z}_I)_{ia\sigma}F_{ia\sigma} - \sum_{pq\sigma,p\le q}(\mathbf{W}_I)_{pq\sigma}(S_{pq\sigma}-\delta_{pq}), \label{Lagrangian}
\end{eqnarray}
and requiring it to be stationary with respect to all its parameters. Here ``U-TDDFT/ROKS'' denotes U-TDDFT calculations upon ROKS references (i.e.~where the ground state is spin-adapted but no RPA correction is added). Most of these stationary conditions merely implement known conditions such as the Casida equation, the Brillouin conditions, the orthonormality of MOs and (for X-TDDFT) the equality of the $\alpha$ and $\beta$ MO coefficients. The only non-trivial condition is the stationarity of $\mathcal{L}_I^{\mathrm{X-TDDFT}}$ (or $\mathcal{L}_I^{\mathrm{U-TDDFT/ROKS}}$/$\mathcal{L}_I^{\mathrm{U-TDDFT}}$) with respect to $\mathbf{C}$, which is usually recast into the equivalent form
\begin{equation}
\sum_\mu \frac{\partial \mathcal{L}}{\partial C_{\mu{}p\sigma}}C_{\mu{}q\sigma} = 0, \quad \forall p,q,\sigma. \label{L_CC}
\end{equation}
The occupied-virtual and the virtual-occupied blocks of Eq.~\eqref{L_CC} yield the Z-vector equation, from which $\mathbf{Z}_I$ can be solved; plugging in $\mathbf{Z}_I$ back to Eq.~\eqref{L_CC} yields $\mathbf{W}_I$. Finally, the gradient of the excitation energy $\omega_I^{\mathrm{X}}$ is given by
\begin{equation}
\frac{d \omega^{\mathrm{X}}_I}{d \xi} = \frac{d \mathcal{L}_I^{\mathrm{X-TDDFT}}}{d \xi} \equiv \frac{\partial \mathcal{L}_I^{\mathrm{X-TDDFT}}}{\partial \xi} = \frac{\partial \mathcal{L}_I^{\mathrm{U-TDDFT}}}{\partial \xi} + \mathbf{X}_I^T\frac{\partial \Delta\mathbf{A}}{\partial \xi}\mathbf{X}_I + \mathbf{Y}_I^T\frac{\partial \Delta\mathbf{A}}{\partial \xi}\mathbf{Y}_I.
\end{equation}
Different from $\omega^{\mathrm{X}}_I$ which is an implicit function of $\xi$ (due to the implicit $\xi$-dependence of $\mathbf{X}_I,\mathbf{Y}_I$ and $\mathbf{C}$), $\mathcal{L}_I^{\mathrm{X-TDDFT}}$ is a fully explicit function of $\xi$, because in the definition of $\mathcal{L}_I^{\mathrm{X-TDDFT}}$, $\mathbf{X}_I,\mathbf{Y}_I$ and $\mathbf{C}$ are treated as free variables just as $\xi$ is. Therefore, while the total derivative of $\omega^{\mathrm{X}}_I$ is cumbersome to evaluate because of contributions from, e.g., $\frac{\partial \omega^{\mathrm{X}}_I}{\partial C_{\mu p\sigma}}$, the total derivative of $\mathcal{L}_I^{\mathrm{X-TDDFT}}$ ($\frac{d \mathcal{L}_I^{\mathrm{X-TDDFT}}}{d \xi}$) is trivially equal to the corresponding partial derivative $\frac{\partial \mathcal{L}_I^{\mathrm{X-TDDFT}}}{\partial \xi}$, greatly simplifying its evaluation. The U-TDDFT gradient is similarly equal to $\frac{\partial \mathcal{L}_I^{\mathrm{U-TDDFT}}}{\partial \xi}$ and is given by
\begin{eqnarray}
\frac{\partial \mathcal{L}_I^{\mathrm{U-TDDFT}}}{\partial \xi} & = & (\begin{array}{cc}\mathbf{X}_I^T & \mathbf{Y}_I^T \end{array}) \left(\frac{\partial}{\partial \xi} \left(\begin{array}{cc}\mathbf{A} &\mathbf{B} \\ \mathbf{B} &\mathbf{A}\end{array}\right) \right) \left(\begin{array}{c}\mathbf{X}_I \\ \mathbf{Y}_I \end{array}\right) \nonumber\\
& & + \sum_{ia\sigma}Z_{ia\sigma}\frac{\partial F_{ia\sigma}}{\partial \xi} - \sum_{pq\sigma, p\le q} W_{pq\sigma} \frac{\partial S_{pq\sigma}}{\partial \xi} . \label{dLUdxi}
\end{eqnarray}

\section{X-TDDFT NACME} \label{sec:XTDDFTNACME}

Like the analytic gradient of the excitation energy of state $I$, which in wavefunction theory reads
\begin{equation}
\frac{d \omega_I}{d \xi} = \frac{d}{d \xi}\left( \langle \Psi_I | H | \Psi_I \rangle - \langle \Psi_0 | H | \Psi_0 \rangle \right),
\end{equation}
the definition of NACMEs (Eq.~\eqref{NACME}) also involves the total derivative operator $\frac{d}{d \xi}$, albeit sandwiched between a bra state and a ket state. Therefore, the theory and implementation of NACMEs at a given level of theory are very similar to the analytic gradients of the same level of theory. In particular, the Lagrangian formalism is required for properly taking into account the contributions due to the nuclear dependence of the basis set, commonly known as the Pulay term.\cite{nacme3-2021} Similar to the derivation of X-TDDFT itself, here we will first derive the SA-RPA NACMEs from EOM theory; the X-TDDFT NACMEs are then given by replacing the U-RPA expressions of $\mathbf{A}$ and $\mathbf{B}$ by their U-TDDFT analogs.

\subsection{EOM theory of X-TDDFT NACMEs}

The point of departure is to reformulate Eq.~\eqref{NACME} in the EOM framework, to get rid of excited state wavefunctions (which do not exist in TDDFT).\cite{nacme1-2014} Instead of the simple-minded substitution of Eq.~\eqref{OIdag} into Eq.~\eqref{NACME}, which will lead to a theory with no contribution from the $\mathbf{Y}$ amplitudes:
\begin{equation}
g_{0I}^\xi = \langle \Psi_0 | \frac{d}{d \xi} O_I^\dag | \Psi_0 \rangle, \label{operator_EOM_0I}
\end{equation}
\begin{equation}
g_{IJ}^\xi = \langle \Psi_0 | O_I \frac{d}{d \xi} O_J^\dag | \Psi_0 \rangle, \label{operator_EOM_IJ}
\end{equation}
we invoke the killer condition Eq.~\eqref{killer} to convert $\frac{d}{d \xi} O_I^\dag$ into a commutator, and $O_I \frac{d}{d \xi} O_J^\dag$ into a double commutator, as is customary in EOM theories:
\begin{eqnarray}
g_{0I}^\xi & = & \langle \Psi_0 | [ \frac{d}{d \xi}, O_I^\dag ] | \Psi_0 \rangle, \label{operator_EOM_0I_commutator} \\
g_{IJ}^\xi & = & \langle \Psi_0 | \frac{1}{2} ([ O_I, [\frac{d}{d \xi}, O_J^\dag] ] + [ [O_I, \frac{d}{d \xi}], O_J^\dag ]) | \Psi_0 \rangle \label{operator_EOM_IJ_commutator}\\
& \stackrel{I\neq J}{=} & \langle \Psi_0 | O_I \frac{d}{d \xi} O_J^\dag + O_J^\dag \frac{d}{d \xi} O_I  | \Psi_0 \rangle. \label{operator_EOM_IJ_commutator2}
\end{eqnarray}
Note that the identity $O_I O_J^\dag = O_J^\dag O_I = 0\quad (I\neq J)$ was used in the last step, which follows from Eq.~\eqref{XXYYI}. Although Eq.~\eqref{operator_EOM_IJ_commutator} is more convenient to use in the following derivations due to the cancellation of many terms, Eq.~\eqref{operator_EOM_IJ_commutator2} highlights that Eq.~\eqref{operator_EOM_IJ_commutator} only differs with Eq.~\eqref{operator_EOM_IJ} by the term $O_J^\dag \frac{d}{d \xi} O_I$. Further manipulations yield\cite{nacme1-2014}
\begin{eqnarray}
g_{0I}^\xi & = & \sum_{pq\sigma} d_{pq\sigma}^\xi \gamma_{pq\sigma}^{0I}, \\
g_{IJ}^\xi & = & g_{IJ}^{\xi,(1)} + g_{IJ}^{\xi,(2)}, \label{gIJ} \\
g_{IJ}^{\xi,(1)} & = & \sum_{pq\sigma} d_{pq\sigma}^\xi \gamma_{pq\sigma}^{IJ},  \label{gIJ1} \\
g_{IJ}^{\xi,(2)} & = & \frac{1}{\omega_J-\omega_I} (\begin{array}{cc}\mathbf{X}_I^T & \mathbf{Y}_I^T \end{array}) \left( \frac{d}{d\xi} \left(\begin{array}{cc}\mathbf{A} &\mathbf{B} \\ \mathbf{B} &\mathbf{A}\end{array}\right) \right)\left(\begin{array}{c}\mathbf{X}_J \\ \mathbf{Y}_J \end{array}\right), \label{gIJ2}
\end{eqnarray}
where the transition density matrices (TDMs) $\gamma_{pq\sigma}^{0I}$ and $\gamma_{pq\sigma}^{IJ}$ are given by
\begin{eqnarray}
\gamma_{pq\sigma}^{0I} & = & \langle \Psi_0 | [a_{p\sigma}^\dag a_{q\sigma}, O_I^\dag] | \Psi_0 \rangle = \langle \Psi_0 | a_{p\sigma}^\dag a_{q\sigma} O_I^\dag - O_I^\dag a_{p\sigma}^\dag a_{q\sigma} | \Psi_0 \rangle, \label{geTDM} \\
\gamma_{pq\sigma}^{IJ} & = & \langle \Psi_0 | \frac{1}{2} ([ O_I, [a_{p\sigma}^\dag a_{q\sigma}, O_J^\dag] ] + [ [O_I, a_{p\sigma}^\dag a_{q\sigma}], O_J^\dag ])  | \Psi_0 \rangle \nonumber\\
& \stackrel{I\neq J}{=} & \langle \Psi_0 | O_I a_{p\sigma}^\dag a_{q\sigma} O_J^\dag + O_J^\dag a_{p\sigma}^\dag a_{q\sigma} O_I | \Psi_0 \rangle, \label{eeTDM}
\end{eqnarray}
and $d_{pq\sigma}^\xi$ is the overlap between MO $p\sigma$ with the full geometric derivative of MO $q\sigma$:
\begin{equation}
d_{pq\sigma}^\xi = \langle \psi_{p\sigma} | \frac{d}{d\xi} | \psi_{q\sigma} \rangle. \label{dpq}
\end{equation}
Roughly speaking, $g_{IJ}^{\xi,(1)}$ is the NACME contribution due to the geometry dependence of the MOs, while $g_{IJ}^{\xi,(2)}$ is due to the geometry dependence of the TDDFT excitation vectors. While $g_{IJ}^{\xi,(1)}$ is strictly a one-electron term (such that it vanishes when the states $I$ and $J$ differ by at least a double excitation), $g_{IJ}^{\xi,(2)}$ contains two-electron contributions due to the geometry derivatives of $\mathbf{A}$ and $\mathbf{B}$, and can thus be non-zero for excited states that differ by double excitations. The ge NACMEs $g_{0I}^\xi$ however only contains one-electron contributions.

With the use of a restricted open-shell reference, the spatial part of MO $\psi_{p}$ is identical to that of MO $\psi_{\bar{p}}$. In this case, the spin index of $d_{pq\sigma}^\xi$ can be dropped:
\begin{equation}
d_{pq\sigma}^{\mathrm{SA},\xi} = \langle \psi_{p} | \frac{d}{d\xi} | \psi_{q} \rangle = \langle \psi_{\bar{p}} | \frac{d}{d\xi} | \psi_{\bar{q}} \rangle \equiv d_{pq}^{\mathrm{SA},\xi}. \label{dSA}
\end{equation}

Apart from the EOM theory, the NACMEs of RPA-like theories can also be derived from time-dependent perturbation theory (TDPT).\cite{nacme1-2014,nacme2-2014,nacme3-2021} The resulting ground state-excited state (ge) NACMEs are identical to the EOM ones, while the excited state-excited state (ee) NACMEs have one additional term, namely the response term. While the TDPT theory of NACMEs is more rigorous than the EOM theory (because the EOM theory uses the killer condition Eq.~\eqref{killer} when it does not rigorously hold), the response term is typically small (usually contributing only 0-10\% to the IC rate\cite{nacme3-2021}); furthermore, the response term can show unphysical divergences when it becomes large, thereby deteriorating rather than improving the NACMEs. Therefore, we have advocated the use of EOM NACMEs in all cases,\cite{nacme3-2021} and unlike our previous papers,\cite{nacme1-2014,nacme2-2014} here we do not re-derive the X-TDDFT NACMEs using TDPT.

Up to now, the expressions are completely general; plugging in the excitation operators $O_I^\dag$ of different theories will yield their corresponding TDMs (and therefore NACMEs). For example, using the U-RPA excitation operator Eq.~\eqref{OI} yields the familiar expressions:\cite{nacme2-2014}
\begin{equation}
\gamma^{0I}_{ia\sigma} = (\mathbf{X}_I)_{ia\sigma}, \quad
\gamma^{0I}_{ai\sigma} = (\mathbf{Y}_I)_{ia\sigma}, \quad
\gamma^{0I}_{ij\sigma} = \gamma^{0I}_{ab\sigma} = 0, \label{U_TDM_0I}
\end{equation}
\begin{eqnarray}
\gamma^{IJ}_{ij\sigma} & = & - \sum_{a}\left((\mathbf{X}_J)_{ia\sigma} \label{U_TDM_IJ1}(\mathbf{X}_I)_{ja\sigma} + (\mathbf{Y}_I)_{ia\sigma} (\mathbf{Y}_J)_{ja\sigma}\right), \\
\gamma^{IJ}_{ab\sigma} & = & \sum_{i}\left((\mathbf{X}_I)_{ia\sigma} (\mathbf{X}_J)_{ib\sigma} + (\mathbf{Y}_J)_{ia\sigma} (\mathbf{Y}_I)_{ib\sigma}\right),\label{U_TDM_IJ2} \\
\gamma^{IJ}_{ia\sigma} & = & \gamma^{IJ}_{ai\sigma} = 0. \label{U_TDM_IJ3}
\end{eqnarray}
Here, the $\mathbf{X}$ contributions of the ge and ee TDMs come from the $a_{p\sigma}^\dag a_{q\sigma} O_I^\dag$ and $O_I a_{p\sigma}^\dag a_{q\sigma} O_J^\dag$ terms of Eq.~\eqref{geTDM} and \eqref{eeTDM}, respectively; the $\mathbf{Y}$ contributions instead come from the $O_I^\dag a_{p\sigma}^\dag a_{q\sigma}$ and $O_J^\dag a_{p\sigma}^\dag a_{q\sigma} O_I$ terms.
Eqs.~\eqref{U_TDM_0I}-\eqref{U_TDM_IJ3} are also applicable to U-TDDFT, except that the $\mathbf{X}$ and $\mathbf{Y}$ vectors must be solved from the U-TDDFT Casida equation instead of the U-RPA one.

Since the SA-RPA excitation operator Eq.~\eqref{OI_SA} only differs from Eq.~\eqref{OI} for CV(1) excitations, the SA-RPA ge TDMs only differ with the U-RPA ones in the CV blocks. Even more gratifyingly, the double excitations in $\Delta O_I^{\mathrm{SA},\dag}$ (i.e., $a_a^\dag a_{\bar{t}}^\dag a_{t} a_{\bar{i}}$) do not contribute:
\begin{equation}
\gamma^{\mathrm{SA},0I}_{ia} = \gamma^{0I}_{ia} + \frac{1}{2}\left(\sqrt{\frac{S}{S+1}}-1\right)((\mathbf{X}_I)_{ia} - (\mathbf{X}_I)_{\bar{i}\bar{a}}),
\end{equation}
\begin{equation}
\gamma^{\mathrm{SA},0I}_{\bar{i}\bar{a}} = \gamma^{0I}_{\bar{i}\bar{a}} - \frac{1}{2}\left(\sqrt{\frac{S}{S+1}}-1\right)((\mathbf{X}_I)_{ia} - (\mathbf{X}_I)_{\bar{i}\bar{a}}),
\end{equation}
\begin{equation}
\gamma^{\mathrm{SA},0I}_{ai} = \gamma^{0I}_{ai} + \frac{1}{2}\left(\sqrt{\frac{S}{S+1}}-1\right)((\mathbf{Y}_I)_{ia} - (\mathbf{Y}_I)_{\bar{i}\bar{a}}),
\end{equation}
\begin{equation}
\gamma^{\mathrm{SA},0I}_{\bar{a}\bar{i}} = \gamma^{0I}_{\bar{a}\bar{i}} - \frac{1}{2}\left(\sqrt{\frac{S}{S+1}}-1\right)((\mathbf{Y}_I)_{ia} - (\mathbf{Y}_I)_{\bar{i}\bar{a}}),
\end{equation}
\begin{equation}
\gamma^{\mathrm{SA},0I}_{\bar{i}\bar{t}} = \gamma^{0I}_{\bar{i}\bar{t}}, \quad \gamma^{\mathrm{SA},0I}_{\bar{t}\bar{i}} = \gamma^{0I}_{\bar{t}\bar{i}}, \quad \gamma^{\mathrm{SA},0I}_{ta} = \gamma^{0I}_{ta}, \quad \gamma^{\mathrm{SA},0I}_{at} = \gamma^{0I}_{at}.
\end{equation}
Finally, the extra contributions of $\gamma^{\mathrm{SA},0I}$ compared to $\gamma^{0I}$ cancel out upon summing over the spin index $\sigma$:
\begin{equation}
\sum_\sigma \gamma^{\mathrm{SA},0I}_{pq\sigma} = \sum_\sigma \gamma^{0I}_{pq\sigma}. \label{gamma0I_spinsum}
\end{equation}
Since $d_{pq}^{\mathrm{SA},\xi}$ is spin-independent, the X-RPA/X-TDDFT ge NACME expression is identical to their U-RPA/U-TDDFT counterparts, except that the SA-RPA (X-TDDFT) $\mathbf{X}_I$ and $\mathbf{Y}_I$ vectors should be used in place of the U-RPA (U-TDDFT) ones (note that here we have utilized the antisymmetry relation $d_{ai}^{\mathrm{SA},\xi} = -d_{ia}^{\mathrm{SA},\xi}$):
\begin{equation}
g_{0I}^{\mathrm{SA},\xi} = \sum_{ia\sigma} \left( (\mathbf{X}_I)_{ia\sigma}d_{ia}^{\mathrm{SA},\xi} + (\mathbf{Y}_I)_{ia\sigma}d_{ai}^{\mathrm{SA},\xi} \right) = \sum_{ia\sigma} d_{ia}^{\mathrm{SA},\xi} (\mathbf{X}_I-\mathbf{Y}_I)_{ia\sigma}.
\end{equation}
Therefore, assuming $d_{pq}^\xi$ is available, existing implementations of U-TDDFT ge NACMEs can be used to calculate X-TDDFT ge NACMEs without modifying a single line of code! Even more, the conclusion holds for other spin-free, one-electron ge transition properties as well, such as transition electric and magnetic multipole moments, as these properties are also given by contracting suitable one-electron operators with the spin-summed ge TDM. Unfortunately, in practical implementations $d_{pq}^{\mathrm{SA},\xi}$ has to be transformed away using the Lagrangian approach, making X-TDDFT ge NACMEs acquire additional computational overhead compared to U-TDDFT (Sec.~\ref{sec:Q}).

In contrast to the ge case, the X-TDDFT ee TDMs differ nontrivially from their U-TDDFT counterparts. Plugging Eq.~\eqref{DeltaOI_SA} into Eq.~\eqref{eeTDM} gives two corrections:
\begin{eqnarray}
\gamma_{pq\sigma}^{\mathrm{SA}, IJ} & = & \gamma_{pq\sigma}^{IJ} + \Delta\gamma_{pq\sigma}^{\mathrm{(1)}, IJ} + \Delta\gamma_{pq\sigma}^{\mathrm{(2)}, IJ}, \\
\Delta\gamma_{pq\sigma}^{\mathrm{(1)}, IJ} & = & \langle \Psi_0 | \frac{1}{2} ([ \Delta O_I, [a_{p\sigma}^\dag a_{q\sigma}, \Delta O_J^\dag] ] + [ [\Delta O_I, a_{p\sigma}^\dag a_{q\sigma}], \Delta O_J^\dag ]) | \Psi_0 \rangle,\\
\Delta\gamma_{pq\sigma}^{\mathrm{(2)}, IJ} & = & \langle \Psi_0 | \frac{1}{2} ([ \Delta O_I, [a_{p\sigma}^\dag a_{q\sigma}, O_J^\dag] ] + [ [\Delta O_I, a_{p\sigma}^\dag a_{q\sigma}], O_J^\dag ]) | \Psi_0 \rangle \nonumber\\
&  & + \langle \Psi_0 | \frac{1}{2} ([ O_I, [a_{p\sigma}^\dag a_{q\sigma}, \Delta O_J^\dag] ] + [ [O_I, a_{p\sigma}^\dag a_{q\sigma}], \Delta O_J^\dag ]) | \Psi_0 \rangle.
\label{gammaSAIJ}
\end{eqnarray}
$\Delta\gamma_{pq\sigma}^{\mathrm{(2)}, IJ}$ can be further separated into CO, OV, CV(0) and CV(1) contributions, based on the excitations that constitute $O_I$ or $O_J^\dag$ (Eq.~\eqref{OI_spintensor}):
\begin{equation}
\Delta\gamma_{pq\sigma}^{\mathrm{(2)}, IJ} = \Delta\gamma_{pq\sigma}^{\mathrm{CO}, IJ} + \Delta\gamma_{pq\sigma}^{\mathrm{OV}, IJ} + \Delta\gamma_{pq\sigma}^{\mathrm{CV(0)}, IJ} + \Delta\gamma_{pq\sigma}^{\mathrm{CV(1)}, IJ}.
\end{equation}
Interestingly, although $\Delta\gamma_{pq\sigma}^{\mathrm{CV(0)}, IJ}$ is non-zero, its $\alpha$ and $\beta$ components exactly cancel, similar to the CV(1) blocks of ge TDMs. Specifically, the CV(0) contributions read
\begin{eqnarray}
\Delta\gamma_{ij}^{\mathrm{CV(0)}, IJ} & = & -\Delta\gamma_{\bar{i}\bar{j}}^{\mathrm{CV(0)}, IJ} \nonumber\\
& = & -\frac{1}{4}\left(\sqrt{\frac{S}{S+1}}-1\right) \sum_{a} \left( ((\mathbf{X}_I)_{ja} + (\mathbf{X}_I)_{\bar{j}\bar{a}}) ((\mathbf{X}_J)_{ia} - (\mathbf{X}_J)_{\bar{i}\bar{a}}) \right. \nonumber\\
& & + ((\mathbf{Y}_I)_{ia} + (\mathbf{Y}_I)_{\bar{i}\bar{a}}) ((\mathbf{Y}_J)_{ja} - (\mathbf{Y}_J)_{\bar{j}\bar{a}}) \nonumber\\
& & + ((\mathbf{X}_I)_{ja} - (\mathbf{X}_I)_{\bar{j}\bar{a}}) ((\mathbf{X}_J)_{ia} + (\mathbf{X}_J)_{\bar{i}\bar{a}}) \nonumber\\
& & + \left.  ((\mathbf{Y}_I)_{ia} - (\mathbf{Y}_I)_{\bar{i}\bar{a}}) ((\mathbf{Y}_J)_{ja} + (\mathbf{Y}_J)_{\bar{j}\bar{a}}) \right), \label{gammaCV0_1} \\
\Delta\gamma_{ab}^{\mathrm{CV(0)}, IJ} & = & -\Delta\gamma_{\bar{a}\bar{b}}^{\mathrm{CV(0)}, IJ} \nonumber\\
& = & \frac{1}{4}\left(\sqrt{\frac{S}{S+1}}-1\right) \sum_{i} \left( ((\mathbf{X}_I)_{ia} + (\mathbf{X}_I)_{\bar{i}\bar{a}}) ((\mathbf{X}_J)_{ib} - (\mathbf{X}_J)_{\bar{i}\bar{b}}) \right. \nonumber\\
& & + ((\mathbf{Y}_I)_{ib} + (\mathbf{Y}_I)_{\bar{i}\bar{b}}) ((\mathbf{Y}_J)_{ia} - (\mathbf{Y}_J)_{\bar{i}\bar{a}}) \nonumber\\
& & + ((\mathbf{X}_I)_{ia} - (\mathbf{X}_I)_{\bar{i}\bar{a}}) ((\mathbf{X}_J)_{ib} + (\mathbf{X}_J)_{\bar{i}\bar{b}}) \nonumber\\
& & + \left. ((\mathbf{Y}_I)_{ib} - (\mathbf{Y}_I)_{\bar{i}\bar{b}}) ((\mathbf{Y}_J)_{ia} + (\mathbf{Y}_J)_{\bar{i}\bar{a}}) \right). \label{gammaCV0_2}
\end{eqnarray}
The CV(1) contribution $\Delta\gamma_{pq\sigma}^{\mathrm{CV(1)}, IJ}$ does not vanish even upon summing the $\alpha$ and $\beta$ parts. However, its sum with $\Delta\gamma_{pq\sigma}^{\mathrm{(1)}, IJ}$ satisfies this property:
\begin{eqnarray}
\Delta\gamma_{ij}^{\mathrm{(1)}, IJ} + \Delta\gamma_{ij}^{\mathrm{CV(1)}, IJ} & = & -(\Delta\gamma_{\bar{i}\bar{j}}^{\mathrm{(1)}, IJ} + \Delta\gamma_{\bar{i}\bar{j}}^{\mathrm{CV(1)}, IJ}) \nonumber\\
& = & \frac{1}{4(S+1)} \sum_{a} \left( ((\mathbf{X}_I)_{ja} - (\mathbf{X}_I)_{\bar{j}\bar{a}}) ((\mathbf{X}_J)_{ia} - (\mathbf{X}_J)_{\bar{i}\bar{a}}) \right. \nonumber\\
& & + \left. ((\mathbf{Y}_I)_{ia} - (\mathbf{Y}_I)_{\bar{i}\bar{a}}) ((\mathbf{Y}_J)_{ja} - (\mathbf{Y}_J)_{\bar{j}\bar{a}}) \right), \label{gammaCV1_1} \\
\Delta\gamma_{ab}^{\mathrm{(1)}, IJ} + \Delta\gamma_{ab}^{\mathrm{CV(1)}, IJ} & = & -(\Delta\gamma_{\bar{a}\bar{b}}^{\mathrm{(1)}, IJ} + \Delta\gamma_{\bar{a}\bar{b}}^{\mathrm{CV(1)}, IJ}) \nonumber\\
& = & -\frac{1}{4(S+1)} \sum_{i} \left( ((\mathbf{X}_I)_{ia} - (\mathbf{X}_I)_{\bar{i}\bar{a}}) ((\mathbf{X}_J)_{ib} - (\mathbf{X}_J)_{\bar{i}\bar{b}}) \right. \nonumber\\
& & + \left. ((\mathbf{Y}_I)_{ib} - (\mathbf{Y}_I)_{\bar{i}\bar{b}}) ((\mathbf{Y}_J)_{ia} - (\mathbf{Y}_J)_{\bar{i}\bar{a}}) \right). \label{gammaCV1_2}
\end{eqnarray}
The CO and OV contributions, however, remain non-zero even after summing the $\alpha$ and $\beta$ components:
\begin{eqnarray}
\Delta\gamma_{ta}^{\mathrm{CO}, IJ} & = & -\frac{1}{2\sqrt{S(S+1)}} \times \nonumber\\
& & \sum_{i} \left( (\mathbf{X}_I)_{\bar{i}\bar{t}} ((\mathbf{X}_J)_{ia} - (\mathbf{X}_J)_{\bar{i}\bar{a}}) + (\mathbf{Y}_J)_{\bar{i}\bar{t}} ((\mathbf{Y}_I)_{ia} - (\mathbf{Y}_I)_{\bar{i}\bar{a}}) \right), \\
\Delta\gamma_{\bar{t}\bar{a}}^{\mathrm{CO}, IJ} & = & -\frac{1}{2} \left(\sqrt{\frac{S}{S+1}}-1\right) \times \nonumber\\
& & \sum_{i} \left( (\mathbf{X}_I)_{\bar{i}\bar{t}} ((\mathbf{X}_J)_{ia} - (\mathbf{X}_J)_{\bar{i}\bar{a}}) + (\mathbf{Y}_J)_{\bar{i}\bar{t}} ((\mathbf{Y}_I)_{ia} - (\mathbf{Y}_I)_{\bar{i}\bar{a}}) \right), \\
\Delta\gamma_{at}^{\mathrm{CO}, IJ} & = & -\frac{1}{2\sqrt{S(S+1)}} \times \nonumber\\
& & \sum_{i} \left( (\mathbf{X}_J)_{\bar{i}\bar{t}} ((\mathbf{X}_I)_{ia} - (\mathbf{X}_I)_{\bar{i}\bar{a}}) + (\mathbf{Y}_I)_{\bar{i}\bar{t}} ((\mathbf{Y}_J)_{ia} - (\mathbf{Y}_J)_{\bar{i}\bar{a}}) \right), \\
\Delta\gamma_{\bar{a}\bar{t}}^{\mathrm{CO}, IJ} & = & -\frac{1}{2} \left(\sqrt{\frac{S}{S+1}}-1\right) \times \nonumber\\
& & \sum_{i} \left( (\mathbf{X}_J)_{\bar{i}\bar{t}} ((\mathbf{X}_I)_{ia} - (\mathbf{X}_I)_{\bar{i}\bar{a}}) + (\mathbf{Y}_I)_{\bar{i}\bar{t}} ((\mathbf{Y}_J)_{ia} - (\mathbf{Y}_J)_{\bar{i}\bar{a}}) \right),
\end{eqnarray}
\begin{eqnarray}
\Delta\gamma_{it}^{\mathrm{OV}, IJ} & = & -\frac{1}{2} \left(\sqrt{\frac{S}{S+1}}-1\right) \times \nonumber\\
& & \sum_{a} \left( (\mathbf{X}_I)_{ta} ((\mathbf{X}_J)_{ia} - (\mathbf{X}_J)_{\bar{i}\bar{a}}) + (\mathbf{Y}_J)_{ta} ((\mathbf{Y}_I)_{ia} - (\mathbf{Y}_I)_{\bar{i}\bar{a}}) \right), \\
\Delta\gamma_{\bar{i}\bar{t}}^{\mathrm{OV}, IJ} & = & -\frac{1}{2\sqrt{S(S+1)}} \times \nonumber\\
& & \sum_{a} \left( (\mathbf{X}_I)_{ta} ((\mathbf{X}_J)_{ia} - (\mathbf{X}_J)_{\bar{i}\bar{a}}) + (\mathbf{Y}_J)_{ta} ((\mathbf{Y}_I)_{ia} - (\mathbf{Y}_I)_{\bar{i}\bar{a}}) \right), \\
\Delta\gamma_{ti}^{\mathrm{OV}, IJ} & = & -\frac{1}{2} \left(\sqrt{\frac{S}{S+1}}-1\right) \times \nonumber\\
& & \sum_{a} \left( (\mathbf{X}_J)_{ta} ((\mathbf{X}_I)_{ia} - (\mathbf{X}_I)_{\bar{i}\bar{a}}) + (\mathbf{Y}_I)_{ta} ((\mathbf{Y}_J)_{ia} - (\mathbf{Y}_J)_{\bar{i}\bar{a}}) \right), \\
\Delta\gamma_{\bar{t}\bar{i}}^{\mathrm{OV}, IJ} & = & -\frac{1}{2\sqrt{S(S+1)}} \times \nonumber\\
& & \sum_{a} \left( (\mathbf{X}_J)_{ta} ((\mathbf{X}_I)_{ia} - (\mathbf{X}_I)_{\bar{i}\bar{a}}) + (\mathbf{Y}_I)_{ta} ((\mathbf{Y}_J)_{ia} - (\mathbf{Y}_J)_{\bar{i}\bar{a}}) \right).
\end{eqnarray}
All the unmentioned blocks of $\Delta\gamma^{\mathrm{(1)}, IJ}$, $\Delta\gamma^{\mathrm{CV(0)}, IJ}$, $\Delta\gamma^{\mathrm{CV(1)}, IJ}$, $\Delta\gamma^{\mathrm{CO}, IJ}$ and $\Delta\gamma^{\mathrm{OV}, IJ}$ are zero. A detailed derivation of the above formulas is given in Appendix~\ref{sec:UXdiff}.

To sum up, the first contribution to the X-TDDFT ee NACMEs (Eq.~\eqref{gIJ}), $g_{IJ}^{\mathrm{SA},\xi,(1)}$, can be written as
\begin{eqnarray}
g_{IJ}^{\mathrm{SA},\xi,(1)} & = & g_{IJ}^{\xi,(1)} + \Delta g_{IJ}^{\xi}, \nonumber\\
\Delta g_{IJ}^{\xi} & = & \sum_{pq} d_{pq}^{\mathrm{SA},\xi} \Delta\gamma_{pq}^{IJ}, \label{g1IJ} \\
\Delta\gamma_{pq}^{IJ} & = & \sum_{\sigma} \left( \Delta\gamma_{pq\sigma}^{\mathrm{(1)}, IJ} + \Delta\gamma_{pq\sigma}^{\mathrm{(2)}, IJ} \right) \nonumber\\
& = & \frac{1}{2}\left(1-\sqrt{\frac{S+1}{S}}\right) \times \nonumber\\
& & \left(\delta_{pt}\delta_{qa} \sum_{i} \left( (\mathbf{X}_I)_{\bar{i}\bar{t}} ((\mathbf{X}_J)_{ia} - (\mathbf{X}_J)_{\bar{i}\bar{a}}) + (\mathbf{Y}_J)_{\bar{i}\bar{t}} ((\mathbf{Y}_I)_{ia} - (\mathbf{Y}_I)_{\bar{i}\bar{a}}) \right) \right. \nonumber\\
& & + \delta_{pa}\delta_{qt} \sum_{i} \left( (\mathbf{X}_J)_{\bar{i}\bar{t}} ((\mathbf{X}_I)_{ia} - (\mathbf{X}_I)_{\bar{i}\bar{a}}) + (\mathbf{Y}_I)_{\bar{i}\bar{t}} ((\mathbf{Y}_J)_{ia} - (\mathbf{Y}_J)_{\bar{i}\bar{a}}) \right) \nonumber\\
& & + \delta_{pi}\delta_{qt} \sum_{a} \left( (\mathbf{X}_I)_{ta} ((\mathbf{X}_J)_{ia} - (\mathbf{X}_J)_{\bar{i}\bar{a}}) + (\mathbf{Y}_J)_{ta} ((\mathbf{Y}_I)_{ia} - (\mathbf{Y}_I)_{\bar{i}\bar{a}}) \right) \nonumber\\
& & \left. + \delta_{pt}\delta_{qi} \sum_{a} \left( (\mathbf{X}_J)_{ta} ((\mathbf{X}_I)_{ia} - (\mathbf{X}_I)_{\bar{i}\bar{a}}) + (\mathbf{Y}_I)_{ta} ((\mathbf{Y}_J)_{ia} - (\mathbf{Y}_J)_{\bar{i}\bar{a}}) \right) \right).
\label{COCV1}
\end{eqnarray}
This correction will also show up in other ee transition properties, such as transition multipole moments. To understand this result further, we consider the special case where the states $I$ and $J$ are respectively dominated by a single spin-tensor basis excitation. $\Delta g_{IJ}^{\xi}$ is only non-zero when one of the two states is a CO or OV state, and the other is a CV(1) state. For example, when state $I$ is the CO excitation $\Psi_{\bar{i}}^{\bar{t}}$ and state $J$ is the CV(1) excitation $T_{ai}^\dag(1,0) \Psi_0$, the U-RPA/U-TDDFT theory gives (note that in this case $(\mathbf{X}_I)_{\bar{i}\bar{t}}=1$, $(\mathbf{X}_J)_{ia}=-(\mathbf{X}_J)_{\bar{i}\bar{a}}=\frac{1}{\sqrt{2}}$, and $\mathbf{Y}_I=\mathbf{Y}_J=0$)
\begin{eqnarray}
\sum_\sigma \gamma_{pq\sigma}^{IJ} & = & -\frac{1}{\sqrt{2}}\delta_{pt}\delta_{qa},
\end{eqnarray}
while the SA-RPA/X-TDDFT correction to it reads
\begin{eqnarray}
\Delta\gamma_{pq}^{IJ} & = & \frac{1}{\sqrt{2}}\left(1-\sqrt{\frac{S+1}{S}}\right)\delta_{pt}\delta_{qa} = \left(\sqrt{\frac{S+1}{S}}-1\right)\sum_\sigma \gamma_{pq\sigma}^{IJ}.
\end{eqnarray}
For TDMs between OV and CV(1) states, it similarly follows that $\Delta\gamma^{IJ}$ is $\left(\sqrt{\frac{S+1}{S}}-1\right)$ times the spin-summed U-TDDFT TDM. Therefore, for transitions between pure CO/OV and pure CV(1) states, the vector $g_{IJ}^{\mathrm{SA},\xi,(1)}$ is longer than $g_{IJ}^{\xi,(1)}$ by a factor of $\sqrt{\frac{S+1}{S}}$, with the direction of the vector unchanged, assuming that the same set of excitation vectors are used in evaluating the two NACMEs.

In contrast to $g_{IJ}^{\mathrm{SA},\xi,(1)}$, $g_{IJ}^{\mathrm{SA},\xi,(2)}$ is formally much simpler, and is given by $g_{IJ}^{\xi,(2)}$ plus a simple contribution due to the SA-RPA correction of $\mathbf{A}$:
\begin{equation}
g_{IJ}^{\mathrm{SA},\xi,(2)} = g_{IJ}^{\xi,(2)} + \frac{1}{\omega_J-\omega_I} ( \mathbf{X}_I^T \frac{d \Delta\mathbf{A}}{d\xi} \mathbf{X}_J + \mathbf{Y}_I^T \frac{d \Delta\mathbf{A}}{d\xi} \mathbf{Y}_J ). \label{g2IJ}
\end{equation}
The numerical difference of the X-TDDFT $g_{IJ}^{\mathrm{SA},\xi,(2)}$ and the U-TDDFT $g_{IJ}^{\xi,(2)}$ can however be more complicated than what Eq.~\eqref{g2IJ} would suggest, because they are calculated from different sets of excitation energies and excitation vectors. In particular, spin-adaptation can have an arbitrarily large effect on the factor $\frac{1}{\omega_J-\omega_I}$ in not only Eq.~\eqref{g2IJ} but also Eq.~\eqref{gIJ2}, as the states $I$ and $J$'s energies can become arbitrarily close or even switch order upon switching from U-RPA/U-TDDFT to SA-RPA/X-TDDFT. Therefore, while X-TDDFT excitation energies of CV(1) states are systematically higher than the U-TDDFT ones,\cite{XTDDFTbenchmarkDD} little can be said in general about how the use of X-TDDFT in place of U-TDDFT can affect the magnitudes of NACMEs.

We finally point out that large relative differences between X-TDDFT and U-TDDFT NACMEs can arise from another mechanism. Although the X-TDDFT and U-TDDFT excitation vectors (as well as the ROKS and UKS MOs) usually do not differ significantly, the difference can induce a large relative change of the NACMEs when the NACMEs are very small. This can especially happen for the NACMEs between the ground state and a CV(1) state, where the $\Psi_i^{a}$ and $\Psi_{\bar{i}}^{\bar{a}}$ contributions of the spin-adapted CV(1) excitation (Eq.~\eqref{Psi1}) exactly cancel out each other in the X-TDDFT NACMEs. However, such cancellation is only approximate for U-TDDFT NACMEs, because the $\alpha$ and $\beta$ UKS MOs are not identical. Thus, X-TDDFT NACMEs between these the ground state and CV(1) states tend to be shorter than the corresponding U-TDDFT ones, assuming U-TDDFT and X-TDDFT predict similar excited state compositions.

\subsection{Lagrangian formulation} \label{sec:Lagrangian}

While the above formulas enable the calculation of X-TDDFT NACMEs, they rely on the knowledge of $d_{pq}^{\mathrm{SA},\xi}$ (Eq.~\eqref{dSA}), which is tedious to calculate (no matter analytically or by numerical differentiation of MOs with respect to the geometry), and numerically unstable when two MOs become (near-)degenerate. The standard way out\cite{nacme2-2014} is to define a Lagrangian with the bra MOs calculated at a geometry $\{\xi^{(0)}\}$, while the ket MOs are calculated at a geometry $\{\xi\}$ that can be different from $\{\xi^{(0)}\}$; after taking derivatives with respect to $\{\xi\}$ (but not with respect to $\{\xi^{(0)}\}$), we set $\xi=\xi^{(0)}$. This allows us to study the derivatives of ket states, without having to involve the derivatives of the bra states.

Like our X-TDDFT gradient paper,\cite{XTDDFTgrad} here we first rederive the U-TDDFT NACMEs, followed by discussing the additional terms due to using X-TDDFT. Writing the MO coefficients at $\{\xi^{(0)}\}$ and $\{\xi\}$ as $\mathbf{C}^{(0)}$ and $\mathbf{C}$, respectively, we have (where the notation $\mathbf{C}(\xi)$ stresses that the dependence of $\mathbf{C}$ on $\xi$ should be taken into account when taking full derivatives w.r.t.~$\xi$)
\begin{eqnarray}
d_{0I}(\xi, \mathbf{C}(\xi)) & = & \sum_{pq\sigma} \gamma^{0I}_{pq\sigma} \langle \psi_{p\sigma}(\xi^{(0)}) | \psi_{q\sigma}(\xi) \rangle = \sum_{pq\sigma\mu\nu} \gamma^{0I}_{pq\sigma} C^{(0)}_{\mu p\sigma} C_{\nu q\sigma}(\xi) \tilde{S}_{\mu\nu}(\xi), \label{d0I} \\
d_{IJ}(\xi, \mathbf{C}(\xi)) & = &  \sum_{pq\sigma} \gamma^{IJ}_{pq\sigma} \langle \psi_{p\sigma}(\xi^{(0)}) | \psi_{q\sigma} (\xi) \rangle + d_{IJ}^{(2)}(\xi, \mathbf{C}(\xi)) \nonumber\\
& = & \sum_{pq\sigma\mu\nu} \gamma^{IJ}_{pq\sigma} C^{(0)}_{\mu p\sigma} C_{\nu q\sigma}(\xi) \tilde{S}_{\mu\nu}(\xi) + d_{IJ}^{(2)}(\xi, \mathbf{C}(\xi)), \\ \label{dIJ}
d_{IJ}^{(2)}(\xi, \mathbf{C}(\xi)) & = & \frac{1}{\omega_J-\omega_I} (\begin{array}{cc}\mathbf{X}_I^T & \mathbf{Y}_I^T \end{array}) \left(\begin{array}{cc}\mathbf{A}(\xi, \mathbf{C}(\xi)) &\mathbf{B}(\xi, \mathbf{C}(\xi)) \\ \mathbf{B}(\xi, \mathbf{C}(\xi)) &\mathbf{A}(\xi, \mathbf{C}(\xi))\end{array}\right) \left(\begin{array}{c}\mathbf{X}_J \\ \mathbf{Y}_J \end{array}\right),
\end{eqnarray}
where $\tilde{\mathbf{S}}(\xi)$ is the mixed-geometry AO overlap integral between the geometries $\{\xi^{(0)}\}$ and $\{\xi\}$:
\begin{equation}
\tilde{S}_{\mu\nu}(\xi) = \langle \mu(\xi^{(0)}) | \nu(\xi) \rangle.
\end{equation}
The Casida matrices $\mathbf{A}$ and $\mathbf{B}$ depend not only on $\xi$ but also on $\mathbf{C}$, because they involve the MO Fock matrices and MO integrals (Eqs.~\eqref{UTD-Amat}-\eqref{Kmat}), which can be expressed as contractions of $\mathbf{C}$ and the respective AO matrices. Furthermore, the XC potential and kernel in the Fock and Casida matrices, respectively, depend on the density and therefore on the MOs; the two-electron contributions of the Fock matrix also depends on the MOs through the density matrix. Note however that the MO excitation amplitudes $\mathbf{X}_I$ and $\mathbf{Y}_I$ are treated as independent variables from $\mathbf{C}$, unlike the Fock matrix and two-electron integrals whose AO representations are viewed as more ``fundamental'' than their MO representations. It is then obvious that
\begin{equation}
\left. \frac{d}{d\xi}d_{0I}(\xi, \mathbf{C}(\xi)) \right|_{\xi=\xi^{(0)}} = g_{0I}^{\xi}, \left. \quad \frac{d}{d\xi}d_{IJ}(\xi, \mathbf{C}(\xi)) \right|_{\xi=\xi^{(0)}} = g_{IJ}^{\xi}.
\end{equation}
To convert the cumbersome full $\xi$ derivatives into partial $\xi$ derivatives, we define a Lagrangian that incorporates all constraints that $\mathbf{C}$ has to satisfy. For U-TDDFT, the conditions are the Brillouin condition $F_{ia\sigma}=0$ and the orthogonality condition $S_{pq\sigma}=\delta_{pq}$, giving (where we have dropped the $\xi$ dependence of $\mathbf{C}$ since now $\mathbf{C}$ is treated as independent of $\xi$)
\begin{eqnarray}
& & \mathcal{L}^{\mathrm{U-TDDFT}}_{0I}[\xi, \mathbf{C}, \mathbf{Z}_{0I}, \mathbf{W}_{0I}] \nonumber\\
& = & d_{0I}(\xi, \mathbf{C}) + \sum_{ia\sigma}(\mathbf{Z}_{0I})_{ia\sigma}F_{ia\sigma} - \sum_{pq\sigma, p\le q} (\mathbf{W}_{0I})_{pq\sigma} (S_{pq\sigma} - \delta_{pq}), \label{Lagrangian-NAC-0I}
\end{eqnarray}
and
\begin{eqnarray}
& & \mathcal{L}^{\mathrm{U-TDDFT}}_{IJ}[\xi, \mathbf{C}, \mathbf{Z}_{IJ}, \mathbf{W}_{IJ}] \nonumber\\
& = & d_{IJ}(\xi, \mathbf{C}) + \sum_{ia\sigma}(\mathbf{Z}_{IJ})_{ia\sigma}F_{ia\sigma} - \sum_{pq\sigma, p\le q} (\mathbf{W}_{IJ})_{pq\sigma} (S_{pq\sigma} - \delta_{pq}). \label{Lagrangian-NAC-IJ}
\end{eqnarray}
In either case, applying Eq.~\eqref{L_CC} followed by imposing $\xi=\xi^{(0)}$ gives
\begin{eqnarray}
W_{ij\sigma}^{\mathrm{s}} & = & \frac{1}{2}\left(Q_{ij\sigma} + H^+_{ij\sigma}[\mathbf{Z}^{\mathrm{s}}]\right), \label{W1} \\
W_{ab\sigma}^{\mathrm{s}} & = & \frac{1}{2}Q_{ab\sigma}, \\
W_{ia\sigma}^{\mathrm{s}} & = & \frac{1}{2}\left(Q_{ia\sigma} + H^+_{ia\sigma}[\mathbf{Z}^{\mathrm{s}}] + \sum_b Z_{ib\sigma}F_{ab\sigma}\right), \\
W_{ai\sigma}^{\mathrm{s}} & = & \frac{1}{2}\left(Q_{ai\sigma} + \sum_j Z_{ja\sigma}F_{ij\sigma}\right), \label{W4}
\end{eqnarray}
where
\begin{eqnarray}
W^{\mathrm{s}}_{pq\sigma} & = & \left\{ \begin{array}{ll} \frac{1}{2}W_{pq\sigma}, & p<q \\ W_{pp\sigma}, & p=q \\ \frac{1}{2}W_{qp\sigma}, & p>q \end{array} \right. , \\ \label{Ws}
Z^{\mathrm{s}}_{ij\sigma} & = & Z^{\mathrm{s}}_{ab\sigma} = 0 ,\\
Z^{\mathrm{s}}_{ia\sigma} & = & Z^{\mathrm{s}}_{ai\sigma} = \frac{1}{2}Z_{ia\sigma},
\end{eqnarray}
and
\begin{equation}
Q_{pq\sigma} = \sum_\mu \frac{\partial d(\xi, \mathbf{C})}{\partial C_{\mu{}p\sigma}}C_{\mu{}q\sigma}. \label{Q}
\end{equation}
Here the quantities $\mathbf{Z}$, $\mathbf{W}$ and $d(\xi, \mathbf{C})$ should be understood to bear the subscripts $0I$ or $IJ$, depending on whether they refer to the ge and ee quantities, respectively. The operator $\mathbf{H}^+$ is the response contribution to the $\mathbf{A}+\mathbf{B}$ matrix:
\begin{eqnarray}
H^+_{pq\sigma}[\mathbf{V}] & = & \sum_{rs\tau}(2(pq\sigma|sr\tau) + 2f_{pq\sigma{}sr\tau}^{\mathrm{xc}}[\rho] - c_{\mathrm{x}}\delta_{\sigma\tau}((pr\sigma|sq\sigma)+(ps\sigma|rq\sigma)))V_{rs\tau}. \label{H+}
\end{eqnarray}
Once $\mathbf{Q}$ is known, $\mathbf{Z}$ can be calculated from the identity $W_{ia\sigma}^{\mathrm{s}} = W_{ai\sigma}^{\mathrm{s}}$, giving the Z-vector equation
\begin{equation}
\sum_b Z_{ib\sigma}F_{ab\sigma} - \sum_j Z_{ja\sigma}F_{ij\sigma} + H^+_{ia\sigma}[\mathbf{Z}^{\mathrm{s}}] = Q_{ai\sigma}-Q_{ia\sigma}, \label{Z0}
\end{equation}
or
\begin{equation}
\sum_{jb\tau}(A+B)_{ia\sigma,jb\tau}Z_{jb\tau} = Q_{ai\sigma}-Q_{ia\sigma}. \label{Z}
\end{equation}
The obtained $\mathbf{Z}$ is then plugged into Eqs.~\eqref{W1}-\eqref{W4} to obtain $\mathbf{W}$. Finally, the NACMEs can be calculated by differentiation of $\mathcal{L}^{\mathrm{U-TDDFT}}_{0I}$ or $\mathcal{L}^{\mathrm{U-TDDFT}}_{IJ}$ with respect to $\xi$, followed by setting $\xi=\xi_0$ so that $\mathbf{C}^{(0)}=\mathbf{C}$:
\begin{eqnarray}
g_{0I}^{\xi} & = & \left. \frac{d \mathcal{L}^{\mathrm{U-TDDFT}}_{0I}}{d \xi} \right|_{\xi=\xi_0} = \left. \frac{\partial \mathcal{L}^{\mathrm{U-TDDFT}}_{0I}}{\partial \xi} \right|_{\xi=\xi_0} \nonumber\\
& = & \sum_{pq\sigma\mu\nu} \gamma^{0I}_{pq\sigma} C_{\mu p\sigma} C_{\nu q\sigma} d^{\xi}_{\mu\nu} + \sum_{\mu\nu\sigma}(\mathbf{Z}_{0I}^{\mathrm{s}})_{\mu\nu\sigma}\frac{\partial F_{\mu\nu\sigma}}{\partial \xi} - \sum_{\mu\nu\sigma} (\mathbf{W}_{0I}^{\mathrm{s}})_{\mu\nu\sigma} \frac{\partial S_{\mu\nu}}{\partial \xi}, \label{dL0Idxi} \\
g_{IJ}^{\xi} & = & \left. \frac{d \mathcal{L}^{\mathrm{U-TDDFT}}_{IJ}}{d \xi} \right|_{\xi=\xi_0} = \left. \frac{\partial \mathcal{L}^{\mathrm{U-TDDFT}}_{IJ}}{\partial \xi} \right|_{\xi=\xi_0} \nonumber\\
& = & \sum_{pq\sigma\mu\nu} \gamma^{IJ}_{pq\sigma} C_{\mu p\sigma} C_{\nu q\sigma} d^{\xi}_{\mu\nu} + \frac{1}{\omega_J-\omega_I}(\begin{array}{cc}\mathbf{X}_I^T & \mathbf{Y}_I^T \end{array}) \left(\frac{\partial}{\partial \xi} \left(\begin{array}{cc}\mathbf{A} &\mathbf{B} \\ \mathbf{B} &\mathbf{A}\end{array}\right) \right) \left(\begin{array}{c}\mathbf{X}_J \\ \mathbf{Y}_J \end{array}\right) \nonumber\\
& & + \sum_{\mu\nu\sigma}(\mathbf{Z}_{IJ}^{\mathrm{s}})_{\mu\nu\sigma}\frac{\partial F_{\mu\nu\sigma}}{\partial \xi} - \sum_{\mu\nu\sigma} (\mathbf{W}_{IJ}^{\mathrm{s}})_{\mu\nu\sigma} \frac{\partial S_{\mu\nu}}{\partial \xi}, \label{dLIJdxi}
\end{eqnarray}
where
\begin{equation}
d^{\xi}_{\mu\nu} = \langle \mu(\xi^{(0)}) | \frac{\partial}{\partial \xi} | \nu(\xi) \rangle.
\end{equation}

One may notice that Eqs.~\eqref{W1}-\eqref{Z} are also found in the derivation of the U-TDDFT gradient,\cite{XTDDFTgrad,FurcheTDDFTgrad} with the only exception that the $d(\xi, \mathbf{C})$ in Eq.~\eqref{Q} should be changed to $\omega$. This is hardly surprising since the NACME Lagrangians (Eqs.~\eqref{Lagrangian-NAC-0I}-\eqref{Lagrangian-NAC-IJ}) share the same $\mathbf{Z}$- and $\mathbf{W}$-dependent Lagrangian multiplier terms as the TDDFT Lagrangian Eq.~\eqref{Lagrangian}.

The extra terms due to using an ROKS reference instead of an UKS reference similarly carry over from TDDFT gradient theory to TDDFT NACME theory. The ROKS Brillouin condition is different from the UKS Brillouin condition ($F_{ia\sigma}=0$), and reads
\begin{equation}
F_{ia}+F_{\bar{i}\bar{a}} = 0, \quad F_{\bar{i}\bar{t}} =0, \quad F_{ta}= 0,\label{Brillouin_ROKS}
\end{equation}
namely, the CV blocks of the $\alpha$ and $\beta$ Fock matrices are not required to be zero; only their sum needs to be zero. As discussed in our X-TDDFT gradient paper,\cite{XTDDFTgrad} this is equivalent to adding a constraint $Z_{ia}=Z_{\bar{i}\bar{a}}$, which introduces another term in the Lagrangian that is proportional to $Z_{ia}-Z_{\bar{i}\bar{a}}$. Meanwhile, the constraint that the ROKS $\alpha$ and $\beta$ orbitals are equal introduces a term proportional to $C_{\mu{}p\alpha}-C_{\mu{}p\beta}$. To sum up, we arrive at the U-TDDFT/ROKS NACME Lagrangian, which in the ge case reads (the ee case is completely analogous)
\begin{eqnarray}
& & \mathcal{L}_{0I}^{\mathrm{U-TDDFT/ROKS}}[\xi,\mathbf{C},\mathbf{Z}_{0I},\mathbf{W}_{0I},\mathbf{\zeta}_{0I},\mathbf{\Lambda}_{0I}] \nonumber\\
& = & \mathcal{L}_{0I}^{\mathrm{U-TDDFT}}[\xi,\mathbf{C},\mathbf{Z}_{0I},\mathbf{W}_{0I}] + \sum_{ia}(\mathbf{\zeta}_{0I})_{ia}((\mathbf{Z}_{0I})_{ia}-(\mathbf{Z}_{0I})_{\bar{i}\bar{a}}) \nonumber\\
& & + \sum_{\mu{}p}(\mathbf{\Lambda}_{0I})_{\mu{}p}(C_{\mu{}p\alpha}-C_{\mu{}p\beta}),
\end{eqnarray}
which exactly parallels Eq.~\eqref{L_ROKS}. The extra Lagrange multiplier $\mathbf{\zeta}_{0I}$ naturally vanishes upon taking either the $\xi$ or the $\mathbf{C}$ derivative of the Lagrangian, but $\mathbf{\Lambda}_{0I}$ survives in the $\mathbf{C}$ derivative (Eq.~\eqref{L_CC}) and only cancels out after summing the $\alpha$ and $\beta$ components, yielding
\begin{equation}
\sum_\sigma\sum_{jb\tau}(A+B)_{ia\sigma,jb\tau}Z_{jb\tau}  + \delta_{i\alpha{}t}\sum_j Z_{jt\beta}F_{ja\beta} - \delta_{a\beta{}\bar{t}}\sum_b Z_{tb\alpha}F_{ib\alpha} = \sum_\sigma(Q_{ai\sigma}-Q_{ia\sigma}). \label{Zvector_X}
\end{equation}
Apart from spin summation, Eq.~\eqref{Zvector_X} involves two corrections on the left hand side, which account for the fact that the CO and OV blocks of the ROKS Brillouin condition (Eq.~\eqref{Brillouin_ROKS}) should not be spin-summed. The equations for $\mathbf{W}$ similarly only hold in spin-summed form, plus corrections that resemble those in Eq.~\eqref{Zvector_X}:
\begin{eqnarray}
\sum_{\sigma}W_{ij\sigma}^{\mathrm{s}} & = & \frac{1}{2}\sum_{\sigma}\left(Q_{ij\sigma} + H^+_{ij\sigma}[\mathbf{Z}^{\mathrm{s}}]\right), \label{W1_ROKS} \\
\sum_{\sigma}W_{ab\sigma}^{\mathrm{s}} & = & \frac{1}{2}\sum_{\sigma}Q_{ab\sigma}, \\
\sum_{\sigma}W_{ia\sigma}^{\mathrm{s}} & = & \frac{1}{2}\sum_{\sigma}\left(Q_{ia\sigma} + H^+_{ia\sigma}[\mathbf{Z}^{\mathrm{s}}] + \sum_b Z_{ib\sigma}F_{ab\sigma}\right) + \frac{1}{2}\delta_{i\alpha{}t}\sum_j Z_{jt\beta}F_{ja\beta}, \\
\sum_{\sigma}W_{ai\sigma}^{\mathrm{s}} & = & \frac{1}{2}\sum_{\sigma}\left(Q_{ai\sigma} + \sum_j Z_{ja\sigma}F_{ij\sigma}\right) + \frac{1}{2}\delta_{a\beta{}\bar{t}}\sum_b Z_{tb\alpha}F_{ib\alpha}. \label{W4_ROKS}
\end{eqnarray}
These equations do not yield the individual spin components of $W_{pq\sigma}^{\mathrm{s}}$, only their sum $\sum_{\sigma}W_{pq\sigma}^{\mathrm{s}}$. However, this does not matter since $\mathbf{W}$ only enters Eqs.~\eqref{dL0Idxi}-\eqref{dLIJdxi} in its spin-summed form.

Finally, the X-TDDFT ge NACMEs are given by:
\begin{eqnarray}
g_{0I}^{\mathrm{SA},\xi} & = & \left. \frac{d \mathcal{L}^{\mathrm{X-TDDFT}}_{0I}}{d \xi} \right|_{\xi=\xi_0} = \left. \frac{\partial \mathcal{L}^{\mathrm{X-TDDFT}}_{0I}}{\partial \xi} \right|_{\xi=\xi_0} \nonumber\\
& = & \sum_{pq\sigma\mu\nu} \gamma^{\mathrm{SA},0I}_{pq\sigma} C_{\mu p\sigma} C_{\nu q\sigma} d^{\xi}_{\mu\nu} + \sum_{\mu\nu\sigma}(\mathbf{Z}_{0I}^{\mathrm{s}})_{\mu\nu\sigma}\frac{\partial F_{\mu\nu\sigma}}{\partial \xi} - \sum_{\mu\nu\sigma} (\mathbf{W}_{0I}^{\mathrm{s}})_{\mu\nu\sigma} \frac{\partial S_{\mu\nu}}{\partial \xi}, \label{NACME_X_0I}
\end{eqnarray}
which is equivalent to the U-TDDFT ge NACME expression Eq.~\eqref{dL0Idxi}, because the correction to the ge TDM vanishes upon spin summation (Eq.~\eqref{gamma0I_spinsum}).
However, this does not mean that X-TDDFT ge NACMEs are numerically identical to U-TDDFT ge NACMEs, because the former uses X-TDDFT excitation amplitudes and Lagrange multipliers, while the latter uses U-TDDFT ones.

The X-TDDFT ee NACME Lagrangian differs from the U-TDDFT/ROKS ee Lagrangian in two ways: in X-TDDFT, one replaces the U-TDDFT Casida matrix $\mathbf{A}$ by $\mathbf{A} + \Delta\mathbf{A}$, and the U-TDDFT ee TDM by its X-TDDFT equivalent. Neither difference introduces corrections to Eqs.~\eqref{Zvector_X}-\eqref{W4_ROKS}, but they do change the intermediate $\mathbf{Q}$ (to be detailed in Section~\ref{sec:Q}) and the final expressions of the NACMEs:
\begin{eqnarray}
g_{IJ}^{\mathrm{SA},\xi} & = & \left. \frac{d \mathcal{L}^{\mathrm{X-TDDFT}}_{IJ}}{d \xi} \right|_{\xi=\xi_0} = \left. \frac{\partial \mathcal{L}^{\mathrm{X-TDDFT}}_{IJ}}{\partial \xi} \right|_{\xi=\xi_0} \nonumber\\
& = & \sum_{pq\sigma\mu\nu} \gamma^{\mathrm{SA},IJ}_{pq\sigma} C_{\mu p\sigma} C_{\nu q\sigma} d^{\xi}_{\mu\nu}  + \sum_{\mu\nu\sigma}(\mathbf{Z}_{IJ}^{\mathrm{s}})_{\mu\nu\sigma}\frac{\partial F_{\mu\nu\sigma}}{\partial \xi} - \sum_{\mu\nu\sigma} (\mathbf{W}_{IJ}^{\mathrm{s}})_{\mu\nu\sigma} \frac{\partial S_{\mu\nu}}{\partial \xi} \nonumber\\
& &  + \frac{1}{\omega_J-\omega_I}(\begin{array}{cc}\mathbf{X}_I^T & \mathbf{Y}_I^T \end{array}) \left(\frac{\partial}{\partial \xi} \left(\begin{array}{cc}\mathbf{A} + \Delta\mathbf{A} &\mathbf{B} \\ \mathbf{B} &\mathbf{A} + \Delta\mathbf{A}\end{array}\right) \right) \left(\begin{array}{c}\mathbf{X}_J \\ \mathbf{Y}_J \end{array}\right). \label{NACME_X_IJ}
\end{eqnarray}

\subsection{The intermediate $\mathbf{Q}$} \label{sec:Q}

Calculation of the Lagrange multipliers $\mathbf{Z}$ and $\mathbf{W}$ requires calculating the intermediate $\mathbf{Q}$ from Eq.~\eqref{Q}. For U-TDDFT and X-TDDFT ge NACMEs, $\mathbf{Q}$ simply reads
\begin{equation}
Q_{ai\sigma} = (\mathbf{X}_I)_{ia\sigma}, \quad Q_{ia\sigma} = (\mathbf{Y}_I)_{ia\sigma}, \quad Q_{ij\sigma} = Q_{ab\sigma} = 0.
\end{equation}
Therefore, for U-TDDFT ge NACME calculations, the Z-vector equation is
\begin{equation}
(\mathbf{A} + \mathbf{B})\mathbf{Z}_{0I} = \mathbf{X}_I - \mathbf{Y}_I, \label{Zvector_0I}
\end{equation}
which has the simple solution
\begin{equation}
\mathbf{Z}_{0I} = \omega_{I}^{-1} (\mathbf{X}_I + \mathbf{Y}_I). \label{U_Z0I}
\end{equation}
Such simplification is however not possible for Eq.~\eqref{Zvector_X}, due to two reasons: (1) the X-TDDFT excitation vectors are given by the X-TDDFT Casida equation Eq.~\eqref{Casida}, which involves the RPA correction $\Delta\mathbf{A}$ (such that $(\mathbf{A} + \Delta\mathbf{A} + \mathbf{B})(\mathbf{X}_I + \mathbf{Y}_I) = \omega_{I}(\mathbf{X}_I - \mathbf{Y}_I)$, but $(\mathbf{A} + \mathbf{B})(\mathbf{X}_I + \mathbf{Y}_I) \neq \omega_{I}(\mathbf{X}_I - \mathbf{Y}_I)$ in general); (2) Eq.~\eqref{Zvector_X} contains two further corrections $\delta_{i\alpha{}t}\sum_j Z_{jt\beta}F_{ja\beta} - \delta_{a\beta{}\bar{t}}\sum_b Z_{tb\alpha}F_{ib\alpha}$ on the left hand side. Therefore, the ge X-TDDFT NACMEs are more expensive than the U-TDDFT ones due to the need to solve the Z-vector equation Eq.~\eqref{Zvector_X}. Despite this, since TDDFT gradient calculations also involve very similar Z-vector equations,\cite{FurcheTDDFTgrad,XTDDFTgrad} ge X-TDDFT NACMEs are only about as expensive as TDDFT gradients, which is acceptable for practical applications.

Turning to ee NACMEs, Eq.~\eqref{Lagrangian-NAC-IJ} is strikingly similar with Eq.~\eqref{Lagrangian}, yielding a $\mathbf{Q}$ that is very similar to that involved in TDDFT gradient. For U-TDDFT ee NACMEs, we have
\begin{equation}
\mathbf{Q} = \mathbf{\gamma}^{IJ} + \frac{1}{\omega_J-\omega_I}\mathbf{\tilde{Q}},
\end{equation}
where $\mathbf{\tilde{Q}}$ reads
\begin{eqnarray}
\tilde{Q}_{ij\sigma} & = & H^+_{ij\sigma}[\mathbf{T}^{IJ}] + 2\sum_{ka\tau{}lb\eta}g^{\mathrm{xc}}_{ij\sigma{}ka\tau{}lb\eta}[\rho](\mathbf{X}_I+\mathbf{Y}_I)_{ka\tau}(\mathbf{X}_J+\mathbf{Y}_J)_{lb\eta} + 2\sum_{p}T^{IJ}_{ip\sigma}F_{jp\sigma} \nonumber\\
& & + \frac{1}{2} \left( \sum_{a}(\mathbf{X}_I+\mathbf{Y}_I)_{ia\sigma}H^+_{ja\sigma}[\mathbf{R}_J^{\mathrm{s}}] + \sum_{a}(\mathbf{X}_I-\mathbf{Y}_I)_{ia\sigma}H^-_{ja\sigma}[\mathbf{L}_J^{\mathrm{a}}] \right. \nonumber\\
& & + \left. (\mathbf{X}_J+\mathbf{Y}_J)_{ia\sigma}H^+_{ja\sigma}[\mathbf{R}_I^{\mathrm{s}}] + \sum_{a}(\mathbf{X}_J-\mathbf{Y}_J)_{ia\sigma}H^-_{ja\sigma}[\mathbf{L}_I^{\mathrm{a}}] \right), \label{Qij} \\
\tilde{Q}_{ia\sigma} & = & H^+_{ia\sigma}[\mathbf{T}^{IJ}] + 2\sum_{jb\tau{}kc\eta}g^{\mathrm{xc}}_{ia\sigma{}jb\tau{}kc\eta}[\rho](\mathbf{X}_I+\mathbf{Y}_I)_{jb\tau}(\mathbf{X}_J+\mathbf{Y}_J)_{kc\eta} + 2\sum_{p}T^{IJ}_{ip\sigma}F_{pa\sigma} \nonumber\\
& & + \frac{1}{2} \left( \sum_{b}(\mathbf{X}_I+\mathbf{Y}_I)_{ib\sigma}H^+_{ab\sigma}[\mathbf{R}_J^{\mathrm{s}}] + \sum_{b}(\mathbf{X}_I-\mathbf{Y}_I)_{ib\sigma}H^-_{ab\sigma}[\mathbf{L}_J^{\mathrm{a}}] \right. \nonumber\\
& & \left. \sum_{b}(\mathbf{X}_J+\mathbf{Y}_J)_{ib\sigma}H^+_{ab\sigma}[\mathbf{R}_I^{\mathrm{s}}] + \sum_{b}(\mathbf{X}_J-\mathbf{Y}_J)_{ib\sigma}H^-_{ab\sigma}[\mathbf{L}_I^{\mathrm{a}}] \right), \label{Qia} \\
\tilde{Q}_{ai\sigma} & = & 2\sum_{p}T^{IJ}_{pa\sigma}F_{ip\sigma} + \frac{1}{2} \left( \sum_{j}(\mathbf{X}_I+\mathbf{Y}_I)_{ja\sigma}H^+_{ij\sigma}[\mathbf{R}_J^{\mathrm{s}}] + \sum_{j}(\mathbf{X}_I-\mathbf{Y}_I)_{ja\sigma}H^-_{ij\sigma}[\mathbf{L}_J^{\mathrm{a}}] \right. \nonumber\\
& & + \left. \sum_{j}(\mathbf{X}_J+\mathbf{Y}_J)_{ja\sigma}H^+_{ij\sigma}[\mathbf{R}_I^{\mathrm{s}}] + \sum_{j}(\mathbf{X}_J-\mathbf{Y}_J)_{ja\sigma}H^-_{ij\sigma}[\mathbf{L}_I^{\mathrm{a}}] \right), \label{Qai} \\
\tilde{Q}_{ab\sigma} & = & 2\sum_{p}T^{IJ}_{pa\sigma}F_{pb\sigma} + \frac{1}{2} \left( \sum_{i}(\mathbf{X}_I+\mathbf{Y}_I)_{ia\sigma}H^+_{ib\sigma}[\mathbf{R}_J^{\mathrm{s}}] + \sum_{i}(\mathbf{X}_I-\mathbf{Y}_I)_{ia\sigma}H^-_{ib\sigma}[\mathbf{L}_J^{\mathrm{a}}] \right. \nonumber\\
& & + \left. \sum_{i}(\mathbf{X}_J+\mathbf{Y}_J)_{ia\sigma}H^+_{ib\sigma}[\mathbf{R}_I^{\mathrm{s}}] + \sum_{i}(\mathbf{X}_J-\mathbf{Y}_J)_{ia\sigma}H^-_{ib\sigma}[\mathbf{L}_I^{\mathrm{a}}] \right) , \label{Qab}
\end{eqnarray}
where $\mathbf{g}^{\mathrm{xc}}[\rho]$ is the XC third-order derivative tensor, and $H^-_{pq\sigma}[\mathbf{V}]$ is the response contribution to $\mathbf{A}-\mathbf{B}$:
\begin{eqnarray}
H^-_{pq\sigma}[\mathbf{V}] & = & \sum_{rs\tau}c_{\mathrm{x}}\delta_{\sigma\tau}((pr\sigma|sq\sigma)-(ps\sigma|rq\sigma))V_{rs\tau}.
\end{eqnarray}
$\mathbf{R}_I^{\mathrm{s}}$ and $\mathbf{L}_I^{\mathrm{a}}$ are the symmetrized and antisymmetrized versions of the $\mathbf{X}_I+\mathbf{Y}_I$ and $\mathbf{X}_I-\mathbf{Y}_I$ vectors, respectively:
\begin{eqnarray}
(\mathbf{R}_I^{\mathrm{s}})_{ia\sigma} & = & (\mathbf{R}_I^{\mathrm{s}})_{ai\sigma} = \frac{1}{2}((\mathbf{X}_I)_{ia\sigma}+(\mathbf{Y}_I)_{ia\sigma}), \\
(\mathbf{L}_I^{\mathrm{a}})_{ia\sigma} & = & (\mathbf{L}_I^{\mathrm{a}})_{ai\sigma} = \frac{1}{2}((\mathbf{X}_I)_{ia\sigma}-(\mathbf{Y}_I)_{ia\sigma}), \\
(\mathbf{R}_I^{\mathrm{s}})_{ij\sigma} & = & (\mathbf{R}_I^{\mathrm{s}})_{ab\sigma} = (\mathbf{L}_I^{\mathrm{a}})_{ij\sigma} = (\mathbf{L}_I^{\mathrm{a}})_{ab\sigma} = 0.
\end{eqnarray}
Finally, $\mathbf{T}^{IJ}$ is the symmetrized ee TDM:
\begin{equation}
\mathbf{T}^{IJ} = \frac{1}{2}(\mathbf{\gamma}^{IJ} + (\mathbf{\gamma}^{IJ})^T).
\end{equation}
Although the formulas may seem complicated, comparison with U-TDDFT gradient theory\cite{FurcheTDDFTgrad,XTDDFTgrad} shows that the intermediate $\mathbf{Q}$ in U-TDDFT gradient is identical with the $\mathbf{\tilde{Q}}$ in U-TDDFT NACME theory after setting $I=J$. This is hardly surprising as the NACME contribution $g_{IJ}^{\xi,(2)}$ (Eq.~\eqref{g2IJ}) is related to the first term of Eq.~\eqref{dLUdxi} by setting $I=J$ and scaling by $\frac{1}{\omega_J-\omega_I}$. Therefore, U-TDDFT ee NACMEs can be implemented with very few modifications upon an existing U-TDDFT gradient code, as remarked before.\cite{nacme2-2014,nacme3-2021}

We are now ready to discuss the X-TDDFT counterpart of $\mathbf{Q}$, $\mathbf{Q}^{\mathrm{SA}}$. $\mathbf{Q}^{\mathrm{SA}}$ differs from $\mathbf{Q}$ due to two reasons: (1) $g_{IJ}^{\mathrm{SA},\xi,(1)}$ differs from $g_{IJ}^{\xi,(1)}$ by the use of the X-TDDFT ee TDM instead of the U-TDDFT ee TDM (Eq.~\eqref{g1IJ}); (2) $g_{IJ}^{\mathrm{SA},\xi,(2)}$ differs from $g_{IJ}^{\xi,(2)}$ due to the RPA correction $\Delta\mathbf{A}$ (Eq.~\eqref{g2IJ}). Therefore, we can write
\begin{eqnarray}
\mathbf{Q}^{\mathrm{SA}} & = & \mathbf{\gamma}^{\mathrm{SA},IJ} + \frac{1}{\omega_J-\omega_I}(\mathbf{\tilde{Q}}+\Delta\mathbf{Q}), \\
\Delta Q_{pq\sigma} & = & \frac{1}{2}\delta_{pt}\sum_{r}(qt|tr)(T'_{qr}+T'_{rq}) + \frac{1}{2}(T'_{pr}+T'_{rp})\sum_{rt}(qt|tr), \label{DeltaQ} \\
T'_{ij} & = & \left(-1+\sqrt{\frac{S+1}{S}}+\frac{1}{2S}\right)\sum_a ((\mathbf{X}_J)_{ia}(\mathbf{X}_I)_{ja} + (\mathbf{Y}_I)_{ia}(\mathbf{Y}_J)_{ja}) \nonumber\\
& & + \left(1-\sqrt{\frac{S+1}{S}}+\frac{1}{2S}\right)\sum_a ((\mathbf{X}_J)_{\bar{i}\bar{a}}(\mathbf{X}_I)_{\bar{j}\bar{a}} + (\mathbf{Y}_I)_{\bar{i}\bar{a}}(\mathbf{Y}_J)_{\bar{j}\bar{a}}) \nonumber\\
& & - \frac{1}{2S}\sum_a ((\mathbf{X}_J)_{ia}(\mathbf{X}_I)_{\bar{j}\bar{a}} + (\mathbf{Y}_I)_{ia}(\mathbf{Y}_J)_{\bar{j}\bar{a}} + (\mathbf{X}_J)_{\bar{i}\bar{a}}(\mathbf{X}_I)_{ja} + (\mathbf{Y}_I)_{\bar{i}\bar{a}}(\mathbf{Y}_J)_{ja}), \\
T'_{ab} & = & \left(1-\sqrt{\frac{S+1}{S}}+\frac{1}{2S}\right)\sum_i ((\mathbf{X}_I)_{ia}(\mathbf{X}_J)_{ib} + (\mathbf{Y}_J)_{ia}(\mathbf{Y}_I)_{ib}) \nonumber\\
& & + \left(-1+\sqrt{\frac{S+1}{S}}+\frac{1}{2S}\right)\sum_i ((\mathbf{X}_I)_{\bar{i}\bar{a}}(\mathbf{X}_J)_{\bar{i}\bar{b}} + (\mathbf{Y}_J)_{\bar{i}\bar{a}}(\mathbf{Y}_I)_{\bar{i}\bar{b}}) \nonumber\\
& & - \frac{1}{2S}\sum_i ((\mathbf{X}_I)_{ia}(\mathbf{X}_J)_{\bar{i}\bar{b}} + (\mathbf{Y}_J)_{ia}(\mathbf{Y}_I)_{\bar{i}\bar{b}} + (\mathbf{X}_I)_{\bar{i}\bar{a}}(\mathbf{X}_J)_{ib} + (\mathbf{Y}_J)_{\bar{i}\bar{a}}(\mathbf{Y}_I)_{ib}), \\
T'_{ia} & = & T'_{ai} = 0.
\end{eqnarray}
Importantly, the symmetrized ee TDM $\mathbf{T}^{IJ}$ should still be calculated from $\mathbf{\gamma}^{IJ}$, not $\mathbf{\gamma}^{\mathrm{SA},IJ}$, because the $\mathbf{T}^{IJ}$-related terms originate from the MO coefficient derivatives of the MO Fock matrix, rather than the Casida matrix.

\subsection{Final NACME expression}

With the intermediate $\mathbf{Q}^{\mathrm{SA}}$ (and therefore the Lagrange multipliers $\mathbf{Z}$ and $\mathbf{W}$) in hand, we can then calculate the X-TDDFT NACMEs from Eqs.~\eqref{dL0Idxi}-\eqref{dLIJdxi}. Since the derivatives of one-electron and two-electron integrals, as well as of the XC potential and kernel appear multiple times in the expression, it is convenient to group them together so that a given integral derivative only needs to be computed once. The X-TDDFT ge NACME expression Eq.~\eqref{dL0Idxi} is already in such a form, and one only needs to recall that
\begin{eqnarray}
\sum_{\mu\nu\sigma}(\mathbf{Z}_{0I}^{\mathrm{s}})_{\mu\nu\sigma}\frac{\partial F_{\mu\nu\sigma}}{\partial \xi} 
& = & \sum_{\mu\nu\sigma}(\mathbf{Z}_{0I}^{\mathrm{s}})_{\mu\nu\sigma}\frac{\partial h_{\mu\nu}}{\partial \xi} + \sum_{\kappa\lambda}\Gamma_{\mu\nu\kappa\lambda}^{0I}\frac{\partial (\mu\nu|\kappa\lambda)}{\partial \xi} + \sum_{\mu\nu\sigma}(\mathbf{Z}_{0I}^{\mathrm{s}})_{\mu\nu\sigma}\frac{\partial v_{\mu\nu\sigma}^{\mathrm{xc}}[\rho]}{\partial \xi}, \\
\Gamma_{\mu\nu\kappa\lambda}^{0I} & = & \sum_{\sigma\tau} (\mathbf{Z}_{0I}^{\mathrm{s}})_{\mu\nu\sigma}D_{\kappa\lambda\tau}  - c_x\delta_{\sigma\tau}(\mathbf{Z}_{0I}^{\mathrm{s}})_{\mu\kappa\sigma}D_{\nu\lambda\sigma},
\end{eqnarray}
where $\mathbf{h}$ and $\mathbf{v}^{\mathrm{xc}}[\rho]$ are the one-electron Hamiltonian and XC potential, respectively, and $\mathbf{D}$ is the ground state density matrix.
For the ee NACMEs, we first construct the relaxed difference density matrix
\begin{equation}
\mathbf{P}^{IJ} = \frac{1}{\omega_J-\omega_I}\mathbf{T}^{IJ} + \mathbf{Z}_{IJ}.
\end{equation}
The NACMEs are then given by
\begin{eqnarray}
g_{IJ}^{\mathrm{SA},\xi} & = & \sum_{pq\sigma\mu\nu} \gamma^{IJ}_{pq\sigma} C_{\mu p\sigma} C_{\nu q\sigma} d^{\xi}_{\mu\nu} + \sum_{\mu\nu\sigma}P^{IJ}_{\mu\nu\sigma}\frac{\partial h_{\mu\nu}}{\partial \xi} + \sum_{\kappa\lambda}\Gamma_{\mu\nu\kappa\lambda}^{\mathrm{SA},IJ}\frac{\partial (\mu\nu|\kappa\lambda)}{\partial \xi} \nonumber\\
& &  + \sum_{\mu\nu\sigma}P^{IJ}_{\mu\nu\sigma}\frac{\partial v_{\mu\nu\sigma}^{\mathrm{xc}}[\rho]}{\partial \xi} + \frac{1}{\omega_J-\omega_I}\sum_{\mu\nu\sigma\kappa\lambda\tau}R^{\mathrm{s}}_{\mu\nu\sigma}R^{\mathrm{s}}_{\kappa\lambda\tau}\frac{\partial f^{\mathrm{xc}}_{\mu\nu\sigma\kappa\lambda\tau}[\rho]}{\partial \xi}, \\
\Gamma_{\mu\nu\kappa\lambda}^{\mathrm{SA},IJ} & = & \Gamma_{\mu\nu\kappa\lambda}^{IJ} + \Delta\Gamma_{\mu\nu\kappa\lambda}^{IJ},\\
\Gamma_{\mu\nu\kappa\lambda}^{IJ} & = & \sum_{\sigma\tau} \left(P^{IJ}_{\mu\nu\sigma}D_{\lambda\kappa\tau} + \frac{1}{\omega_J-\omega_I}(\mathbf{R}_I^{\mathrm{s}})_{\mu\nu\sigma}(\mathbf{R}_J^{\mathrm{s}})_{\lambda\kappa\tau} \right.\nonumber\\
& & \left. - c_x\delta_{\sigma\tau}(P^{IJ}_{\mu\kappa\sigma}D_{\lambda\nu\sigma} + \frac{1}{\omega_J-\omega_I}\left((\mathbf{R}_I^{\mathrm{s}})_{\mu\kappa\sigma}(\mathbf{R}_J^{\mathrm{s}})_{\lambda\nu\sigma} + (\mathbf{L}_I^{\mathrm{a}})_{\mu\kappa\sigma}(\mathbf{L}_J^{\mathrm{a}})_{\lambda\nu\sigma})\right) \right), \\
\Delta\Gamma_{\mu\nu\kappa\lambda} & = & -\frac{1}{\omega_J-\omega_I}D^S_{\mu\kappa}T'_{\nu\lambda}, \label{DeltaGamma} \\
D^S_{\mu\nu} & = & \frac{1}{2}(D_{\mu\nu\beta} - D_{\mu\nu\alpha}).
\end{eqnarray}
The expression is identical to the case of U-TDDFT NACMEs,\cite{nacme2-2014,nacme3-2021} except for the correction $\Delta\Gamma$ which is absent in U-TDDFT.

\subsection{Electron translation factor (ETF)}

The NACMEs calculated from the above formulas are generally not translationally invariant; naive use of these NACMEs would predict that the molecule can undergo IC via global translation alone, which is unphysical. To see this, recall that $g_{IJ}^{\xi}$ is the overlap of the nuclear derivative of $|\Psi_J\rangle$ with $\langle\Psi_I|$ (Eq.~\eqref{NACME}). When the molecule undergoes translational motion, the change of $|\Psi_J\rangle$ is non-zero, and in general has a non-zero overlap with $\langle\Psi_I|$. However, this certainly does not mean that $|\Psi_J\rangle$ will undergo IC to $|\Psi_I\rangle$ due to translation, as $|\Psi_J\rangle$ continues to be orthogonal with the $|\Psi_I\rangle$ state of the \emph{moving} molecule.

A well-known remedy of the problem\cite{ETF} is to treat each AO as co-moving with its nucleus, followed by taking the slow-moving limit of the nuclei. It has been shown that, regardless of the electronic structure theory used, this treatment amounts to substituting the $d_{\mu\nu}^{\xi}$ in the theory (for X-TDDFT, they are Eqs.~\eqref{NACME_X_0I}-\eqref{NACME_X_IJ}) by $\frac{1}{2}\frac{\partial S_{\mu\nu}}{\partial \xi}$. The resulting correction to the NACMEs is called the ETF. In the numerical tests mentioned below, all NACMEs refer to those that contain the ETF.

\section{Benchmark calculations and applications} \label{sec:benchmark}

To benchmark the accuracies of the X-TDDFT and U-TDDFT NACMEs, we compare the NACMEs of formaldehyde radical cation (\ch{CH2O+}) against high-level multireference (extended multi-state complete active space second order perturbation theory, XMS-CASPT2\cite{XMSCASPT2,XMSCASPT2NACME}) calculations. We then apply X-TDDFT NACMEs to the excited state relaxation pathway calculations of copper(II) porphyrins. All NACME calculations were performed using a development version of BDF,\cite{BDF1,BDF2,BDF3,BDF4,BDF5} except for XMS-CASPT2 NACME calculations which were performed using BAGEL.\cite{bagel} When the angle $\theta$ between two NACME vectors is larger than 90$^\circ$, we convert it to 180$^\circ-\theta$, as flipping the direction of an NACME vector does not change any physical observable.

\subsection{Small molecule: \ch{CH2O+}} \label{sec:CH2O}

We have previously reported the X-TDDFT vertical\cite{XTDDFTbenchmarkDD} and adiabatic\cite{XTDDFTgrad} excitation energies, as well as excited state geometries\cite{XTDDFTgrad} of \ch{CH2O+}, whose errors are only about half of the errors of U-TDDFT. Building upon our previous understanding of \ch{CH2O+}, here we use \ch{CH2O+} as a prototypical small molecule test case. All calculations were performed at the B3LYP\cite{B3LYP,Becke93}/def2-SVP\cite{DEF2SVP} ground state equilibrium geometry.

The compositions of the first seven excited states of \ch{CH2O+} at the X-TDDFT B3LYP/aug-cc-pVTZ\cite{Dunning1989} and state-averaged (SA)-CASSCF(11,10)/aug-cc-pVTZ levels are listed in Table~\ref{table:compositions} (such a large basis set is not necessary in real calculations, but is used here to minimize basis set effects). The CASSCF active space consists of all the valence orbitals of \ch{CH2O+} (ground state occupation: $(3a_1)^2$$(4a_1)^2$$(1b_1)^2$$(1b_2)^2$$(5a_1)^2$$(2b_1)^1$$(2b_2)^0$$(6a_1)^0$$(3b_1)^0$$(7a_1)^0$), and the 10 lowest states were averaged. All methods predict that the four low-lying states are CO and OV states, with minimal spin contamination at the U-TDDFT level (deviation of $\langle S^2 \rangle$ from its ideal value $\le 0.1$). The next three states are all CV states with significant spin contamination at the U-TDDFT level ($\langle S^2 \rangle$ error 1.95, 1.90 and 1.43, respectively) due to the lack of double excitations, which are implicitly accounted for by X-TDDFT. To add dynamic correlation on top of the CASSCF wavefunctions, we performed XMS-CASPT2 calculations with a real IPEA shift of 0.2 a.u. With the XMS-CASPT2 excitation energies as the reference, X-TDDFT fixes the systematic underestimation of U-TDDFT excitation energies of CV states, as noted in our previous works.\cite{XTDDFTbenchmarkDD,XTDDFTgrad}

\begin{sidewaystable}
\centering
\caption{Excitation energies and dominant compositions of the first seven states of \ch{CH2O+}, computed at the XMS-CASPT2/aug-cc-pVTZ, U-TDDFT B3LYP/aug-cc-pVTZ, and X-TDDFT B3LYP/aug-cc-pVTZ levels. For XMS-CASPT2, the CASSCF state compositions are given. Only compositions with coefficients $|c|>0.2$ are shown. ``CV*'' denotes double excitations of the form $\Psi_{\bar{i}t}^{\bar{t}a}$. The ground state is $1^2 B_1$.}
\begin{tabular}{@{}c@{}c@{}c@{}c@{}c@{}c@{}c@{}}
\hline
State & \multicolumn{2}{c}{XMS-CASPT2} & \multicolumn{2}{c}{U-TDDFT} & \multicolumn{2}{c}{X-TDDFT} \\
\cmidrule(lr){2-3} \cmidrule(lr){4-5} \cmidrule(lr){6-7}
 & $\omega$ (eV) & Composition &  $\omega$ (eV) & Composition & $\omega$ (eV) & Composition \\
\midrule
 $1^2 B_2$ & 3.93 & CO($1b_2\to 2b_1$) ($c=0.98$) & 3.93 & CO($1b_2\to 2b_1$) ($c=0.98$) & 3.78 & CO($1b_2\to 2b_1$) ($c=0.98$) \\
 \midrule
 $1^2 A_1$ & 5.27 & CO($5a_1\to 2b_1$) ($c=0.96$) & 5.38 & CO($5a_1\to 2b_1$) ($c=0.99$) & 5.24 & CO($5a_1\to 2b_1$) ($c=0.98$) \\
 \midrule
 $2^2 B_2$ & 5.51 & OV($2b_1\to 2b_2$) ($c=0.95$) & 5.76 & OV($2b_1\to 2b_2$) ($c=0.99$) & 5.70 & OV($2b_1\to 2b_2$) ($c=0.98$) \\
 \midrule
 $2^2 B_1$ & 6.25 & \begin{tabular}{c} CO($1b_1\to 2b_1$) ($c=0.88$) \\ CV($\beta\beta$)($1b_2\to 2b_2$) ($c=-0.27$) \\ CV($\alpha\alpha$)($1b_2\to 2b_2$) ($c=0.26$) \end{tabular} & 6.53 & \begin{tabular}{c} CO($1b_1\to 2b_1$) ($c=0.91$) \\ CV($\alpha\alpha$)($1b_2\to 2b_2$) ($c=-0.37$) \end{tabular} & 6.42 & \begin{tabular}{c} CO($1b_1\to 2b_1$) ($c=0.92$) \\ CV($\alpha\alpha$)($1b_2\to 2b_2$) ($c=-0.28$) \\ CV($\beta\beta$)($1b_2\to 2b_2$) ($c=-0.22$) \end{tabular} \\
 \midrule
 $3^2 B_1$ & 7.98 & \begin{tabular}{c} CV*($1b_2\to 2b_2$) ($c=0.79$) \\ CV($\beta\beta$)($1b_2\to 2b_2$) ($c=0.46$) \\ CV($\alpha\alpha$)($1b_2\to 2b_2$) ($c=-0.34$) \end{tabular} & 6.66 & \begin{tabular}{c} CV($\beta\beta$)($1b_2\to 2b_2$) ($c=0.75$) \\ CV($\alpha\alpha$)($1b_2\to 2b_2$) ($c=-0.64$) \end{tabular} & 7.64 & \begin{tabular}{c} CV*($1b_2\to 2b_2$) ($c=0.81$) \\ CV($\beta\beta$)($1b_2\to 2b_2$) ($c=0.45$) \\ CV($\alpha\alpha$)($1b_2\to 2b_2$) ($c=-0.36$) \end{tabular} \\
 \midrule
 $1^2 A_2$ & 9.62 & \begin{tabular}{c} CV*($5a_1\to 2b_2$) ($c=0.76$) \\ CV($\alpha\alpha$)($5a_1\to 2b_2$) ($c=0.52$) \\ CV($\beta\beta$)($5a_1\to 2b_2$) ($c=0.24$) \end{tabular} & 8.83 & \begin{tabular}{c} CV($\beta\beta$)($5a_1\to 2b_2$) ($c=0.84$) \\ CV($\alpha\alpha$)($5a_1\to 2b_2$) ($c=-0.54$) \end{tabular} & 9.72 & \begin{tabular}{c} CV($\beta\beta$)($5a_1\to 2b_2$) ($c=0.80$) \\ CV*($5a_1\to 2b_2$) ($c=0.50$) \\ CV($\alpha\alpha$)($5a_1\to 2b_2$) ($c=0.30$) \end{tabular} \\
 \midrule
 $3^2 B_2$ & 9.88 & \begin{tabular}{c} CV($\beta\beta$)($1b_1\to 2b_2$) ($c=0.80$) \\ CV($\alpha\alpha$)($1b_1\to 2b_2$) ($c=0.41$) \\ CV*($1b_1\to 2b_2$) ($c=0.39$) \end{tabular} & 9.26 & \begin{tabular}{c} CV($\beta\beta$)($1b_1\to 2b_2$) ($c=0.96$) \\ CV($\alpha\alpha$)($1b_1\to 2b_2$) ($c=-0.24$) \end{tabular} & 9.65 & \begin{tabular}{c} CV($\beta\beta$)($1b_1\to 2b_2$) ($c=0.81$) \\ CV($\alpha\alpha$)($1b_1\to 2b_2$) ($c=0.44$) \\ CV*($1b_1\to 2b_2$) ($c=0.37$) \end{tabular} \\
\hline
\end{tabular}
\\
\label{table:compositions}
\end{sidewaystable}

\begin{figure}[htbp]
\centering
\includegraphics[width=\textwidth]{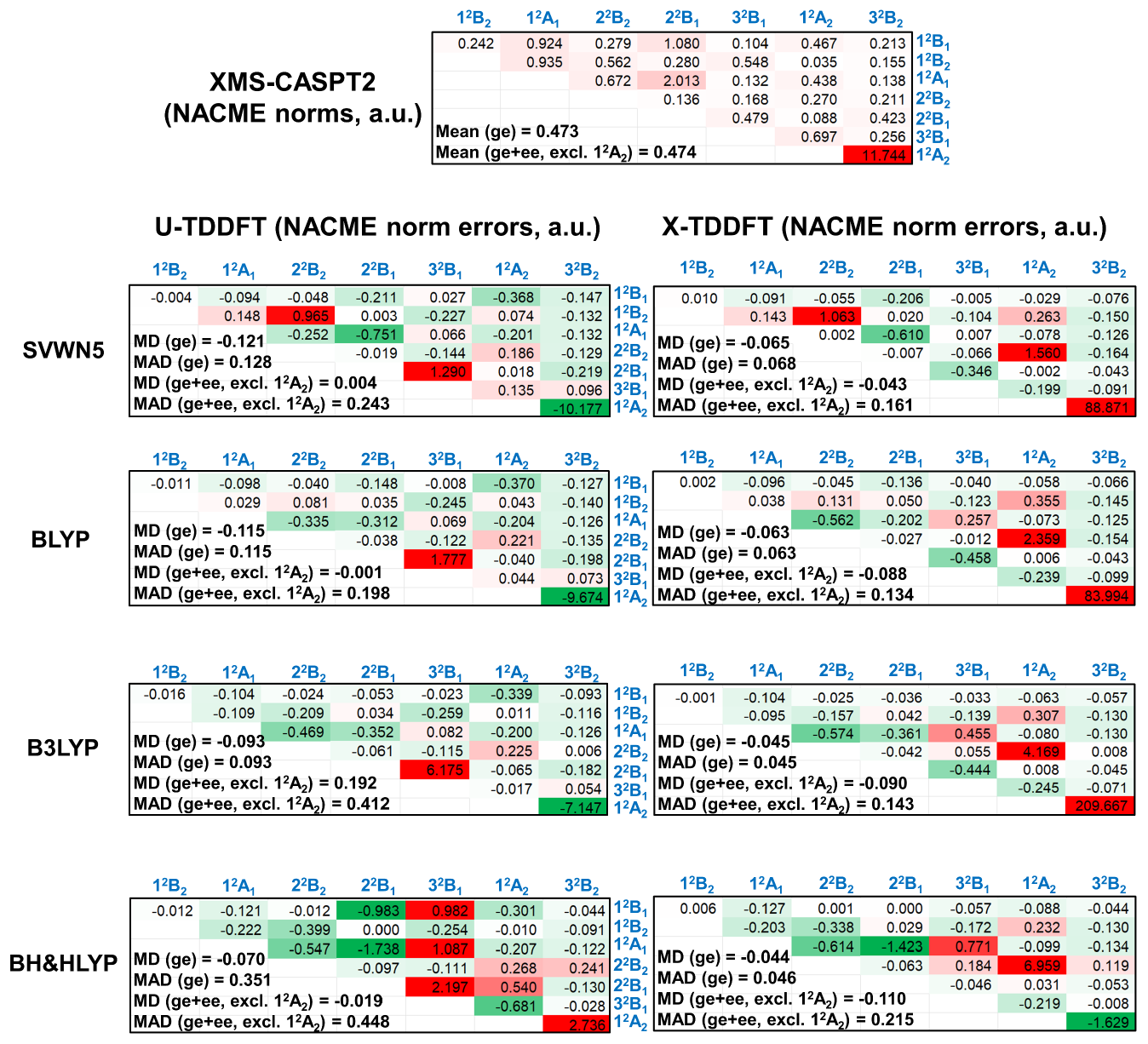}
\caption{Heatmaps of the norm errors of the U-TDDFT and X-TDDFT NACMEs of \ch{CH2O+} computed by different functionals, compared to XMS-CASPT2. The norms of the XMS-CASPT2 NACMEs are shown for comparison. The mean deviations (MD) and mean absolute deviations (MAD) of each method (ge NACMEs and ge+ee NACMEs) are shown. The $1^2 A_2$ state was excluded from the statistics of ge+ee NACMEs.}
\label{fig:ch2o_nacme_norm}
\end{figure}

\begin{figure}[htbp]
\centering
\includegraphics[width=\textwidth]{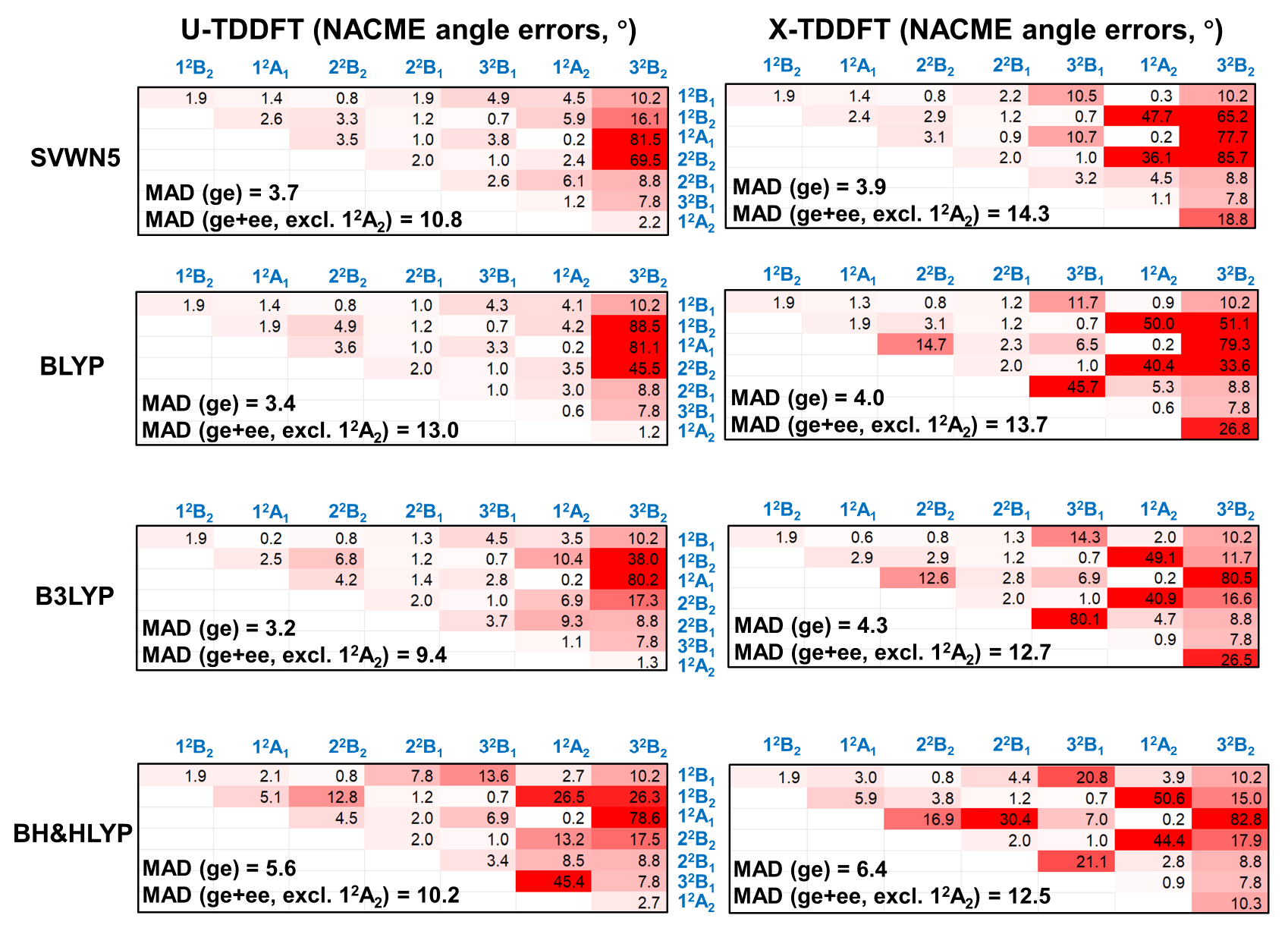}
\caption{Heatmaps of the angles between the U-TDDFT/X-TDDFT NACMEs computed by different functionals, with the XMS-CASPT2 NACMEs of \ch{CH2O+}. The MADs of each method (ge NACMEs and ge+ee NACMEs) are shown (as all angles are positive, the MDs are equal to the MADs and are thus not shown). The $1^2 A_2$ state was excluded from the statistics of ge+ee NACMEs.}
\label{fig:ch2o_nacme_angle}
\end{figure}

We now compare the U-TDDFT and X-TDDFT NACMEs calculated with different functionals (SVWN5,\cite{VWN5} BLYP,\cite{B88x,LYP} B3LYP, BHandHLYP\cite{B88x,LYP,Becke93}) against XMS-CASPT2 results. As shown in Figures~\ref{fig:ch2o_nacme_norm} and \ref{fig:ch2o_nacme_angle}, X-TDDFT reduces the errors of many NACMEs compared to U-TDDFT, at the expense of deteriorating others. In particular, X-TDDFT systematically improves the norms of the ge NACMEs, though slightly worsening their directions. The most consistent improvements are found in the $1^2 B_1$-$1^2 A_2$ and $1^2 B_1$-$3^2 B_2$ NACMEs. An analysis of the excited state compositions of $1^2 A_2$ and $3^2 B_2$ shows that while U-TDDFT predicts opposite signs for the CV($\alpha\alpha$) and CV($\beta\beta$) contributions, both XMS-CASPT2 and X-TDDFT predict that they have the same sign (Table~\ref{table:compositions}). Therefore, the CV($\alpha\alpha$) and CV($\beta\beta$) contributions cancel out in U-TDDFT but add up in both XMS-CASPT2 and X-TDDFT. Such a trend is different from those expected for pure CV(0) or pure CV(1) states: the ge NACMEs of pure CV(0) states are expected to change minimally from U-TDDFT to X-TDDFT, where those of pure CV(1) states should be shortened. By contrast, both $1^2 A_2$ and $3^2 B_2$ are heavy mixtures of CV(0) and CV(1) states, whose coefficients change significantly upon spin contamination, resulting in longer X-TDDFT NACMEs than their U-TDDFT counterparts.

Regarding ee NACMEs, the most significant improvements of X-TDDFT relative to U-TDDFT are found in the $2^2 B_1$-$3^2 B_1$ NACMEs, where U-TDDFT predicts NACMEs that are one order of magnitude too long, while X-TDDFT gives a reasonable estimate. This is obviously attributed to the drastic underestimation of the U-TDDFT $3^2 B_1$ excitation energies due to spin contamination, leading to an underestimation of $\omega(3^2 B_1)-\omega(2^2 B_1)$ by an order of magnitude. Some other NACMEs involving the $3^2 B_1$ state, especially $1^2 B_2$-$3^2 B_1$ and $2^2 B_2$-$3^2 B_1$, are also consistently improved. However, the improvement of the $1^2 B_2$-$3^2 B_1$ NACMEs cannot be explained by excitation energy differences, and must be due to the improved excited state compositions of $3^2 B_1$, as $\omega(3^2 B_1)-\omega(1^2 B_2)$ increases (Table~\ref{table:compositions}), but the NACME vector is elongated, upon going from U-TDDFT to X-TDDFT.

We then move to the NACMEs for which X-TDDFT is less accurate than U-TDDFT. The worst cases are the NACMEs between the $1^2 A_2$ state and another excited state, e.g.~the $1^2 A_2$-$2^2 B_2$ NACME for which X-TDDFT predicts a much longer NACME vector compared to XMS-CASPT2, while U-TDDFT performs much better. However, a deeper look into the excited state compositions (Table~\ref{table:compositions}) shows that U-TDDFT got the right answer for the wrong reason. The XMS-CASPT2 wavefunction of $1^2 A_2$ is dominated by the $[5a_1(\beta)\to 2b_1(\beta), 2b_1(\alpha)\to 2b_2(\alpha)]$ double excitation, followed by the CV($\alpha\alpha$)($5a_1\to 2b_2$) excitation and then the CV($\beta\beta$)($5a_1\to 2b_2$) excitation. All U-TDDFT and X-TDDFT calculations however predict a higher CV($\beta\beta$)($5a_1\to 2b_2$) composition than CV($\alpha\alpha$)($5a_1\to 2b_2$), probably due to inherent limitations of the density functionals. As the OV($2b_1\to 2b_2$) excitation (the dominant excitation of the $2^2 B_2$ state) is a single excitation with respect to both $[5a_1(\beta)\to 2b_1(\beta), 2b_1(\alpha)\to 2b_2(\alpha)]$ and CV($\alpha\alpha$)($5a_1\to 2b_2$), but a double excitation with respect to CV($\beta\beta$)($5a_1\to 2b_2$), the one-electron NACME contribution $g_{IJ}^{\xi,(1)}$ to the U-TDDFT NACMEs is expected to be small; the two-electron contribution $g_{IJ}^{\xi,(2)}$ is probably also small due to a large $1^2 A_2$-$2^2 B_2$ gap ($\omega_J - \omega_I = 4.11~\mathrm{eV}$). This may explain why U-TDDFT predicts a short NACME vector between the two states. X-TDDFT incorporates $[5a_1(\beta)\to 2b_1(\beta), 2b_1(\alpha)\to 2b_2(\alpha)]$ and therefore gives a much longer NACME vector. The short XMS-CASPT2 NACME vector is thus probably due to a cancellation of contributions of $[5a_1(\beta)\to 2b_1(\beta), 2b_1(\alpha)\to 2b_2(\alpha)]$ and CV($\alpha\alpha$)($5a_1\to 2b_2$). Thus, U-TDDFT fortuitously gives good results due to two errors with opposite signs (neglect of double excitations and underestimation of CV($\alpha\alpha$) contributions), and X-TDDFT fixes the former but not the latter. Other outliers of NACMEs involving the $1^2 A_2$ state can be explained analogously. It should be noted, however, that the pronounced error of the X-TDDFT $1^2 A_2$-$3^2 B_2$ NACME is in part due to a wrong energy separation of the two involved states.

We finish this subsection by two statistical analyses, one including all ge NACMEs, and the other including all ge and ee NACMEs but excluding those involving the $1^2 A_2$ state. For all four tested functionals, X-TDDFT reduces the MADs of the U-TDDFT NACME norms by 32\%$\sim$65\% (47\%$\sim$87\% for ge NACMEs) depending on the functional, at the expense of a slight systematic underestimation of the norms (Figure~\ref{fig:ch2o_nacme_norm}), and a slight worsening of the NACME vectors' directions (Figure~\ref{fig:ch2o_nacme_angle}). As the NACME norm correlates well with the IC rate (Fermi's Golden Rule predicts a quadratic dependence of the latter on the former, which we will numerically verify in the next subsection), we conclude that X-TDDFT should improve the IC rates noticeably compared to U-TDDFT.

\subsection{Realistic systems: copper(II) porphyrins} \label{sec:CuP}

With the encouraging results of \ch{CH2O+}, we move on to larger open-shell molecules (i.e.~copper(II) porphyrins) for which high-level reference calculations are not affordable.
Equilibrium geometries and harmonic vibrational frequencies were calculated in the gas phase at the X-TDDFT PBE0\cite{PBE0}-D3BJ\cite{DFT-D3,DFT-D3BJ}/x2c-SVPall\cite{X2CBas2017} level, while excited-state properties, including vertical excitation energies ($E_\text{ex}$), transition dipole moments, spin–orbit coupling matrix elements (SOCMEs) and NACMEs, were obtained using X-TDDFT at the PBE0/x2c-TZVPall\cite{X2CBas2017} level.
Scalar relativistic effects were taken into account using the spin-free exact two-component (sf-X2C) approach,\cite{X2C2009,X2C-2016,X2CSOC1,sf-X2C-2018,sf-X2C-2020} while the so-DKH1 spin-orbit Hamiltonian\cite{X2CSOC1,X2CSOC2} was used for all SOCME calculations.
ISC rate constants were calculated using the the ESD module\cite{ESD2018} of the ORCA 5.0.4 program\cite{orca-5.0,neese2012orca2,neese2018software3,neese2020orca4} with the thermal vibration correlation function (TVCF) method, based on BDF-calculated SOCMEs and a multimode harmonic oscillator model.
IC rate constants involved in the excited state relaxation process were computed using the MOMAP package,\cite{momap} version 2022A, also employing the TVCF method and a harmonic approximation of the potential energy surfaces. 
Default program parameters were used in all TVCF calculations, except for the ``tmax'' parameter in MOMAP calculations, which controls the propagation time of the TVCF and was set to 3000 fs to guarantee full convergence of the TVCF. 
The adiabatic excitation energies ($E_\text{ad}$) used in the rate constant calculations for the $\mathrm{^2T_1}$ $\rightarrow$ $\mathrm{^2S_0}$ processes of all three molecules were evaluated using X-TDDFT at the PBE0-D3BJ/x2c-TZVPall level.

The static-dynamic-static second-order perturbation theory (SDSPT2)\cite{SDS,SDSPT2,SDSRev} calculations were performed using an active space of CAS(13,14) (selected using the imposed CAS (iCAS) method\cite{iCAS}), including the Cu 3d and 4d orbitals together with the four frontier porphyrin $\pi$ orbitals that constitute the Gouterman four-orbital model (the occupied $a_\text{1u}$, $a_\text{2u}$ and a pair of unoccupied $e_\text{g}$ orbitals). The reference wavefunction for the SDSPT2 calculations was optimized using the iCISCF(2) (iterative configuration interaction (iCI)-based multiconfigurational self-consistent field theory) method\cite{iCISCF} with the configuration interaction (CI) truncation coefficient $C_\mathrm{min}$ set to $1.0\times10^{-5}$. In the subsequent perturbative treatment, the CI truncation coefficient $P_\mathrm{min}$ was set to $1.0\times10^{-3}$. The Pople correction was included in all SDSPT2 calculations. For the basis sets, x2c-TZVPall was employed for the Cu and N atoms, whereas x2c-SVPall was used for the remaining atoms.

For calculations involving d-d states, X-TDDFT was found to predict an incorrect state ordering. Therefore, the $E_\text{ad}$ associated with the d–d states were evaluated according to (where $E_\text{ex}$ stands for vertical excitation energy)
\begin{eqnarray}
& & E_\mathrm{ad}(\mathrm{^2T_1 \text{-} {^2dd_n}}) \nonumber\\
& = & E_\text{ex}(\mathrm{^2T_1, SDSPT2, {^2S_0}~geometry}) - E_\text{ex}(\mathrm{^2dd_n, SDSPT2, {^2S_0}~geometry}) \nonumber\\
& & + [E_\text{ex}(\mathrm{^2T_1, X\textrm{-}TDDFT, {^2T_1}~geometry}) - E_\text{ex}(\mathrm{^2T_1, X\textrm{-}TDDFT, {^2S_0}~geometry})] \nonumber\\
& & - [E_\text{ex}(\mathrm{^2dd_n, X\textrm{-}TDDFT, {^2dd_n}~geometry}) - E_\text{ex}(\mathrm{^2dd_n, X\textrm{-}TDDFT, {^2S_0}~geometry})].
\end{eqnarray}
This allows us to obtain close-to-SDSPT2 quality adiabatic energy differences from a single state-averaged SDSPT2 calculation at the ground state geometry.
Specifically, for the $\mathrm{^2T_1}$ state, a total of 21 doublet states were averaged, comprising one $A_\text{g}$ state, ten $B_\text{2u}$ states, and ten $B_\text{3u}$ states. For the $\mathrm{^2dd_1}$, $\mathrm{^2dd_{2/3}}$, and $\mathrm{^2dd_4}$ states, 2, 3, and 2 doublet states were averaged, respectively, corresponding to the lowest $A_\text{g}$ and $B_\text{1g}$ states for $\mathrm{^2dd_1}$, the lowest $A_\text{g}$, $B_\text{2g}$, and $B_\text{3g}$ states for $\mathrm{^2dd_{2/3}}$, and the two lowest $A_\text{g}$ states for $\mathrm{^2dd_4}$.
Note that multiple states were included for each relevant irreducible representation in the state averaged CASSCF calculation of the $\mathrm{^2T_1}$ state, to ensure successful convergence to the target $\mathrm{^2T_1}$ state. In contrast, only one excited state (two for $\mathrm{^2dd_{2/3}}$ due to symmetry reasons) was included for the 
$\mathrm{^2dd_n}$ states, since the higher roots under these irreducible representations lie substantially higher in energy. This setup has been validated as reasonable in our previous work.\cite{wang2023CuP}

Kinetics simulations were performed using a modified version of our MATLAB script.\cite{wang2025molecules,zhou2026imi} Since all reactions are first order, the kinetics has closed form and can be simulated across multiple timescales (from picosecond to microsecond) without numerical integration error. The luminescence lifetime is defined as the time required for $1-1/e \approx 63.2\%$ of the total luminescence photons to emit.\cite{wang2023CuP}

All orbital diagrams were drawn and visualized with VMD,\cite{VMD} version 1.9.3, based on cube files generated by Multiwfn 3.8(dev).\cite{multiwfn2012,multiwfn2024}

\begin{figure}[htbp]
\centering
\includegraphics[width=0.9\textwidth]{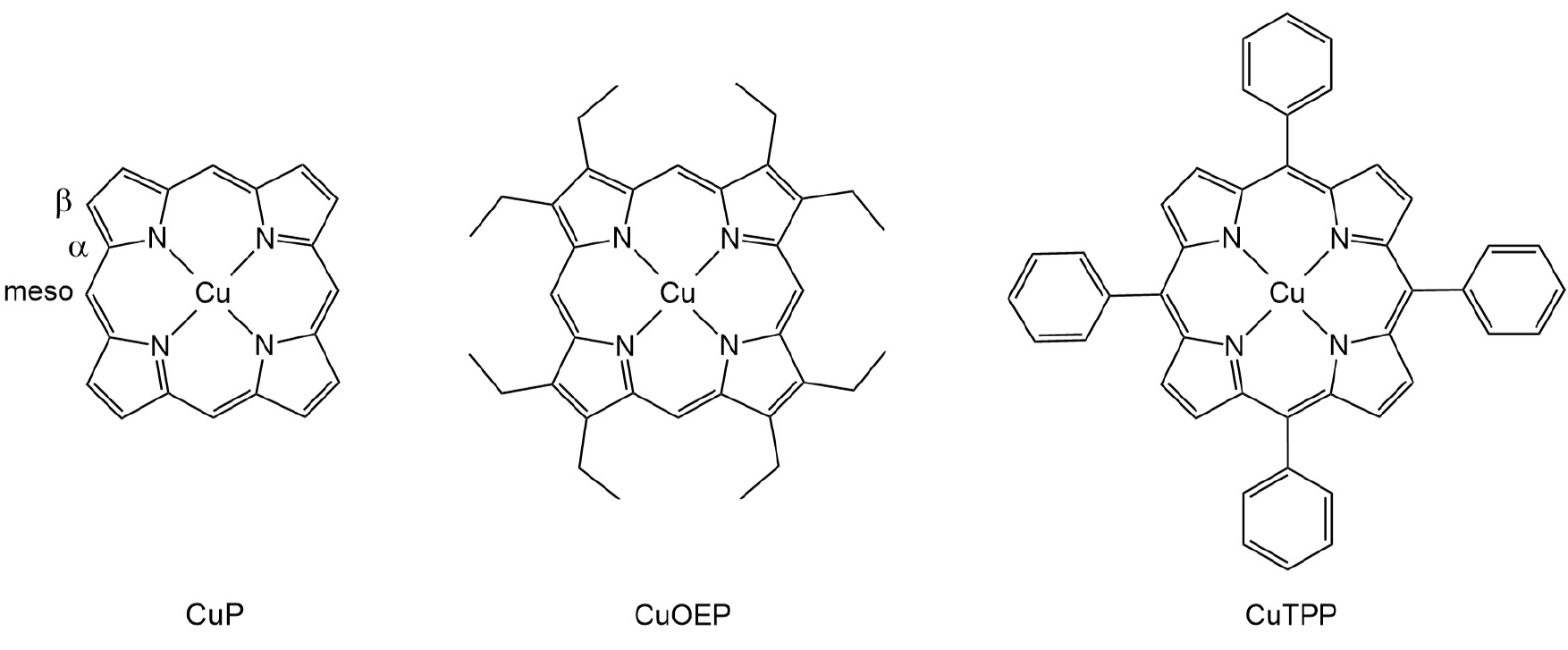}
\caption{Molecular structures of CuP, CuOEP, and CuTPP.}
\label{Cu-porphyrin}
\end{figure}

Table~\ref{nacmes} shows the computed photophysical parameters of the $\mathrm{^2T_1}$ state (here the 2 represent the overall spin multiplicity of the complex, and T denotes that the ``local'' spin multiplicity of the porphyrin ring is triplet) of the unsubstituted copper(II) porphyrin CuP, as well as its octaethyl and tetraphenyl substituted analogs (Figure~\ref{Cu-porphyrin}).
For the transition from $\mathrm{^2T_1}$ to the ground state ($\mathrm{^2S_0}$), the spin contamination of U-TDDFT leads to an underestimation of the vertical emission energies by 0.04$\sim$0.06 eV relative to X-TDDFT, in accord with previous observations.\cite{wang2023CuP} This is because the $\mathrm{^2T_1}$ state, being a ``tripdoublet'' state obtained from antiferromagnetic coupling of a triplet ligand and a doublet metal, has almost pure CV(1) character, and therefore its U-TDDFT calculations suffer maximally from spin contamination.

\begin{table}[t]
\centering
\caption{Parameters for the $\mathrm{^2T_1}$ $\rightarrow$ $\mathrm{^2S_0}$ process calculated using the U-TDDFT and X-TDDFT methods, including the vertical emission energy differences based on the $\mathrm{^2T_1}$ geometries ($\Delta E$, in eV), the norms of the NACMEs (Norm), the internal conversion rates ($k_\text{IC}$, in $\mathrm{s^{-1}}$), and the angles between the U-TDDFT and X-TDDFT NACMEs (Angle, in $^{\circ}$).}
\begin{tabular}{cccccccccc}
\hline
\multirow{2}{*}{Mol.} & \multicolumn{3}{c}{U-TDDFT} & \multicolumn{3}{c}{X-TDDFT} & \multirow{2}{*}{Angle} & \multirow{2}{*}{$\left(\frac{\text{Norm}_\text{X}}{\text{Norm}_\text{U}}\right)^2$} & \multirow{2}{*}{$\frac{k_\text{IC, X}}{k_\text{IC, U}}$}\\
\cmidrule(lr){2-4} \cmidrule(lr){5-7}
& $\Delta E$ & Norm & $k_\text{IC}$ & $\Delta E$ & Norm & $k_\text{IC}$ & \\
\midrule
 CuP   & 1.48 & 0.0206 & $2.00\times10^8$ & 1.53 & 0.0034 & $5.31\times10^6$ & 25.9 & 0.027 & 0.027\\
 CuOEP & 1.50 & 0.0293 & $1.71\times10^6$ & 1.54 & 0.0127 & $2.71\times10^5$ & 21.8 & 0.188 & 0.159\\
 CuTPP & 1.32 & 0.0283 & $2.17\times10^7$ & 1.38 & 0.0030 & $2.17\times10^5$ & 36.5 & 0.011 & 0.010\\
\hline
\end{tabular}
\label{nacmes}
\end{table}

\begin{figure}[htbp]
\centering
\includegraphics[width=0.8\textwidth]{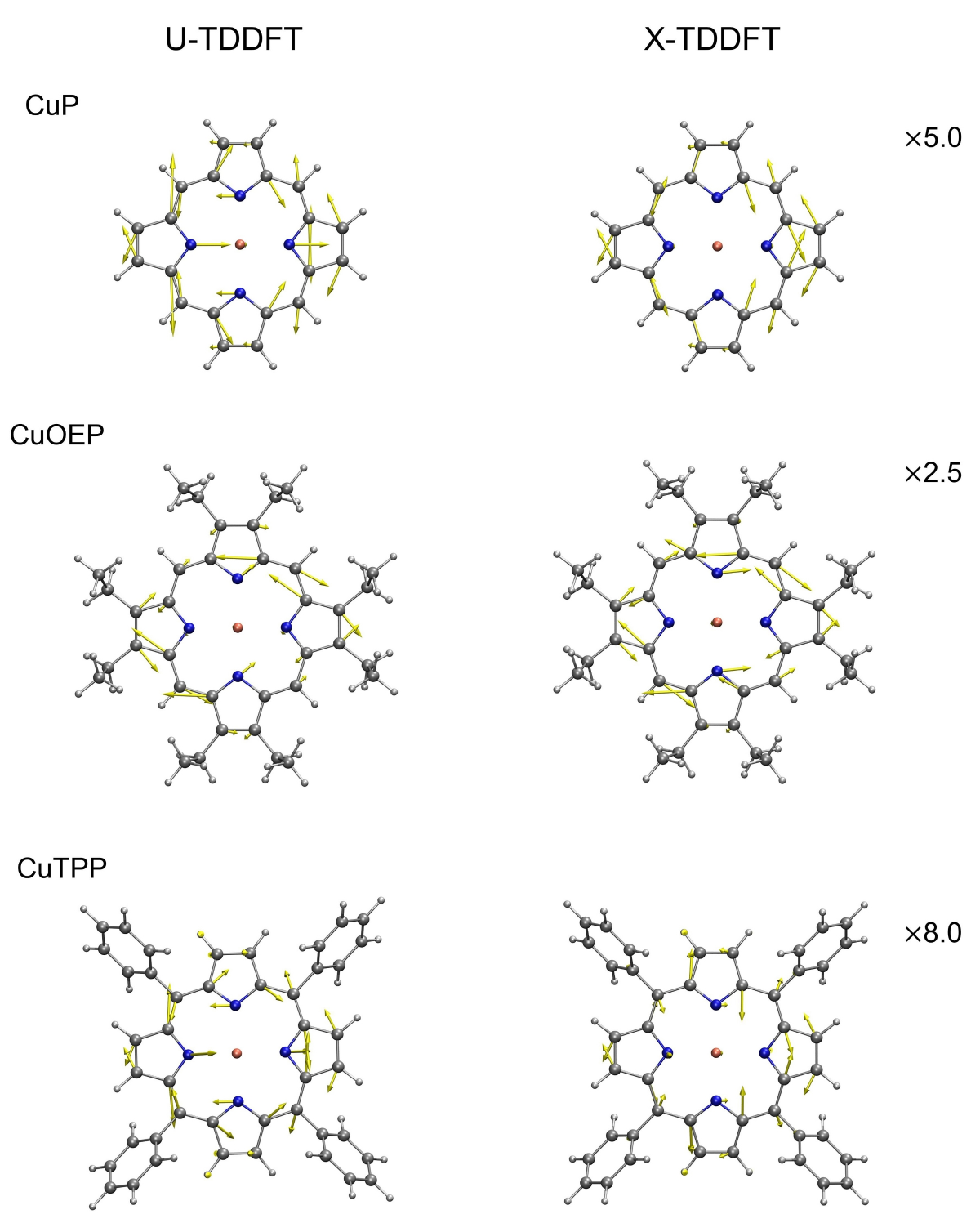}
\caption{NACME vectors obtained from U-TDDFT and X-TDDFT for the molecules CuP, CuOEP, and CuTPP. As the X-TDDFT NACMEs are much shorter than the U-TDDFT ones, the scale used for X-TDDFT is multiplied by a factor relative to that of U-TDDFT, as indicated in the upper right corner of each subfigure. (The NACME vector plots were rendered in VMD using scripts available at http://sobereva.com/attach/567/file.zip.)}
\label{U-X NACMEs}
\end{figure}

In our previous work,\cite{wang2023CuP} all IC rates were calculated using U-TDDFT NACMEs, though with X-TDDFT geometries/Hessians and SDSPT2 adiabatic excitation energies.
We now recompute the NACMEs using X-TDDFT.
As shown in Table~\ref{nacmes}, X-TDDFT yields systematically smaller NACME magnitudes than U-TDDFT; while for CuOEP only a 2.3-fold reduction is observed, for CuP and CuTPP the difference amounts to almost an order of magnitude. In any case, the reduction is too large to be explained by the change of the initial state-final state energy difference. Instead, we note that pure CV(1) excitations give zero contribution to X-TDDFT ge NACMEs, whereas this is not expected to hold exactly in U-TDDFT, as the UKS $\alpha$ and $\beta$ orbitals have different shapes, such that their contributions to NACMEs are unlikely to exactly cancel out. The residual non-zero X-TDDFT NACMEs therefore must be due to tiny admixtures of CV(0), CO and OV excitations of the $\mathrm{^2T_1}$ state.
Meanwhile, the NACME vectors obtained from U-TDDFT and X-TDDFT exhibit a moderate angular deviation ($21.8^\circ \sim 36.5^\circ$), indicating that although the dominant non-adiabatic coupling mechanism predicted by the two methods remains essentially the same, the relative contributions of individual nuclear degrees of freedom to the total coupling vector are redistributed after spin-adaptation. Such redistribution is most obvious for CuP and CuTPP, where U-TDDFT NACMEs have sizable contributions from the four nitrogen atoms, while X-TDDFT NACMEs are almost free of such contributions (Figure~\ref{U-X NACMEs}). Despite this, the ratio of the X-TDDFT and U-TDDFT IC rates remains in good accord with the square of the ratio of the NACME norms (Table~\ref{nacmes}), indicating that the directions of the NACME vectors do not significantly influence the IC rates in this system. Due to the much smaller spin-adaptation correction of CuOEP's IC rate compared to those of CuP and CuTPP, while U-TDDFT predicts the IC rates to follow the trend CuOEP < CuTPP < CuP, X-TDDFT predicts CuTPP < CuOEP < CuP, i.e.~U-TDDFT is unreliable not only for absolute IC rates, but even for the relative IC rates of closely related molecules.

\begin{figure}[htbp]
\centering
\includegraphics[width=0.75\textwidth]{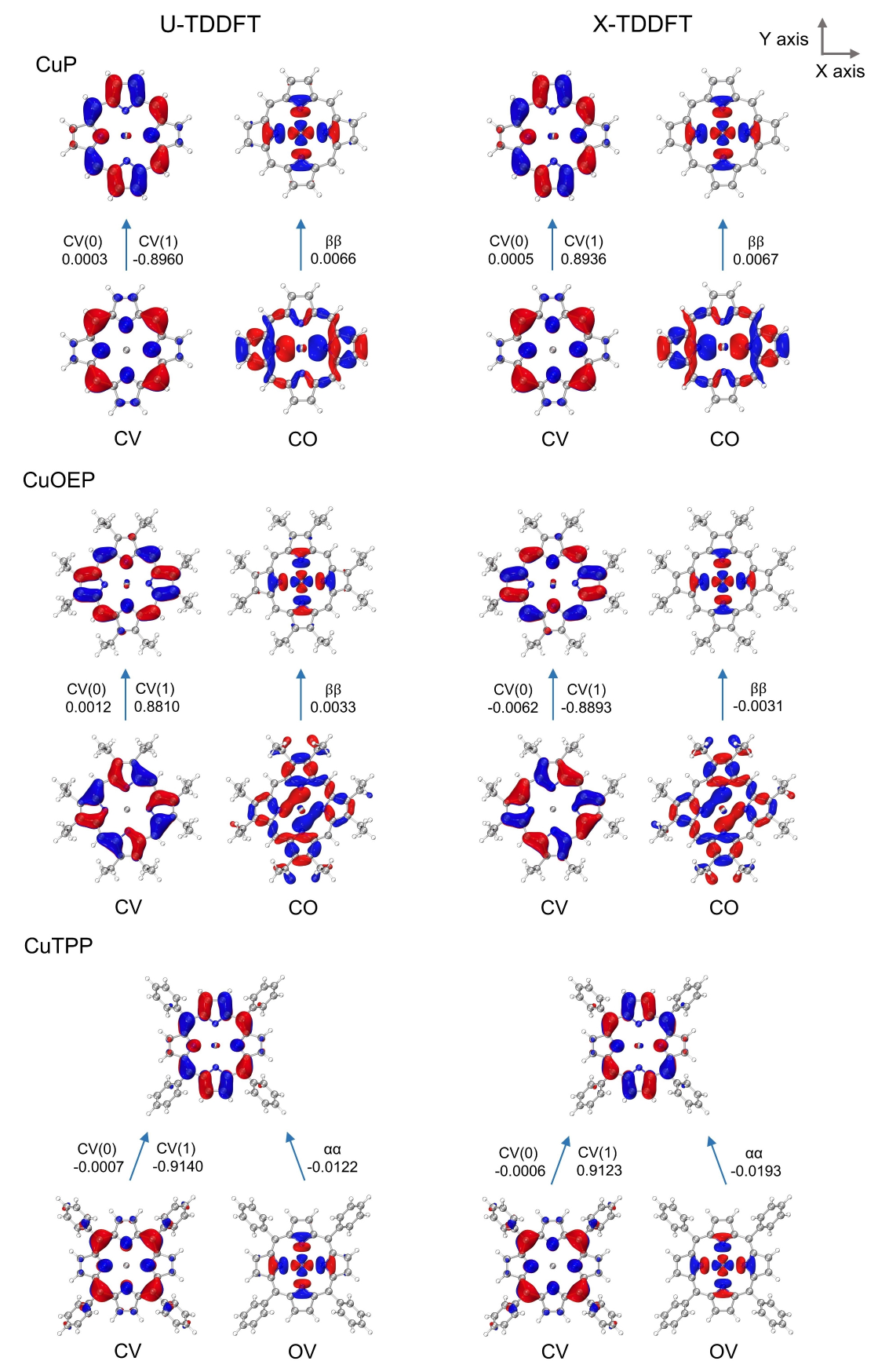}
\caption{$\mathrm{^2T_1}$ state configuration coefficients of selected excitations of CuP, CuOEP, and CuTPP at the $\mathrm{^2T_1}$ geometries, and the plots of the involved canonical orbitals (PBE0/x2c-TZVPall, isovalue: 0.03). The dominant excitation (which is always a CV(1) excitation), the accompanying CV(0) excitation, and the non-CV excitation with the largest contribution percentage are shown. For CV excitations, the plots of $\alpha$ orbitals are shown; the shapes of the corresponding $\beta$ orbitals do not differ noticeably.}
\label{2T1-orbitals}
\end{figure}

\begin{figure}[htbp]
\centering
\includegraphics[width=0.85\textwidth]{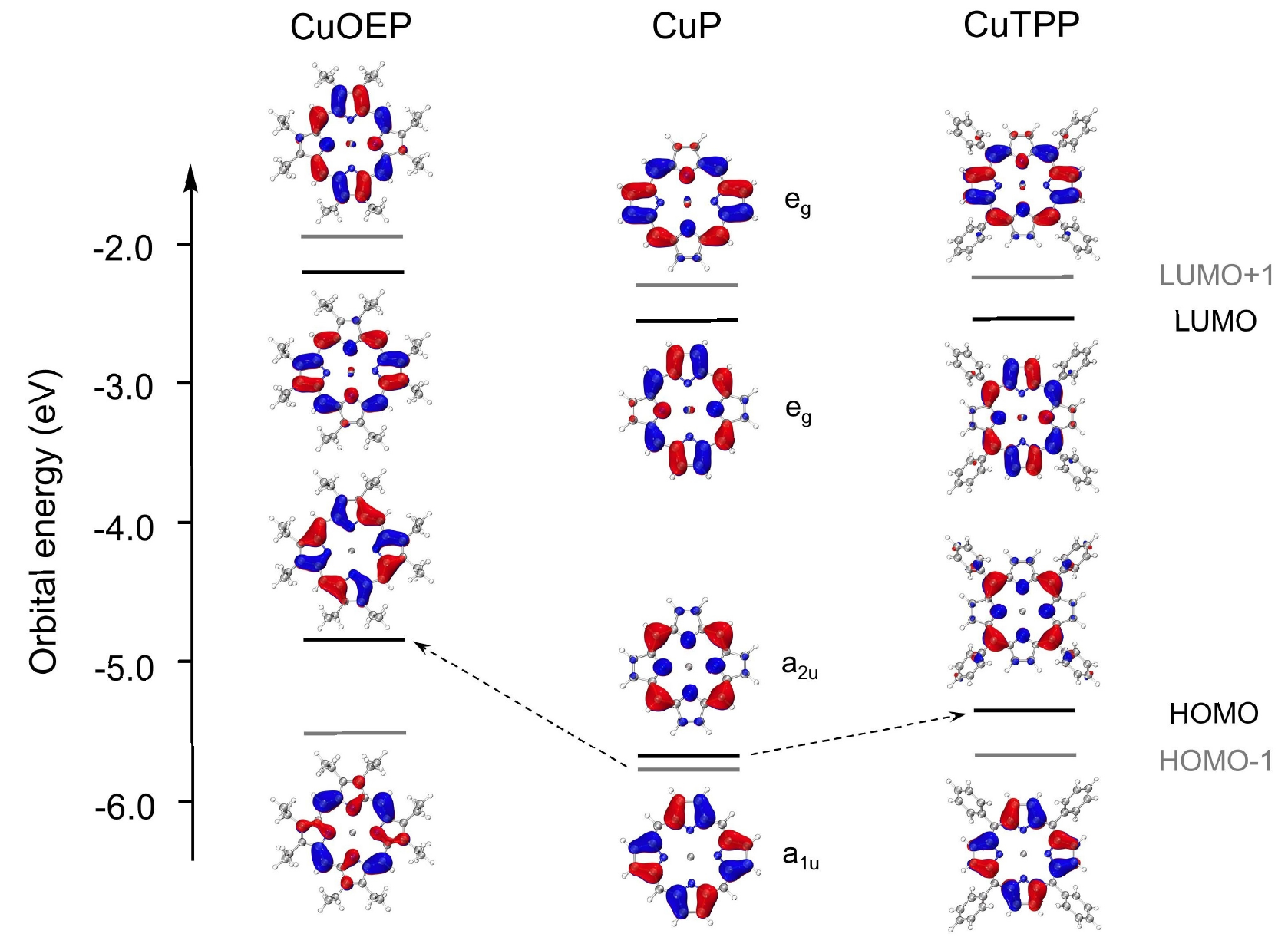}
\caption{ROKS frontier molecular orbital energy levels and isosurface plots of CuOEP, CuP and CuTPP at the $\mathrm{^2T_1}$ geometries (PBE0/x2c-TZVPall, isovalue: 0.03). The SOMO (Cu 3d$_{x^2-y^2}$) orbitals, which lie between the ligand HOMO and LUMO, are omitted for clarity. The correspondence of the ligand HOMO and HOMO-1 orbitals between different molecules is highlighted with dashed arrows.}
\label{H-L energy}
\end{figure}

As shown in Figure~\ref{2T1-orbitals}, to further elucidate the origin of the differences in the NACMEs calculated using U-TDDFT and X-TDDFT, we analyzed the orbitals involved in the dominant configurations of the $\mathrm{^2T_1}$ states of CuP, CuOEP and CuTPP. The UKS and ROKS orbitals are similar for all three molecules. CuP and CuTPP exhibit similar orbital shapes, whereas CuOEP shows noticeable differences, particularly for the occupied orbitals. This behavior can be understood in terms of substituent effects. Ethyl groups, which act as weak electron-donating substituents, occupy the eight $\beta$ positions of the porphyrin ring in CuOEP. Because the $a_\text{1u}$ orbital has a large distribution at the $\beta$ positions, ethyl substitution selectively raises the energy of the $a_\text{1u}$ orbital, while keeping the $a_\text{2u}$ orbital energy largely intact, giving a HOMO that is dominated by $a_\text{1u}$ character with small $a_\text{2u}$ contributions (Figure~\ref{H-L energy}). In contrast, in CuTPP, the four \emph{meso} phenyl groups mainly influence the $a_\text{2u}$ orbital (which has significant amplitudes at the \emph{meso} positions), while exerting a relatively minor effect on the $a_\text{1u}$ orbital. As a result, CuTPP preserves the same energy level ordering and similar orbital shapes as CuP.

We then investigate the $\mathrm{^2T_1}$ state compositions of the copper(II) porphyrins (Figure~\ref{2T1-orbitals}). All excited states, no matter computed via U-TDDFT or X-TDDFT, are dominated by the HOMO$\to$LUMO transition, but as the HOMO of CuOEP is a mixture of $a_\text{1u}$ and $a_\text{2u}$ character, while the HOMOs of the other two molecules have almost pure $a_\text{2u}$ character, the $\mathrm{^2T_1}$ state of CuOEP is characterized by a more severe symmetry breaking than those of CuP and CuTPP. This encourages a heavier admixture of CV(0) excitations and therefore a longer NACME vector at the X-TDDFT level (Figure~\ref{2T1-orbitals}). Considering that the $\mathrm{^2T_1}$ state can also gain a non-zero ge NACME by mixing with CO or OV excitations, we have also plotted the CO/OV excitations that contributed the most to the $\mathrm{^2T_1}$ states of the three molecules (Figure~\ref{2T1-orbitals}). However, the contribution of this excitation is smaller for CuOEP than for CuP and CuTPP, suggesting that mixture with CO/OV excitations may not be the dominant contribution to the smaller X-TDDFT/U-TDDFT difference of CuOEP compared to CuP and CuTPP. In sum, the smaller improvement of X-TDDFT $\mathrm{^2T_1}$-$\mathrm{^2S_0}$ NACMEs for CuOEP is probably due to its larger CV(0) excitation contributions. More generally, our results suggest that X-TDDFT can give radically different corrections upon U-TDDFT NACMEs for closely related molecules (CuP, CuOEP, CuTPP), even when the corrections to the excitation energies are negligible (Table~\ref{nacmes}).

\begin{table}[ht]
\centering
\caption{Transition parameters for the $\mathrm{^2T_1}$ $\rightarrow$ $\mathrm{^2dd_n}$ process of CuP calculated using the U-TDDFT and X-TDDFT methods, including the excitation energies ($E$, in eV) of the $\mathrm{^2dd_n}$ state, the TDDFT vertical emission energy differences ($\Delta E$, in eV) between the $\mathrm{^2T_1}$ and $\mathrm{^2dd_n}$ states based on the $\mathrm{^2T_1}$ geometries, the norms of the NACMEs (Norm, in a.u.), the IC rates ($k_\text{IC}$, in $\mathrm{s^{-1}}$), and the angles between the U-TDDFT and X-TDDFT NACMEs (Angle, in $^{\circ}$). Note that $\Delta E$ is only used to compare against the NACME trends; the adiabatic excitation energies used in IC rate calculations are corrected by SDSPT2 results.}
\footnotesize
\setlength{\tabcolsep}{5.6pt}
\begin{tabular}{cccccccccccc}
\hline
\multirow{2}{*}{State$^a$} & \multicolumn{4}{c}{U-TDDFT} & \multicolumn{4}{c}{X-TDDFT} & \multirow{2}{*}{Angle} & \multirow{2}{*}{$\left(\frac{\text{Norm}_\text{X}}{\text{Norm}_\text{U}}\right)^2$} & \multirow{2}{*}{$\frac{k_\text{IC, X}}{k_\text{IC, U}}$} \\
\cmidrule(lr){2-5} \cmidrule(lr){6-9}
& $E$ & $\Delta E$ & Norm & $k_\text{IC}$ & $E$ & $\Delta E$ & Norm & $k_\text{IC}$ & \\
\midrule
 $\mathrm{^2dd_1}$ & 2.52 & 1.04 & 0.0103 & $9.69\times10^7$ & 2.45 & 0.92 & 0.0114 & $9.73\times10^7$ & 6.5 & 1.235 & 1.004\\ 
 $\mathrm{^2dd_2}$ & 2.29 & 0.81 & 0.0313 & $4.77\times10^8$ & 2.24 & 0.70 & 0.0222 & $1.33\times10^8$ & 41.1 & 0.505 & 0.278\\
 $\mathrm{^2dd_3}$ & 2.24 & 0.76 & 0.0078 & $4.15\times10^7$ & 2.18 & 0.65 & 0.0110 & $7.44\times10^7$ & 5.9 & 2.023 & 1.794\\
 $\mathrm{^2dd_4}$ & 2.28 & 0.80 & 0.0077 & $2.68\times10^7$ & 2.23 & 0.70 & 0.0109 & $5.84\times10^7$ & 4.8 & 2.015 & 2.177\\
Total & & & & $6.43\times10^8$ & & & & $3.63\times10^8$ & & & \\
\hline
\end{tabular}\\
\textsuperscript{\emph{a}} $\mathrm{^2dd_1}$: 3d$_{xy}$$\rightarrow$3d$_{x^2-y^2}$, $\mathrm{^2dd_2}$: 3d$_{xz}$$\rightarrow$3d$_{x^2-y^2}$, $\mathrm{^2dd_3}$: 3d$_{yz}$$\rightarrow$3d$_{x^2-y^2}$, $\mathrm{^2dd_4}$: 3d$_{z^2}$$\rightarrow$3d$_{x^2-y^2}$. \\
\label{nacmes2}
\end{table}

Another important relaxation pathway of the $\mathrm{^2T_1}$ states is the $\mathrm{^2T_1 \to {^2dd_n}}$ IC process. These are CV(1)$\to$CO transitions, such that if $g_{IJ}^{\mathrm{SA},\xi,(1)}$ dominates over $g_{IJ}^{\mathrm{SA},\xi,(2)}$, we would expect the X-TDDFT NACME norm to be \emph{longer} than the U-TDDFT one by a factor of $\sqrt{3}$, corresponding to an increase of the IC rate by a factor of 3. In reality, the NACME changes are smaller than expected, and for the $\mathrm{^2dd_2}$ state X-TDDFT even predicts a shorter NACME than U-TDDFT; these suggest that the contribution of $g_{IJ}^{\mathrm{SA},\xi,(2)}$ may be significant.
The total $\mathrm{^2T_1 \to {^2dd_n}}$ IC rate thus increases only marginally from U-TDDFT to X-TDDFT.
Although we cannot separate the $g_{IJ}^{\mathrm{SA},\xi,(1)}$ and $g_{IJ}^{\mathrm{SA},\xi,(2)}$ terms under the Lagrangian formalism, we were able to turn off the extra term of $g_{IJ}^{\mathrm{SA},\xi,(1)}$ compared to $g_{IJ}^{\xi,(1)}$ (i.e.~$\Delta g_{IJ}^{\xi}$, Eq.~\eqref{COCV1}), and the extra terms of $g_{IJ}^{\mathrm{SA},\xi,(2)}$ compared to $g_{IJ}^{\xi,(2)}$ (i.e.~$\Delta\mathbf{Q}$ and $\Delta\mathbf{\Gamma}$, Eqs.~\eqref{DeltaQ} and \eqref{DeltaGamma}), to elucidate their contributions to the NACMEs (Table~\ref{nacmes3}). Removing $\Delta g_{IJ}^{\xi}$ almost leaves the NACMEs unchanged, except for the direction of the $\mathrm{^2T_1 \to {^2dd_2}}$ NACMEs; this is consistent with the fact that the $\mathrm{^2T_1}$ and $\mathrm{^2dd_n}$ states differ by a double excitation. Further removing $\Delta\mathbf{Q}$ and $\Delta\mathbf{\Gamma}$ makes the $\mathrm{^2T_1 \to {^2dd_2}}$ X-TDDFT NACMEs similar to the corresponding U-TDDFT NACMEs, while still having only small effects on the remaining NACMEs. Finally, the differences of X-TDDFT and U-TDDFT energy gaps ($\omega_J - \omega_I$) are invariably small (within 15\% of the energy gaps themselves; Table~\ref{nacmes2}) and contribute negligibly to the NACME norms. Thus, while the X-TDDFT-related extra terms in $g_{IJ}^{\mathrm{SA},\xi,(2)}$ are important for the accuracy of the $\mathrm{^2T_1 \to {^2dd_n}}$ NACMEs, other factors such as the reference wavefunction (ROKS vs.~UKS) and the differences of X-TDDFT and U-TDDFT excitation vectors may also contribute to the differences of the X-TDDFT and U-TDDFT NACMEs.

\begin{table}[h]
\centering
\caption{Norms (a.u.) of the $\mathrm{^2T_1}$ $\rightarrow$ $\mathrm{^2dd_n}$ NACMEs of CuP, calculated using the X-TDDFT methods but with certain terms turned off. The angles ($^{\circ}$) between these NACMEs and the U-TDDFT/X-TDDFT NACMEs are also shown.}
\small
\begin{tabular}{cccc}
\hline
 State & Norm & Angle (U-TDDFT) & Angle (X-TDDFT) \\
\midrule
 $\mathrm{^2dd_1}$ (turn off $\Delta g_{IJ}^{\xi}$) & 0.0115 & 6.5 & 0.1 \\
 $\mathrm{^2dd_2}$ (turn off $\Delta g_{IJ}^{\xi}$) & 0.0239 & 30.7 & 10.6 \\
 $\mathrm{^2dd_3}$ (turn off $\Delta g_{IJ}^{\xi}$) & 0.0107 & 4.3 & 1.8 \\
 $\mathrm{^2dd_4}$ (turn off $\Delta g_{IJ}^{\xi}$) & 0.0109 & 4.6 & 0.3 \\
 $\mathrm{^2dd_1}$ (turn off $\Delta g_{IJ}^{\xi}$, $\Delta\mathbf{Q}$ and $\Delta\mathbf{\Gamma}$) & 0.0115 & 6.4 & 0.5 \\
 $\mathrm{^2dd_2}$ (turn off $\Delta g_{IJ}^{\xi}$, $\Delta\mathbf{Q}$ and $\Delta\mathbf{\Gamma}$) & 0.0323 & 5.6 & 36.8 \\
 $\mathrm{^2dd_3}$ (turn off $\Delta g_{IJ}^{\xi}$, $\Delta\mathbf{Q}$ and $\Delta\mathbf{\Gamma}$) & 0.0095 & 1.2 & 5.0\\
 $\mathrm{^2dd_4}$ (turn off $\Delta g_{IJ}^{\xi}$, $\Delta\mathbf{Q}$ and $\Delta\mathbf{\Gamma}$) & 0.0115 & 5.8 & 1.5 \\

\hline
\end{tabular}\\
\label{nacmes3}
\end{table}

For completeness, we also calculated the IC rates between the different $\mathrm{^2dd_n}$ states, as well as between them and the ground state $\mathrm{^2S_0}$ (Table~\ref{IC-2dd-2S0}). As expected, in most cases, the X-TDDFT and U-TDDFT IC rates differ minimally, as all the $\mathrm{^2dd_n}$ states are CO states and are thus almost free of spin contamination. The large difference in the U-TDDFT and X-TDDFT $\mathrm{^2dd_4 \text{-} ^2dd_{2/3}}$ IC rates can be attributed to the small vertical energy differences of these transitions (0.0022 and 0.0066 eV for U-TDDFT and X-TDDFT, respectively). Due to their smallness, even a small spin-adaptation correction changes the energy difference dramatically, which translates to large changes of the NACMEs. However, these changes are hardly practically relevant, as the TVCF method is unreliable for such ultrafast IC processes anyway, and as long as these rates are predicted to be very fast, the overall kinetics of the excited state relaxation should be largely unaffected.

\begin{table}[h]
\centering
\caption{IC rates (in $\mathrm{s^{-1}}$) for the $\mathrm{^2dd_n}$ $\rightarrow$ $\mathrm{^2S_0}$ and $\mathrm{^2dd_m}$-$\mathrm{^2dd_n}$ transitions, calculated with U-TDDFT and X-TDDFT. The rates involving $\mathrm{^2dd_3}$ are equal to the corresponding rates of $\mathrm{^2dd_2}$ by symmetry.}
\begin{tabular}{ccc}
\hline
 State & U-TDDFT & X-TDDFT \\
\midrule
 $\mathrm{^2dd_1 \text{-} ^2S_0}$  & $2.22\times10^{10}$ & $2.15\times10^{10}$ \\ 
 $\mathrm{^2dd_2 \text{-} ^2S_0}$  & $6.37\times10^8$    & $6.41\times10^8$ \\
 $\mathrm{^2dd_3 \text{-} ^2S_0}$  & $6.37\times10^8$    & $6.41\times10^8$ \\
 $\mathrm{^2dd_4 \text{-} ^2S_0}$  & $2.11\times10^9$    & $2.12\times10^9$ \\
 $\mathrm{^2dd_2 \text{-} ^2dd_1}$ & $7.69\times10^{10}$ & $8.27\times10^{10}$ \\
 $\mathrm{^2dd_3 \text{-} ^2dd_1}$ & $7.69\times10^{10}$ & $8.27\times10^{10}$ \\
 $\mathrm{^2dd_3 \text{-} ^2dd_2}$ & $1.25\times10^{11}$ & $1.26\times10^{11}$ \\
 $\mathrm{^2dd_2 \text{-} ^2dd_3}$ & $1.25\times10^{11}$ & $1.26\times10^{11}$ \\
 $\mathrm{^2dd_4 \text{-} ^2dd_1}$ & $2.21\times10^{11}$ & $2.54\times10^{11}$ \\
 $\mathrm{^2dd_4 \text{-} ^2dd_2}$ & $1.37\times10^{15}$ & $1.73\times10^{14}$ \\
 $\mathrm{^2dd_4 \text{-} ^2dd_3}$ & $1.37\times10^{15}$ & $1.73\times10^{14}$ \\
\hline
\end{tabular}\\
\label{IC-2dd-2S0}
\end{table}

\begin{figure}[h]
\begin{tabular}{cc}
U-TDDFT & X-TDDFT \\
\resizebox{0.45\textwidth}{!}{\includegraphics{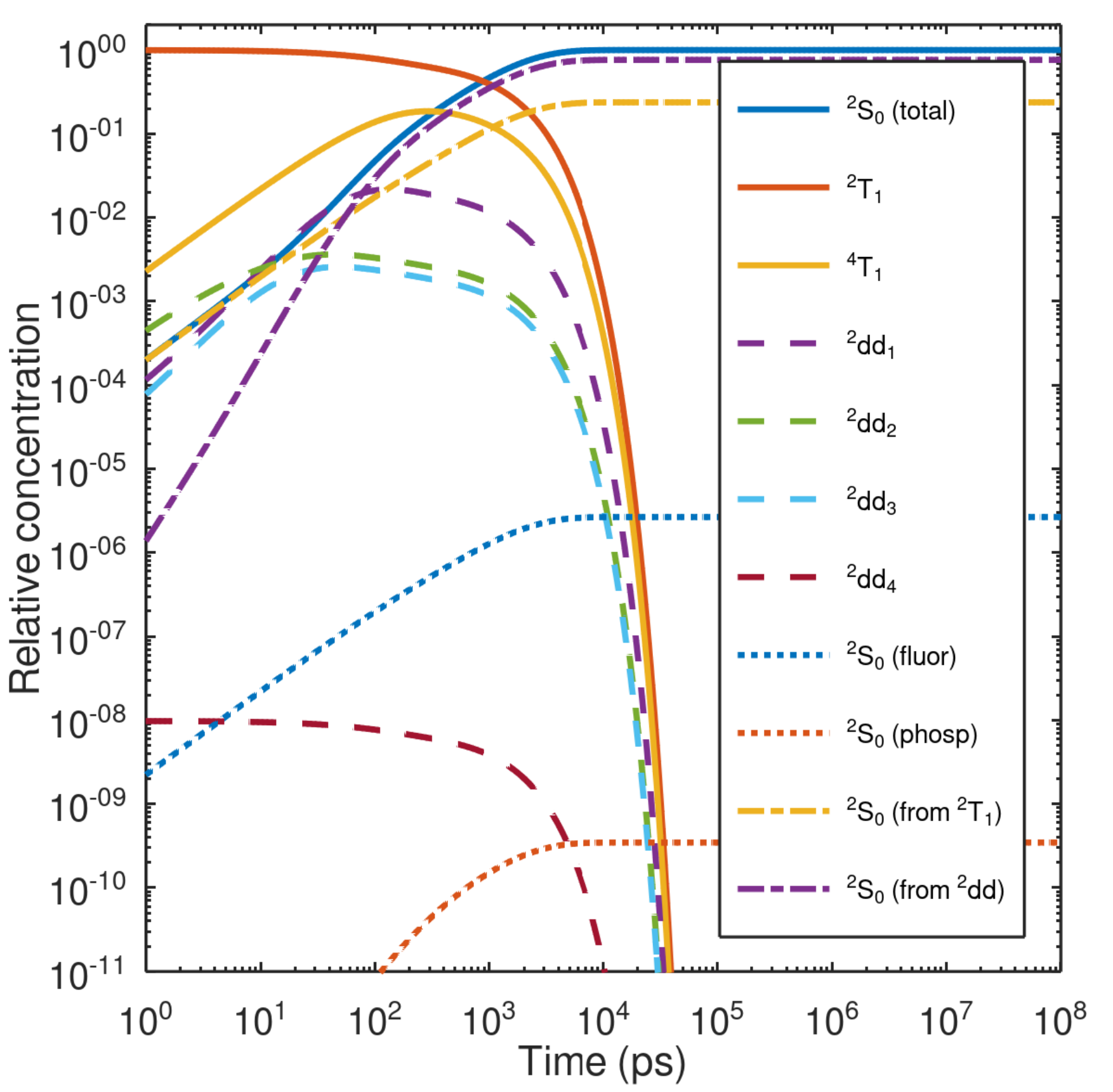}} &
\resizebox{0.45\textwidth}{!}{\includegraphics{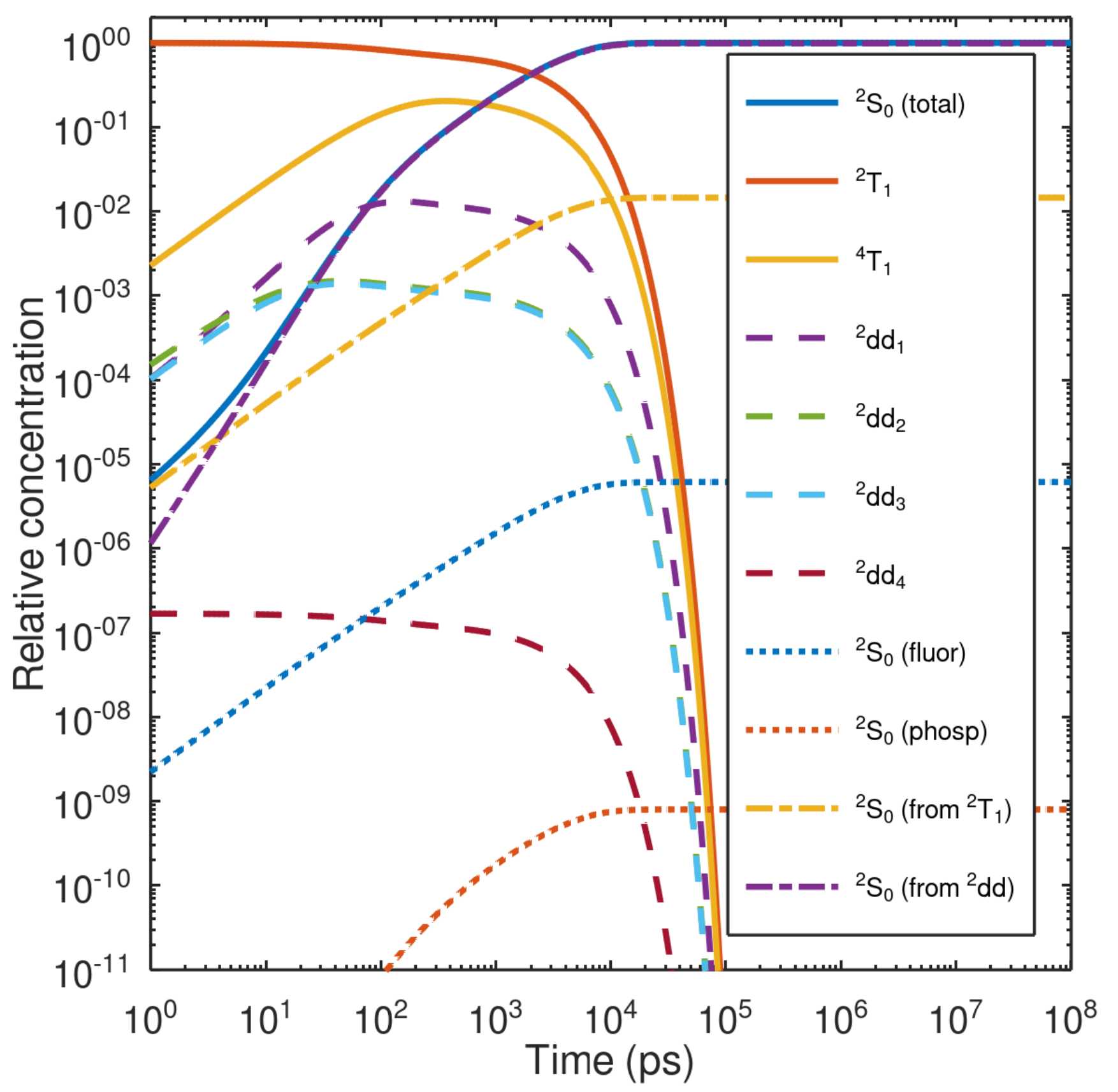}} \\
\end{tabular}
\caption{Relative concentrations of different states as a function of time, starting from the $\mathrm{^2T_1}$ state. The kinetic simulations were performed using IC rate constants from U-TDDFT NACMEs and X-TDDFT NACMEs, respectively. ``$\mathrm{^2S_0}$ (fluor)'', ``$\mathrm{^2S_0}$ (phosp)'', ``$\mathrm{^2S_0}$ (from $\mathrm{^2T_1}$)'' and ``$\mathrm{^2S_0}$ (from $\mathrm{^2dd}$)'' denote the $\mathrm{^2S_0}$ state reached via fluorescence emission from $\mathrm{^2T_1}$, phosphorescence emission from $\mathrm{^4T_1}$, direct IC from $\mathrm{^2T_1}$ and IC from one of the $\mathrm{^2dd_n}$ states, respectively.}
\label{fig:kinetics}
\end{figure}

We can now simulate the excited state kinetics using the above IC rate constants (Figure~\ref{fig:kinetics}); the required ISC ($\mathrm{^2T_1 \text{-} ^4T_1}$) and reverse ISC ($\mathrm{^4T_1 \text{-} ^2T_1}$) rates have been recomputed at the current level of theory as $7.80\times 10^9~\mathrm{s}^{-1}$ and $2.30\times 10^9~\mathrm{s}^{-1}$, respectively (here $\mathrm{^4T_1}$ is the first quartet state), while the fluorescence ($\mathrm{^2T_1 \text{-} ^2S_0}$) and phosphorescence ($\mathrm{^4T_1 \text{-} ^2S_0}$) rates were taken from Ref.~\citenum{wang2023CuP}.
Compared to U-TDDFT, X-TDDFT predicts a later rise of the ground state concentration (obviously due to the slower $\mathrm{^2T_1 \text{-} ^2S_0}$ IC), a lower steady state concentration of $\mathrm{^2dd_{1/2/3}}$, as well as a higher concentration of $\mathrm{^2dd_4}$. Overall, X-TDDFT predicts a longer luminescence lifetime (3.6 ns), and a higher luminescence quantum yield ($6.1\times 10^{-6}$), than U-TDDFT (1.5 ns and $2.7\times 10^{-6}$, respectively).
Moreover, as X-TDDFT predicts a moderately lower $\mathrm{^2T_1 \to {^2dd_n}}$ IC rate but a much lower $\mathrm{^2T_1 \to {^2S_0}}$ IC rate than U-TDDFT, the two methods predict radically different branching ratios for the IC decay channels of CuP: while U-TDDFT predicts that 23.7\% of $\mathrm{^2T_1}$ decays directly to $\mathrm{^2S_0}$, X-TDDFT predicts a direct decay ratio of only 1.5\% (in both cases almost all of the remaining $\mathrm{^2T_1}$ decay to the ground state via the intermediacy of $\mathrm{^2dd_n}$). We therefore conclude that X-TDDFT can give very different pictures of the relaxation pathways of open-shell molecules compared to U-TDDFT, which further highlights the necessity of spin-adaptation in the study of IC processes.

\section{Conclusions and outlook} \label{sec:conclusion}

In conclusion, we have developed and implemented the analytic NACMEs of the X-TDDFT method, which is the first analytic NACME implementation of a spin-adapted TDDFT method, and completes the last piece (after X-TDDFT itself,\cite{XTDDFT1,XTDDFT2,XTDDFT3} X-TDDFT SOCMEs\cite{XTDDFT-SOC} and X-TDDFT analytic gradient\cite{XTDDFTgrad}) of our toolbox for studying arbitrary photophysical and photochemical processes of open-shell molecules using X-TDDFT. The X-TDDFT NACMEs can be implemented easily in any program with an existing implementation of U-TDDFT NACMEs, and do not involve significant time or memory overhead. Theoretical considerations and numerical tests reveal that while X-TDDFT NACMEs are sometimes similar to U-TDDFT NACMEs, it is equally common for them to differ considerably, resulting in drastically different IC rates. More importantly, even the relative trend of the IC rates, no matter for the same states of closely related molecules or for different states of the same molecule, can often change as a result of spin-adaptation. We therefore recommend to use X-TDDFT NACMEs for open-shell systems in general, even when the U-TDDFT and X-TDDFT excitation energies are similar (as is the case for the copper(II) porphyrins studied here). Future possibilities may include extending the present formalism to spin-flip excitations,\cite{XSF-TDA} as well as combining X-TDDFT with non-adiabatic molecular dynamics approaches to study the ultrafast phototophysics and photochemistry of open-shell systems.

\section*{Acknowledgement}

W.L. was supported by the National Natural Science Foundation of China (T2441001). Z.W.~acknowledges support by the Qilu Young Scholar project of Shandong University, as well as helpful discussions with Prof.~Zhendong Li (Beijing Normal University).

\section*{Data Availability Statement}

The data that support the findings of this study are available within the article.

\appendix

\section{Individual contributions to the U-TDDFT/X-TDDFT difference of ee TDMs} \label{sec:UXdiff}

In this Appendix, we derive the contributions of CO, OV, CV(0) and CV(1) excitations to the X-TDDFT correction of ee TDMs, Eq.~\eqref{gammaSAIJ}.

We first consider the sum of the two corrections
\begin{eqnarray}
& & \Delta\gamma_{pq\sigma}^{\mathrm{(1)}, IJ} + \Delta\gamma_{pq\sigma}^{\mathrm{CV(1)}, IJ} \nonumber\\
& = & \langle \Psi_0 | \frac{1}{2} ([ \Delta O_I, [a_{p\sigma}^\dag a_{q\sigma}, \Delta O_J^\dag] ] + [ [\Delta O_I, a_{p\sigma}^\dag a_{q\sigma}], \Delta O_J^\dag ]) | \Psi_0 \rangle \nonumber\\
& + & \left( \langle \Psi_0 | \frac{1}{2} ([ \Delta O_I, [a_{p\sigma}^\dag a_{q\sigma}, O_J^{\mathrm{CV(1)},\dag}] ] + [ [\Delta O_I, a_{p\sigma}^\dag a_{q\sigma}], O_J^{\mathrm{CV(1)},\dag} ]) | \Psi_0 \rangle \right. \nonumber\\
&  & + \left. \langle \Psi_0 | \frac{1}{2} ([ O_I^{\mathrm{CV(1)}}, [a_{p\sigma}^\dag a_{q\sigma}, \Delta O_J^\dag] ] + [ [O_I^{\mathrm{CV(1)}}, a_{p\sigma}^\dag a_{q\sigma}], \Delta O_J^\dag ]) | \Psi_0 \rangle \right). \label{deltagamma}
\end{eqnarray}
It is obvious that $\Delta\gamma_{pq\sigma}^{\mathrm{(1)}, IJ} + \Delta\gamma_{pq\sigma}^{\mathrm{CV(1)}, IJ}$ is only non-zero when $p\sigma$ and $q\sigma$ are both closed-shell orbitals or both vacant-shell orbitals, because both $\Delta O_J^\dag$ and $O_J^{\mathrm{CV(1)},\dag}$ excite closed-shell electrons to vacant-shell orbitals, and both $\Delta O_I$ and $O_I^{\mathrm{CV(1)}}$ de-excite vacant-shell electrons to closed-shell orbitals; orbitals that are not involved in these excitations and de-excitations give zero contributions after applying the nested commutators. Now consider the closed-closed matrix element $\Delta\gamma_{ij}^{\mathrm{(1)}, IJ} + \Delta\gamma_{ij}^{\mathrm{CV(1)}, IJ}$ as an example (the vacant-vacant block can be derived analogously). The first term $\Delta\gamma_{ij}^{\mathrm{(1)}, IJ}$ evaluates to
\begin{eqnarray}
\Delta\gamma_{ij}^{\mathrm{(1)}, IJ} & = & \frac{1}{4} \sum_{a} \langle \Psi_0 | \left(\sqrt{\frac{S}{S+1}}-1\right)^2 a_j^\dag a_a a_i^\dag a_j a_a^\dag a_i | \Psi_0 \rangle ((\mathbf{X}_I)_{ja} - (\mathbf{X}_I)_{\bar{j}\bar{a}}) ((\mathbf{X}_J)_{ia} - (\mathbf{X}_J)_{\bar{i}\bar{a}}) \nonumber\\
& + & \frac{1}{4} \sum_{a} \langle \Psi_0 | \left(\sqrt{\frac{S}{S+1}}-1\right)^2 a_j^\dag a_a a_i^\dag a_j a_a^\dag a_i  | \Psi_0 \rangle ((\mathbf{Y}_I)_{ia} - (\mathbf{Y}_I)_{\bar{i}\bar{a}}) ((\mathbf{Y}_J)_{ja} - (\mathbf{Y}_J)_{\bar{j}\bar{a}}) \nonumber\\
& = & - \frac{1}{4} \left(\sqrt{\frac{S}{S+1}}-1\right)^2 \sum_{a} \left( ((\mathbf{X}_I)_{ja} - (\mathbf{X}_I)_{\bar{j}\bar{a}}) ((\mathbf{X}_J)_{ia} - (\mathbf{X}_J)_{\bar{i}\bar{a}}) \right. \nonumber\\
& &  \left. + ((\mathbf{Y}_I)_{ia} - (\mathbf{Y}_I)_{\bar{i}\bar{a}}) ((\mathbf{Y}_J)_{ja} - (\mathbf{Y}_J)_{\bar{j}\bar{a}}) \right).
\end{eqnarray}
The result has the same structure as the U-RPA ee TDM (Eq.~\eqref{U_TDM_IJ1}): the overall minus sign comes from permuting the creation and annihilation operators. The contributions of all other creation and annihilation operators in Eq.~\eqref{DeltaOI_SA} vanish, because they only involve $\beta$ closed-shell orbitals but not $\alpha$ ones. The $\beta$ TDM matrix elements $\Delta\gamma_{\bar{i}\bar{j}}^{\mathrm{(1)}, IJ}$ are more complicated, because double excitations like $a_a^\dag a_{t} a_{\bar{t}}^\dag a_{\bar{i}}$ now contribute. The final result is
\begin{eqnarray}
\Delta\gamma_{\bar{i}\bar{j}}^{\mathrm{(1)}, IJ} & = & \frac{1}{4}\left(2\sqrt{\frac{S}{S+1}}-\frac{2S+3}{S+1}\right) \sum_{a} \left( ((\mathbf{X}_I)_{ja} - (\mathbf{X}_I)_{\bar{j}\bar{a}}) ((\mathbf{X}_J)_{ia} - (\mathbf{X}_J)_{\bar{i}\bar{a}}) \right. \nonumber\\
& & + \left. ((\mathbf{Y}_I)_{ia} - (\mathbf{Y}_I)_{\bar{i}\bar{a}}) ((\mathbf{Y}_J)_{ja} - (\mathbf{Y}_J)_{\bar{j}\bar{a}}) \right),
\end{eqnarray}
where each $a_{\bar{j}}^\dag a_{\bar{t}} a_{t}^\dag a_a a_i^\dag a_j a_a^\dag a_{u} a_{\bar{u}}^\dag a_{\bar{i}}$ term contributes $-\frac{1}{4S(S+1)}\delta_{tu}$ to the prefactor, and there are $2S$ open-shell orbitals, yielding the final prefactor 
\begin{equation}
- \frac{1}{4} \left(\sqrt{\frac{S}{S+1}}-1\right)^2-2S\frac{1}{4S(S+1)} = \frac{1}{4}\left(2\sqrt{\frac{S}{S+1}}-\frac{2S+3}{S+1}\right).
\end{equation}

Meanwhile, in $\Delta\gamma_{ij\sigma}^{\mathrm{CV(1)}, IJ}$, the contributions of all double excitations vanish, yielding the following result for both $\alpha$ and $\beta$ ee TDM elements:
\begin{eqnarray}
\Delta\gamma_{ij}^{\mathrm{CV(1)}, IJ} = \Delta\gamma_{\bar{i}\bar{j}}^{\mathrm{CV(1)}, IJ} & = & -\frac{1}{2}\left(\sqrt{\frac{S}{S+1}}-1\right) \sum_{a} \left( ((\mathbf{X}_I)_{ja} - (\mathbf{X}_I)_{\bar{j}\bar{a}}) ((\mathbf{X}_J)_{ia} - (\mathbf{X}_J)_{\bar{i}\bar{a}}) \right. \nonumber\\
& & + \left. ((\mathbf{Y}_I)_{ia} - (\mathbf{Y}_I)_{\bar{i}\bar{a}}) ((\mathbf{Y}_J)_{ja} - (\mathbf{Y}_J)_{\bar{j}\bar{a}}) \right).
\end{eqnarray}
Summing $\Delta\gamma_{ij}^{\mathrm{CV(1)}, IJ}$ with $\Delta\gamma_{ij}^{\mathrm{(1)}, IJ}$ yields Eq.~\eqref{gammaCV1_1},
i.e.~the $\alpha$ and $\beta$ TDMs receive exactly opposite corrections.
The vacant-vacant block of $\Delta\gamma^{\mathrm{(1)}, IJ} + \Delta\gamma^{\mathrm{CV(1)}, IJ}$ can be derived in a completely analogous way, yielding Eq.~\eqref{gammaCV1_2}.

The CV(0) contribution to Eq.~\eqref{gammaSAIJ}, $\Delta\gamma^{\mathrm{CV(0)}, IJ}$, can be derived in the same manner as $\Delta\gamma^{\mathrm{CV(1)}, IJ}$. But since $\Delta\gamma^{\mathrm{CV(0)}, IJ}$ changes sign with respect to swapping the $\alpha$ and $\beta$ indices (in contrast to $\Delta\gamma^{\mathrm{CV(1)}, IJ}$, which is invariant with respect to the swap), the $\alpha$ and $\beta$ components of $\Delta\gamma^{\mathrm{CV(0)}, IJ}$ differ by a minus sign, instead of being equal (Eq.~\eqref{gammaCV0_1} and Eq.~\eqref{gammaCV0_2}). Therefore, their contributions to the NACMEs cancel.

We finally derive the CO term $\Delta\gamma^{\mathrm{CO}, IJ}$, noting that the OV term can be calculated completely analogously:
\begin{eqnarray}
\Delta\gamma_{pq\sigma}^{\mathrm{CO}, IJ} & = & \langle \Psi_0 | \frac{1}{2} ([ \Delta O_I, [a_{p\sigma}^\dag a_{q\sigma}, O_J^{\mathrm{CO},\dag}] ] + [ [\Delta O_I, a_{p\sigma}^\dag a_{q\sigma}], O_J^{\mathrm{CO},\dag} ]) | \Psi_0 \rangle \nonumber\\
&  & + \langle \Psi_0 | \frac{1}{2} ([ O_I^{\mathrm{CO}}, [a_{p\sigma}^\dag a_{q\sigma}, \Delta O_J^\dag] ] + [ [O_I^{\mathrm{CO}}, a_{p\sigma}^\dag a_{q\sigma}], \Delta O_J^\dag ]) | \Psi_0 \rangle. \label{deltagamma_CO}
\end{eqnarray}
Since $\Delta O_I$, $\Delta O_J$, $\Delta O_I^\dag$ and $\Delta O_J^\dag$ only perform excitations or de-excitations between closed-shell and vacant-shell orbitals, and CO excitations only involve $\beta$ closed-shell and open-shell orbitals, Eq.~\eqref{deltagamma_CO} is zero except for the OV and VO blocks. Specifically, the CV($\beta\beta$) excitation $a_{\bar{a}}^\dag a_{\bar{i}}$ contributes to the $\beta$ matrix elements $\Delta\gamma_{\bar{t}\bar{a}}^{\mathrm{CO}, IJ}$ and $\Delta\gamma_{\bar{a}\bar{t}}^{\mathrm{CO}, IJ}$, while the double excitation $a_a^\dag a_{t} a_{\bar{t}}^\dag a_{\bar{i}}$ contributes to the $\alpha$ matrix elements  $\Delta\gamma_{ta}^{\mathrm{CO}, IJ}$ and $\Delta\gamma_{at}^{\mathrm{CO}, IJ}$:
\begin{eqnarray}
\Delta\gamma_{\bar{t}\bar{a}}^{\mathrm{CO}, IJ} & = & -\frac{1}{2} \left(\sqrt{\frac{S}{S+1}}-1\right) \sum_{a} \left( \langle \Psi_0 |  a_{\bar{i}}^\dag a_{\bar{t}} a_{\bar{t}}^\dag a_{\bar{a}} a_{\bar{a}}^\dag a_{\bar{i}} | \Psi_0 \rangle (\mathbf{X}_I)_{\bar{i}\bar{t}} ((\mathbf{X}_J)_{ia} - (\mathbf{X}_J)_{\bar{i}\bar{a}}) \right. \nonumber\\
& & + \left. \langle \Psi_0 | a_{\bar{i}}^\dag a_{\bar{t}} a_{\bar{t}}^\dag a_{\bar{a}} a_{\bar{a}}^\dag a_{\bar{i}} | \Psi_0 \rangle (\mathbf{Y}_J)_{\bar{i}\bar{t}} ((\mathbf{Y}_I)_{ia} - (\mathbf{Y}_I)_{\bar{i}\bar{a}}) \right) \nonumber\\
& = & -\frac{1}{2} \left(\sqrt{\frac{S}{S+1}}-1\right) \times \nonumber\\
& & \sum_{a} \left( (\mathbf{X}_I)_{\bar{i}\bar{t}} ((\mathbf{X}_J)_{ia} - (\mathbf{X}_J)_{\bar{i}\bar{a}}) + (\mathbf{Y}_J)_{\bar{i}\bar{t}} ((\mathbf{Y}_I)_{ia} - (\mathbf{Y}_I)_{\bar{i}\bar{a}}) \right), \\
\Delta\gamma_{ta}^{\mathrm{CO}, IJ} & = & -\frac{1}{2\sqrt{S(S+1)}} \sum_{a} \left( \langle \Psi_0 |  a_{\bar{i}}^\dag a_{\bar{t}} a_t^\dag a_a a_a^\dag a_t a_{\bar{t}}^\dag a_{\bar{i}} | \Psi_0 \rangle (\mathbf{X}_I)_{\bar{i}\bar{t}} ((\mathbf{X}_J)_{ia} - (\mathbf{X}_J)_{\bar{i}\bar{a}}) \right. \nonumber\\
& & + \left. \langle \Psi_0 |  a_{\bar{i}}^\dag a_{\bar{t}} a_t^\dag a_a a_a^\dag a_t a_{\bar{t}}^\dag a_{\bar{i}} | \Psi_0 \rangle (\mathbf{Y}_J)_{\bar{i}\bar{t}} ((\mathbf{Y}_I)_{ia} - (\mathbf{Y}_I)_{\bar{i}\bar{a}}) \right) \nonumber\\
& = & -\frac{1}{2\sqrt{S(S+1)}} \times \nonumber\\
& & \sum_{a} \left( (\mathbf{X}_I)_{\bar{i}\bar{t}} ((\mathbf{X}_J)_{ia} - (\mathbf{X}_J)_{\bar{i}\bar{a}}) + (\mathbf{Y}_J)_{\bar{i}\bar{t}} ((\mathbf{Y}_I)_{ia} - (\mathbf{Y}_I)_{\bar{i}\bar{a}}) \right).
\end{eqnarray}
The VO elements $\Delta\gamma_{\bar{a}\bar{t}}^{\mathrm{CO}, IJ}$ and $\Delta\gamma_{at}^{\mathrm{CO}, IJ}$ are obtained similarly.

The OV contributions $\Delta\gamma_{pq\sigma}^{\mathrm{OV}, IJ}$ can be calculated analogously by swapping the roles of $\alpha$ particles and $\beta$ holes, as well as between $\beta$ particles and $\alpha$ holes. Similarly, only the CO and OC blocks of $\Delta\gamma_{pq\sigma}^{\mathrm{OV}, IJ}$ survive.


\bibliography{XTDDFTNAClib}

\end{document}